\documentclass[aps,prb,nopacs,floatfix,12pt]{revtex4}

\usepackage{graphicx}
\usepackage{amssymb}
\usepackage{amsmath}
\usepackage{bm}
\usepackage{mathtools}
\usepackage{color}
\usepackage{makecell}   
\usepackage{mathrsfs}
\usepackage{amsfonts}
\usepackage{textcomp}
\usepackage{rotating}
\usepackage{comment}
\usepackage{rotating}

\newcommand{\bra}{\langle}
\newcommand{\ket}{\rangle}
\newcommand{\gr}[1]{\mathbf{#1}}

\newcommand{\Eph}{E_\mathrm{ph}}

\newcommand{\QQ}{\mathbf{Q}}
\newcommand{\RR}{\mathbf{R}}
\newcommand{\bmu}{\boldsymbol{\mu}}
\newcommand{\mini}{\mathrm{min}}

\newcommand{\MatVar}[1]{\ensuremath{\underline{#1}}}%

\begin{document}
\bibliographystyle{jcp}

\title{Photodissociation dynamics in the
first absorption band of pyrrole: I. Molecular Hamiltonian and the 
Herzberg-Teller absorption spectrum
for the $^1\!A_2(\pi \sigma^*) \leftarrow \widetilde{X}^1\!A_1(\pi\pi)$
transition}  
\author{David Picconi\footnote{Electronic mails:\,\,David.Picconi@ch.tum.de and david.picconi@gmail.com}}
\author{Sergy\ Yu.\ 
Grebenshchikov\footnote{Electronic mails: Sergy.Grebenshchikov@ch.tum.de and sgreben@gwdg.de}}
\affiliation{Department of Chemistry, Technical University of Munich,
  Lichtenbergstr. 4, 85747 Garching, Germany}

\begin{abstract}
This paper opens a series in which 
the photochemistry of the two lowest $\pi \sigma^*$ states of pyrrole
and their interaction with each other and with the ground electronic state
$\widetilde{X}$ are studied using 
 ab initio quantum mechanics. New 24-dimensional potential
energy surfaces 
for the photodissociation of the N---H bond 
and the formation of the pyrrolyl radical 
are calculated using the CASPT2 method 
for the electronic states 
$\widetilde{X}(\pi\pi)$, $1^1\!A_2(\pi\sigma^*)$ and
 $1^1\!B_1(\pi \sigma^*)$ 
and locally diabatized. 
In the first paper, the ab initio calculations are described and 
the photodissociation in the state
$1^1\!A_2(\pi\sigma^*)$ is analyzed. 
The excitation $1^1\!A_2 \leftarrow \widetilde{X}$
is mediated by the coordinate dependent transition dipole moment functions
constructed using the Herzberg-Teller expansion. 
Nuclear dynamics, including 6, 11, and 15 active degrees of freedom, 
is studied using  the 
multi-configurational time-dependent Hartree method.
The focus is on the frequency resolved 
absorption spectrum, as well as on the dissociation time scales and the 
resonance lifetimes. 
Calculations are compared with available experimental data. An approximate
convolution method is developed and validated, with which absorption 
spectra can be calculated and assigned in terms of vibrational quantum
numbers. The method represents the total absorption spectrum as
a convolution of the diffuse spectrum of the detaching H-atom and the
Franck-Condon spectrum of the heteroaromatic ring. Convolution calculation
requires a minimal quantum chemical input and is a promising tool for 
studying the
 $\pi \sigma^*$ photodissociation in model biochromophores.

\end{abstract}

\maketitle

\section{Introduction}
\label{intro}

This and the subsequent two papers 
(termed \lq Paper II'\cite{PG17B} and 
\lq Paper III'\cite{PDG17}) describe the results of
an ab initio quantum dynamical study of the absorption spectra, non-adiabatic
dynamics, and the photofragment distributions of pyrrole, C$_4$H$_4$NH, 
photoexcited with ultraviolet
(UV) light with the wavelengths $\lambda$ between
250\,nm and 220\,nm. Pyrrole, which acts as a chromophore unit in porphyrins 
and in the amino acid tryptophan,  is a characteristic 
example of a molecule exhibiting the so-called $\pi\sigma^*$ 
photochemistry.\cite{SDDJ02,MKAD16} 
Photodissociation in the considered wavelength range
occurs in the first two excited singlet electronic states of 
$^1\pi\sigma^*$ character which are 
dissociative along the N---H bond and form conical intersections  
with the ground electronic state. 
This paper gives our motivation behind the
study, describes the construction of the full-dimensional molecular
Hamiltonian for the three electronic states involved, 
and discusses the absorption spectra due to the lowest $^1\pi\sigma^*$ state.
Paper II focuses on the photofragment 
distributions in this state. Paper III analyzes the non-adiabatic 
photodissociation dynamics in the two $^1\pi\sigma^*$ states coupled to the
ground electronic state. 

The atomistic mechanisms of non-radiative decay of the initial electronic
excitation in aromatic molecules, serving as models of
broad classes of UV biochromophores, are actively studied using
experiment and theory (see Refs.\
\onlinecite{WKRRT03,LRHR04,DCNA06,HWRBPS11,MCOFCL12,CBYEU11,RWPUS12,ADDKNO08,ACDDN06,NDCDA06,NDCA06,GYBU15,OKA11,ZTBPT15,AKMNOS10,RS14,MKAD16,SDDJ02,LDVSM05,VLMSD05,BLSM09,YXZT14,ZY16,XMZYXG16,BKPMKTS10,KPNWF17} and references therein).
In the gas phase, the internal energy of the UV excited pyrrole is large
enough to allow 
dissociation into several chemically distinct channels. One group of
channels is populated on a nanosecond time scale\cite{MICHLBONACIC90} and
is reached via electronic relaxation to the
ground electronic state $\widetilde{X}$ along the ring-deformation and ring-opening
pathways.\cite{BNL94,WKRRT03,RS14,BLSM09,ZTBPT15,MKAD16} In contrast,
dissociation into hydrogen atom and pyrrolyl radical via
$^1\pi\sigma^*$ electronic states occurs on a sub-picosecond time 
scale.\cite{LRHR04,ACDDN06,NDCDA06,AKMNOS10,OKA11,RS14}  
The $^1\pi\sigma^*$ states often lack oscillator strength (especially
if compared to the optically \lq bright' $^1\pi\pi^*$ states), but
their potential energy surfaces are repulsive or only weakly bound, and this
topographic feature allows them to control
ultrafast electronic relaxation pathways.\cite{SDDJ02,AKMNOS10,RS14}
 The $^1\pi\sigma^*$ state mediated 
H-atom detachment in heteroaromatic molecules 
is an actively expanding research field, and 
a range of powerful spectroscopic techniques --- 
both high resolution frequency resolved\cite{AKMNOS10} and 
ultrafast time resolved\cite{RS14,S14,ZTBPT15,KPNWF17} 
--- are applied to monitor
the reaction fragments. The experimental studies are complemented by
ab initio theoretical methods\cite{SDDJ02,LDVSM05,CBYEU11,MSMS15,
XMZYXG16,KPNWF17} 
providing a convincing 
interpretation in terms of the electronic structure and quantum 
dynamics calculations.

\begin{figure}[h!]
\caption{(a) Overview of the experimental absorption spectrum in the 
first absorption band of pyrrole (adapted from Ref. 
\onlinecite{RWYCYUS13}). (b) 
Theoretical total absorption spectrum due to the 
transition $1^1\!A_2(\pi \sigma^*) \leftarrow \widetilde{X}^1\!A_1(\pi\pi)$
calculated as described in Sect. \ref{res_abs_3} and corresponding to the 
energy range to the left of the dashed line in panel (a). 
} 
\label{F00}
\bigskip
\includegraphics[scale=0.33]{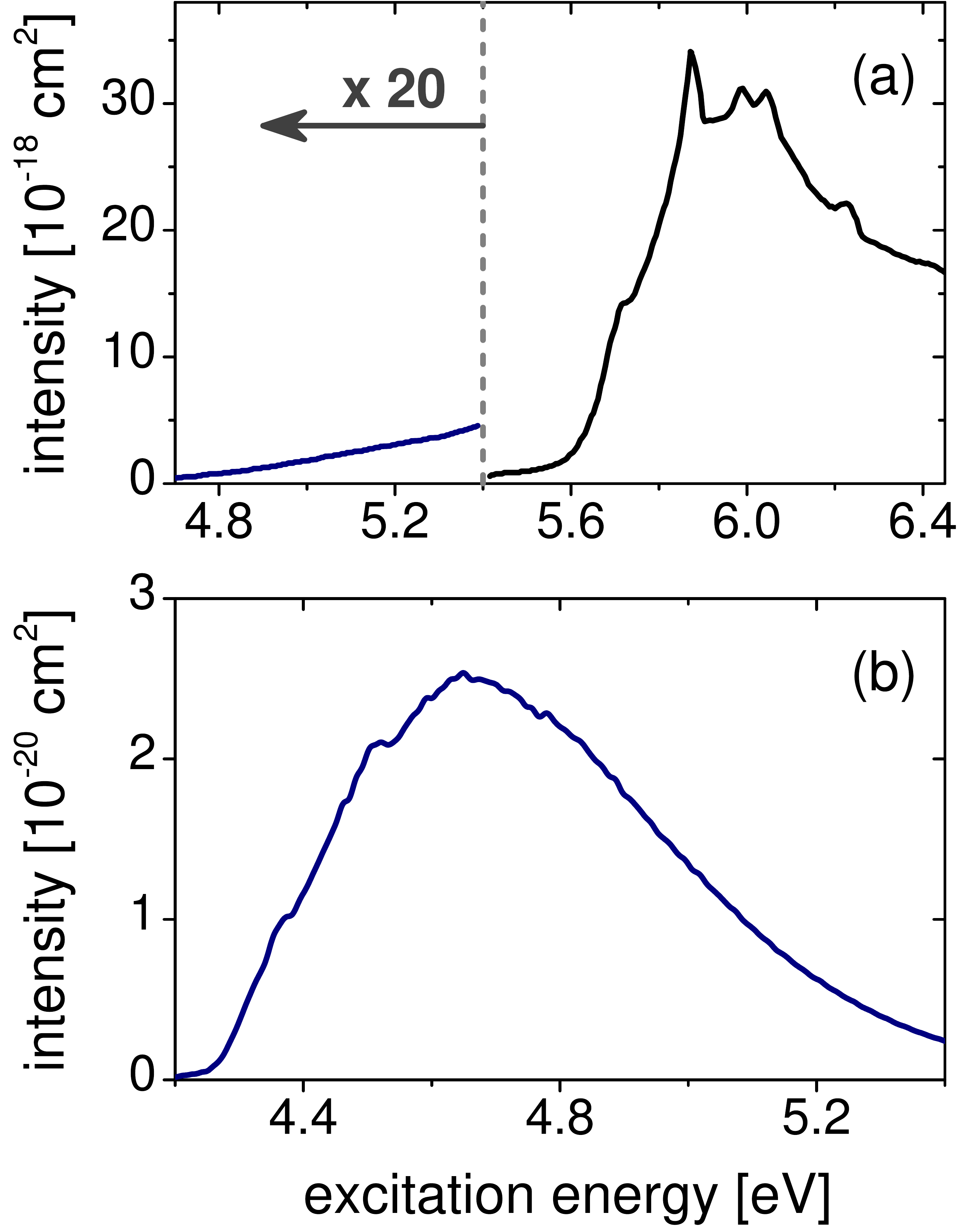}
\end{figure}

\begin{figure}[h!]
\caption{One-dimensional potential energy cuts of the diabatic states
$1^1\!A_1$ (blue), $1^1\!A_2$ (red), $1^1\!B_1$ (green)
as functions of Jacobi coordinates $R$ (a,b), $\theta$ (c), and
$\phi$ (d) on the ab initio grid (indicated with dots). 
Cuts at the ground state 
pyrrole equilibrium [panel (a)] and at the pyrrolyl fragment equilibrium
 [panel (b)] are shown. Conical intersections are marked with
circles, and the respective local diabatic couplings are exemplified with
dashed lines (enlarged --- not to scale). 
In panel (a), the vibrational ground state wave function in $\widetilde{X}$
is sketched. In panels (c) and (d), $R = 4.15\,a_0$ and the 
vibrational modes of pyrrolyl are fixed to their values on the MEP in
the $1^1\!A_2$ state. The complementary angles in (c) and (d) are
fixed to $\phi = 0^\circ$ in (c) and $\theta = 45^\circ$ in (d). 
} 
\label{F01}
\bigskip
\includegraphics[scale=0.47]{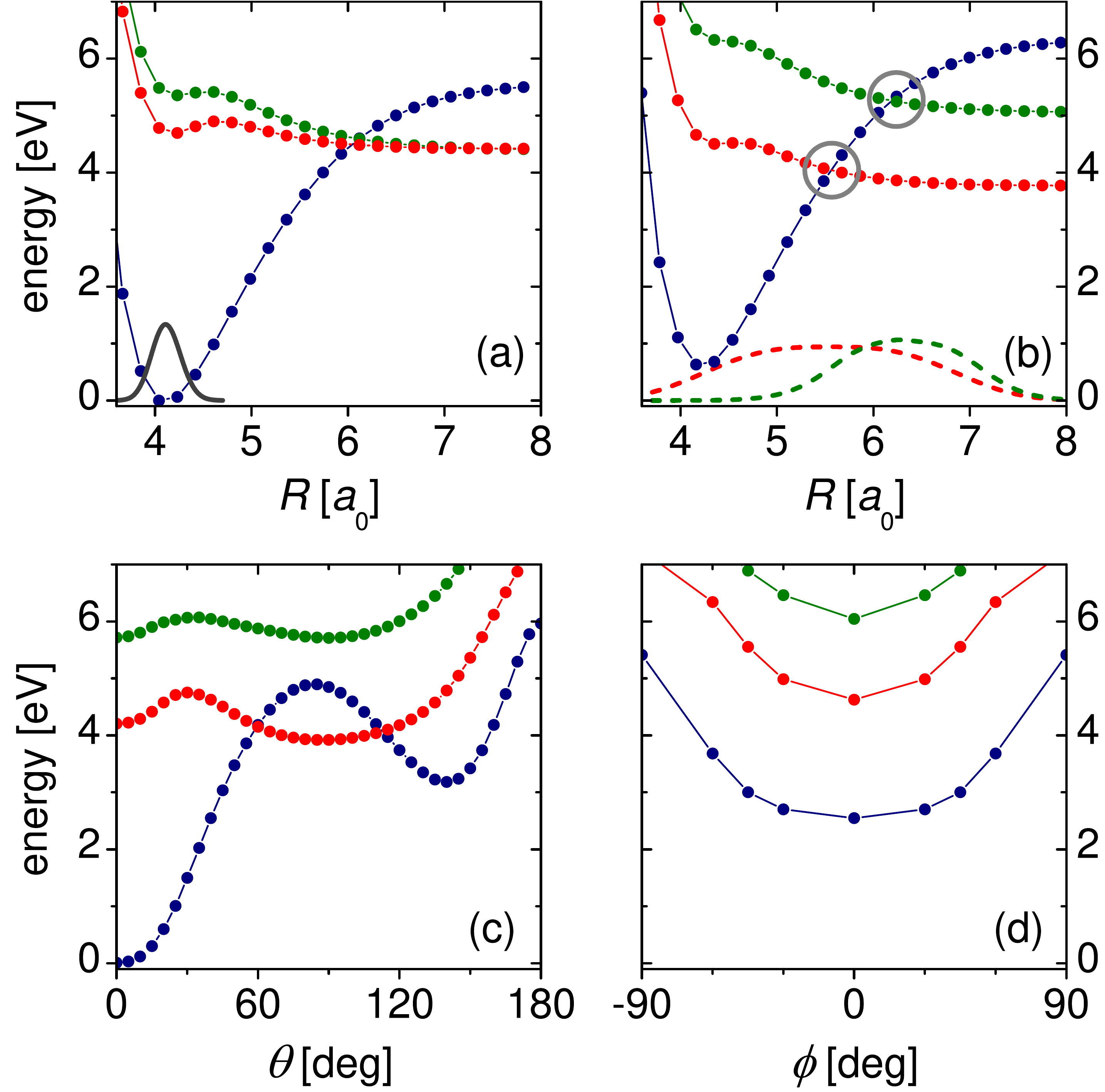}
\end{figure}

The first absorption band of pyrrole [see Fig.\ \ref{F00}(a)], 
extending  from 5.0\,eV to 
6.5\,eV, was extensively studied in the past. While its complete
electronic assignment remains a
subject of hot discussions, most researches currently 
agree that the weak structureless
absorption at the longest wavelengths, highlighted
in the left part of panel (a), is due to the 
two lowest excited $^1\pi\sigma^*$ states. Their one-dimensional (1D) 
potential energy curves along the dissociation coordinate are
illustrated in Fig.\ \ref{F01}(a,b). The lowest excited singlet
 state $1^1\!A_2(\pi \sigma^*)$ (with the vertical excitation
energy $T_v$ of about 5.0\,eV) 
arises mainly from the promotion of an electron from the $1a_2(\pi)$ to the
$10a_1(3s/\sigma^*)$ Rydberg 
molecular orbital; the transition from the ground
electronic state $\widetilde{X}^1\!A_1(\pi\pi)$ 
(hereafter referred to as $\widetilde{X}$) is electric dipole forbidden and is
accomplished via the vibronic intensity borrowing. The second
$\pi\sigma^*$ state, located $\sim 1.0$\,eV above $1^1\!A_2$,
is the state $1^1\!B_1(\pi\sigma^*)$. It originates mainly from the 
$2b_1(\pi) \rightarrow 10a_1(3s/\sigma^*)$ orbital excitation and, although
its excitation from $\widetilde{X}$ is electric dipole allowed, the oscillator
strength of this transition is tiny. It is therefore not 
surprising that the absorption cross 
sections of both $\pi\sigma^*$ states are small. Panel (b) of Fig.\ \ref{F00}
illustrates the absorption band calculated in this work
for the state $1^1\!A_2(\pi \sigma^*)$ 
coupled to $\widetilde{X}$. The 
calculated peak intensity at 4.6\,eV 
does not exceed $2.5\times 10^{-20}$\,cm$^{2}$, and in experiment the
band is overlayed by the shoulder of a much stronger absorption of the
higher lying $^1\pi\pi^*$ states.\cite{CNQA04,NW14} The absorption of the
state $1^1\!B_1(\pi \sigma^*)$, discussed in paper III and not shown in 
Fig.\ \ref{F00}(b), is somewhat stronger (it 
reaches $10^{-19}$\,cm$^{2}$ around 5.8\,eV)  but is still very weak.

The $\pi \sigma^*$ states in Fig.\ \ref{F01}(a,b)
are repulsive with respect to
the extension of the N---H bond; the state $1^1\!A_2(\pi \sigma^*)$ correlates 
with the photofragments in their electronic ground states, 
${\rm H}(^1\!S) + {\rm pyrrolyl}(\widetilde{X}^2\!A_2)$, while the state 
$1^1\!B_1(\pi\sigma^*)$ correlates with the first excited state
$1^2\!B_1$ of pyrrolyl.  Both $\pi\sigma^*$ states  
feature shallow local minima in the Franck-Condon (FC) zone separated
from the asymptotic region by low barriers, indicating that a tunneling 
contribution to the dissociation can be 
expected.\cite{VLMSD05,WNSSAWS15,XDFG16}
In the exit channel, away from the FC zone, 
both $\pi \sigma^*$ states 
form conical intersections (CIs) with the ground electronic state;
the CIs are marked with circles in Fig.\ \ref{F01}(a,b). 
The intersections are symmetry allowed: The exact degeneracies are found 
in $C_{2v}$ symmetric configurations and involve states belonging to
different irreps of the $C_{2v}$ point group. 
The totally symmetric dissociation 
coordinate $R$ acts as a tuning mode, while the coupling modes have $a_2$
symmetry for the $\widetilde{X}/A_2$ CI and $b_1$ symmetry for the
$\widetilde{X}/B_1$ CI. These CIs, predicted
by Sobolewski, Domcke, and co-workers on the basis of general symmetry
arguments,\cite{SDDJ02} are the most salient features of the
$\pi\sigma^*$ states in pyrrole and, as argued in Ref.\ \onlinecite{MKAD16}, 
in many other model biochromophores. Located in the 
exit dissociation channel, these CIs strongly affect 
the molecular photoreactivity and, in particular, 
influence the vibrational state distributions and the kinetic energy release
of the photofragments.\cite{PG14,PG15} In a recent quantum mechanical study,
we discovered their fingerprints in the absorption  spectra
as strongly asymmetric Fano resonances.\cite{GP17A} 

With the aim to understand the dynamics at 
these exit channel CIs, 
the weakest portion of the first absorption band, located 
below 5.7\,eV, has been extensively studied 
in the frequency- and time domain by 
many experimental and theoretical groups. 
The H-atom Rydberg tagging photofragment 
translational spectroscopy studies of 
Ashfold and co-workers\cite{CNQA04,CDNA06,AKMNOS10}
interrogated the formation of the fragment hydrogen atom 
for a series of photolysis wavelengths $\lambda$ between 254\,nm and 190\,nm. 
The key observable in these experiments 
is the total kinetic energy release (TKER) in the 
photodissociation reaction. In 
the two-fragment channel ${\rm H} + {\rm pyrrolyl}$, the TKER spectra are
equivalent to the rovibrational distributions of the pyrrolyl radical. 
The observed kinetic energy distributions are bimodal,\cite{CNQA04} 
with the fast (average kinetic energy $E_{\rm kin} \ge 4000$\,cm$^{-1}$)
and the slow ($E_{\rm kin} \sim 1000$\,cm$^{-1}$)
components well resolved for most
wavelengths. The angular distributions of the fast
fragments are typically
anisotropic (non-zero recoil anisotropy parameter), while the
slow components correspond to isotropically distributed
fragments. These observations led experimentalists
\cite{AKMNOS10,CNQA04}
to associate the fast products with the direct dissociation
in the excited $\pi\sigma^*$ 
electronic states and to assume that the slow products
emerge as a result of statistical decomposition in
the ground electronic state reached via internal conversion. 
In line with this assignment, two well separated 
dissociation time constants were established in the 
pump-probe measurements performed 
close to the absorption origin near 
$\lambda = 250$\,nm:\cite{LRHR04} The time  
$\tau_d = 110$\,fs was  attributed to the direct reaction, 
and the time $\tau_s = 1.1$\,ps was associated with the statistical 
dissociation. Recent ultrafast time-resolved experiments at the same 
wavelength of 250\,nm  confirmed the fast time scale 
(the reported value was $\tau_d =126$\,fs) and also argued that this is a 
tunneling lifetime.\cite{RWYCYUS13} The same group
found a much shorter dissociation time of $\tau_d = 46$\,fs for 
$\lambda = 238$\,nm. Even shorter lifetimes  
of $19$\,fs and $12$\,fs were found 
for $\lambda = 236$\,nm and 242\,nm, respectively, in the 
time resolved photoelectron measurements of Wu et al.\cite{WNSSAWS15} 
The measured dependence of $\tau_d$ on the photolysis wavelength is 
non-monotonous, and this conclusion is reinforced by the recent 
time resolved study of Ref.\ \onlinecite{KPNWF17}. 

Theoretical studies of the electronic structure, spectroscopy, and
photochemistry of pyrrole are numerous. Accurate multireference ab
initio calculations of vertical excitation energies
clarified the ordering of the low lying valence and Rydberg
excited states.\cite{VLMSD05,BVAEML06,FVSELK11,RMMSM02,CW03,SBL09}
The branching
ratios of several arrangement channels, including ring deformation
and ring opening reactions, were simulated using 
classical trajectory surface hopping algorithm.\cite{BVAEML06,SBL09}  
Multi-dimensional multi-state quantum dynamics
investigations of the ultrafast
electronic population dynamics in the excited states of
pyrrole were performed by K\"oppel, Lischka, and co-workers.\cite{FVSELK11}
Recently, Neville and Worth\cite{NW14} studied the first absorption band
using ab initio Hamiltonian constructed
as an extension of the quadratic vibronic coupling model and
including the first seven electronic states. 
These studies yielded important insight into the
radiationless decay dynamics of pyrrole, provided new
vibronic assignments of
the intense features between 5.5\,eV and 6.5\,eV,
and gave theoretical estimates of the dissociation lifetimes in the 
$1^1\!A_2(\pi\sigma^*)$ state\cite{NW14,WNSSAWS15}  
between 35\,fs and 133\,fs, in general agreement
with experiment.

Despite these efforts, 
several key questions related to the $\pi\sigma^*$ states of pyrrole 
require further investigation. In the frequency
domain, the absorption spectrum of the two lowest $\pi\sigma^*$ states 
remains virtually unexplored; the only published spectra of these states were
synthesized by Roos et al. from harmonic frequencies and vertical
excitation energies 15 years ago.\cite{RMMSM02} The wavelength resolved TKER 
distributions measured by Ashfold and co-workers\cite{CNQA04} 
have neither been calculated nor assigned theoretically. 
In the time domain, the tunneling contribution to the sub-picosecond 
dissociation in the $1^1\!A_2(\pi\sigma^*)$ state has not been quantified. 
A univocal 
assignment of the measured reaction time scales remains outstanding, and 
the fluctuations in the dependence of the measured dissociation times 
$\tau_d$ on the photon energy are not explained.

We recently launched a systematic investigation of the photodissociation 
of pyrrole photoexcited into the states $1^1\!A_2(\pi\sigma^*)$  and
$1^1\!B_1(\pi\sigma^*)$. The primary goal is to provide a
comprehensive picture of the photodissociation dynamics: 
To analyze the weak diffuse absorption bands, to assign them in terms of
the vibrational quantum numbers, 
to evaluate the photodissociation time scales, 
and to compare the resulting 
TKER distributions with the frequency resolved measurements of
Ashfold and co-workers. In this opening paper of the series, 
new high level ab initio calculations are described which  are
performed on a coordinate grid uniformly
covering the inner FC zone and the asymptotically separated H-atom
and pyrrolyl. These calculations serve as a basis for construction of a
new 24-dimensional (24D)
molecular Hamiltonian of pyrrole in a local quasi-diabatic representation.
The Hamiltonian is next used in the quantum dynamical calculations of
photodissociation. 
The Hamiltonian is inspired by the reaction path 
formalism\cite{MHA80,CM84} extended to three electronic states; 
the constructed diagonal quasi-diabatic potentials are chosen 
 harmonic only in the degrees of freedom of the pyrrolyl fragment. 
Further, the coordinate dependences of the transition dipole
moment (TDM) vectors are explicitly included within the framework of the
Herzberg-Teller expansion.\cite{BUNKERJENSEN06} In this way, 
direct excitation of these optically dark states is enabled. Paper I 
focuses on the absorption spectrum of the state 
$1^1\!A_2(\pi\sigma^*)$, both isolated and coupled to the ground electronic
state $\widetilde{X}$. The TKER distributions in this state are analyzed in 
paper II.\cite{PG17B} It is 
paper III\cite{PDG17} which extensively deals with the three-state 
effects and with the photodissociation dynamics in the 
second excited state $1^1\!B_1(\pi\sigma^*)$. 

The second, methodological, goal of this study is the development of 
simplified computational
schemes within which the diffuse absorption in the repulsive $\pi\sigma^*$ 
states and the subsequent sub-picosecond formation of the photofragments can
be quantitatively analyzed. The aim is to construct computational tools 
for dissociating systems which require a minimal ab initio
input and a numerical effort not exceeding that of a standard FC factor
calculation of bound-bound transitions in polyatomic molecules. 
In paper I, a 
convolution approximation for the absorption spectra is introduced, which 
casts the total spectrum as a convolution of 
the absorption spectra due to the departing H-atom and due to the 
pyrrolyl ring. In paper II, a related overlap integral-based mapping 
method for the TKER distributions is proposed. 

The remainder of the paper is organized as follows: The details
of the electronic structure calculations and 
the design of the 
24D molecular Hamiltonian comprising the states $\widetilde{X}$, 
$1^1\!A_2(\pi\sigma^*)$,  and $1^1\!B_1(\pi\sigma^*)$ are presented
in Sect.\ \ref{const}. The quantum mechanical methods and the convolution 
approximation are summarized in Sect.\ \ref{qm} and in two Appendices. 
The resulting absorption spectra are discussed in Sect.\ \ref{res_abs}. 
While the emphasis here is on the dissociation
in the isolated state $1^1\!A_2(\pi\sigma^*)$, dissociation in the
coupled pair $\widetilde{X}/1^1\!A_2(\pi\sigma^*)$ is also considered and the
effects inherent to the two-state dynamics are highlighted. 
Conclusions are given  in Sect.\ \ref{sum}. 


\section{Construction of the molecular Hamiltonian}
\label{const}

\subsection{The form of the 24 dimensional Hamiltonian}
\label{form24}

\begin{figure}[h!]
\caption{The SA-CASSCF(8,7)/CASPT2/aug-cc-pVTZ+ equilibrium geometries of the ground states of pyrrole and pyrrolyl are shown in panel (a) and (b). 
Bond distances are in $a_0$. Also shown are the Cartesian
coordinate system used in the calculation of the TDMs (panel (a); the axis $x$
is perpendicular to the molecular plane)
and the definition of the disappearing modes $R$ and $\theta$
[panel (b)]. The 
azimuthal angle $\phi$ is defined relative to the $x$ axis in the $xy$ 
plane. $(\theta,\phi) = (0^\circ,0^\circ)$ corresponds to the H-atom on the
axis $z$; $(\theta \ne 0,\phi=0^\circ)$ describes the H-atom in the plane
$xz$;  $(\theta \ne 0,\phi=90^\circ)$ describes the H-atom in the plane
$yz$.  
} 
\label{F01A}
\bigskip
\includegraphics[scale=0.39]{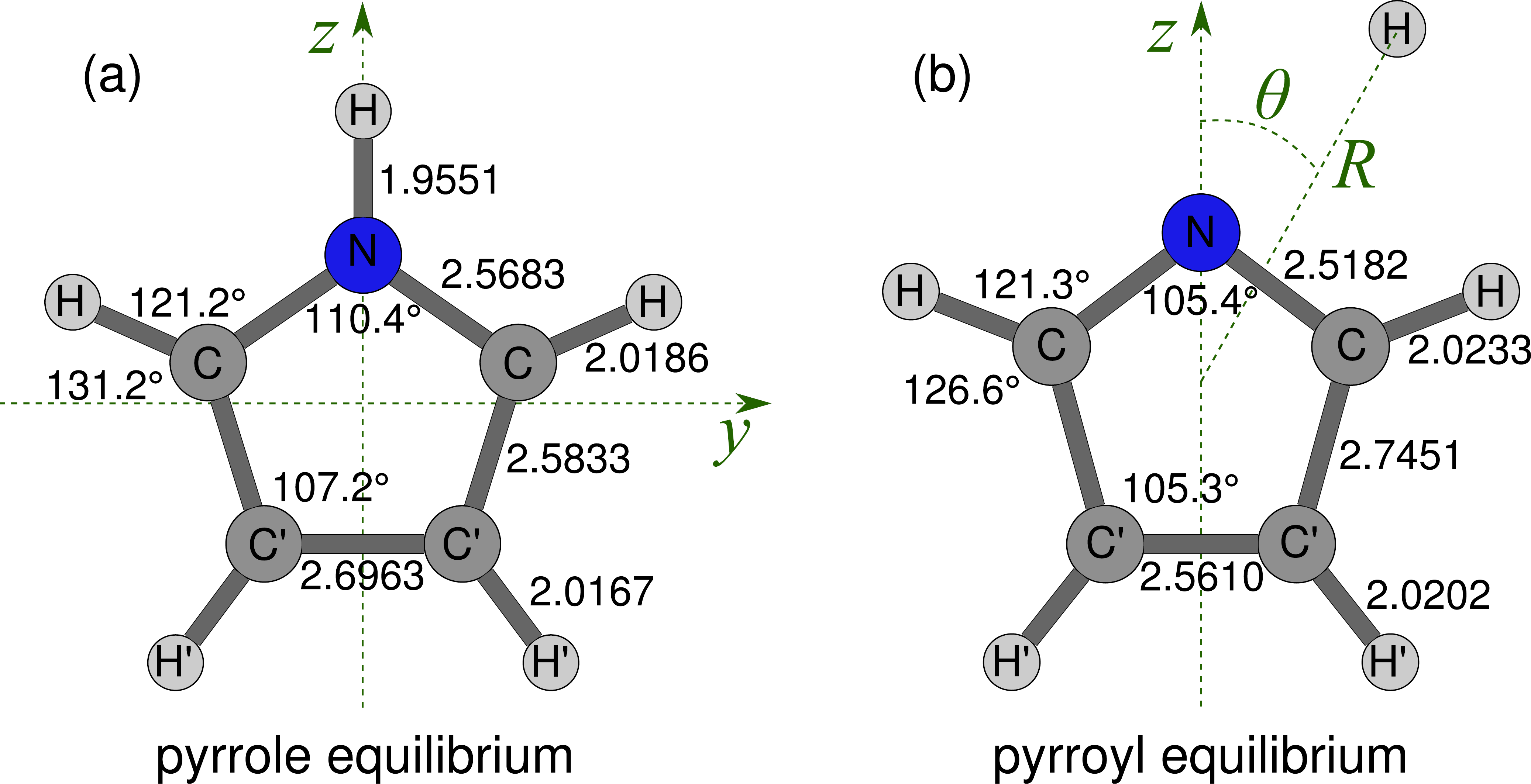}
\end{figure}

The molecular Hamiltonian,
\begin{equation}
  \hat{\MatVar{{\bf H}}} = \hat{T}\,\MatVar{{\bf 1}} +
\left( \begin{array}{ccc} V^{X} & V^{X A_2} & V^{X B_1} \\
  V^{X A_2} & V^{A_2} & V^{A_2 B_1} \\
  V^{X B_1} & V^{A_2 B_1} & V^{B_1} \end{array} \right) \, ,
\label{Ham}
\end{equation}
is set in the basis of three
locally diabatic electronic states $\widetilde{X}$ 
(its $C_{2v}$ symmetry label is $A_1$), 
$A_2$, and $B_1$ illustrated in Fig.\ \ref{F01}(a,b). Bold faced
underlined symbols are used for $3\times 3$ matrices ($\MatVar{{\bf 1}}$ is the
unit matrix). Pyrrole is described using (i)
three Jacobi coordinates $\RR \equiv (R,\theta,\phi)$ of the
dissociating H-atom with respect to the center of mass of the pyrrolyl
fragment [the so-called \lq disappearing modes'; see Fig. \ref{F01A}(b)]
and (ii) 21
dimensionless normal modes $\QQ$ of pyrrolyl, calculated at the
equilibrium geometry of the fragment (the so-called \lq
non-disappearing modes').  The normal modes $\QQ$ are partitioned into
four blocks according to the irreps $\Gamma$ of
the $C_{2v}$ point group, $\QQ = \{\QQ_{a_1}, \QQ_{a_2}, \QQ_{b_1},
\QQ_{b_2} \}$.

The kinetic energy operator in Eq.\ (\ref{Ham})
is set for the zero total angular momentum of pyrrole
in the frame of  the body-fixed principal axes 
(atomic units are used hereafter): 
\begin{eqnarray}
\hat{T} & = & \hat{T}_R + \hat{T}_Q \nonumber \\
& = & - \frac{1}{2m_R}\frac{\partial^2}{\partial R^2}
+ \frac{\hat{{\bm j}}^2}{2 m_R R^2} +
\frac{1}{2}\left(\frac{\hat{j}_x^2}{I_x} +
\frac{\hat{j}_y^2}{I_y} + \frac{\hat{j}_z^2}{I_z} \right)
-\frac{1}{2} \sum_{\substack{\Gamma = a_1, a_2, \\ \ \ b_1, b_2}} {\sum_{i}}^{\Gamma} \omega_\Gamma(i)
\frac{\partial^2}{\partial {Q_\Gamma(i)}^2} \, .
\label{T_op}
\end{eqnarray}
The kinetic energy $\hat{T}_R$ of the disappearing modes comprises the
first three terms including 
the kinetic energy of the relative
motion of H-atom and pyrrolyl ($m_R$ is the corresponding reduced mass),  
the orbital motion of the H-atom, and
the rotational motion of the rigid
pyrrolyl ring, respectively; $\hat{\bm{j}} =
(\hat{j}_x, \hat{j}_y, \hat{j}_z)$ is the pyrrolyl angular momentum 
operator and the inertia
constants $I_x$, $I_y$ and $I_z$ are evaluated at pyrrolyl equilibrium. 
The term $\hat{T}_Q$ refers to pyrrolyl vibrations; 
the sum $\sum_i^\Gamma$ is over the vibrational modes $i$
belonging to an irrep $\Gamma$.
The $C_{2v}$ symmetric pyrrolyl ring lies in the $yz$-plane, with $z$
being the $C_2$ axis (see Fig. \ref{F01A}).

The quasi-diabatic potential energy matrix in Eq.\ (\ref{Ham})
is constructed as a sum of two matrices:
\begin{equation}
\MatVar{{\bf V}}(\RR,\QQ) = \MatVar{{\bf U}}_R(\RR) +  
\MatVar{{\bf W}}_Q(\QQ|R) \, ;
\label{V_gen}
\end{equation}
Elements of the matrix  $\MatVar{{\bf U}}_R(\RR)$ 
depend on the three disappearing modes
$(R,\theta,\phi)$ only; they are spline
interpolations of the quasi-diabatized energies on a dense ab initio 
three-dimensional (3D) grid in $\RR$.  Elements of the 
matrix $\MatVar{{\bf W}}_Q(\QQ|R)$ depend on 
the 21 non-disappearing modes $\QQ$. These 21D functions are constructed 
in the spirit of the vibronic coupling
model of Ref.\ \onlinecite{KDC84}. The pyrrolyl ring is treated using
quadratic Hamiltonians, and the parameters of the vibronic coupling model
depend on the interfragment
distance $R$ (but not on the angles $\theta$ and $\phi$). 
The structure of the  Hamiltonian of Eq.\ (\ref{Ham}) is similar to that 
adopted in the work of Neville and Worth.\cite{NW14} The difference is in
the choice of the disappearing modes $\RR$ (here: 
three Jacobi coordinates), 
in the choice of the coordinate grids (here: the coordinate grid
uniformly covers the complete dissociation path), and in 
the construction of the matrices $\MatVar{{\bf U}}_R(\RR)$ and 
$\MatVar{{\bf W}}_Q(\QQ|R)$ (here: spline interpolations on a regular grid). 

The diagonal elements of the potential matrix 
have the form ($\alpha = X, A_2, B_1$):
\begin{eqnarray}
V^\alpha(R,\theta,\phi,\QQ) & = &
U_R^\alpha(\RR) + W_Q^\alpha(\QQ|R) \nonumber \\ 
&= & U^\alpha_{\rm 1D}(R) + U^\alpha_{\rm ang}(R,\theta,\phi) \nonumber \\ 
& & + {\sum_{i}}^{a_1} \kappa^\alpha_i(R) Q_{a_1}(i) + \frac{1}{2}
\sum_\Gamma {\sum_{i,j}}^{\Gamma} \gamma^\alpha_{\Gamma,ij}(R)
Q_\Gamma(i) Q_\Gamma(j) \, .
\label{V_op}
\end{eqnarray}
$U^\alpha_{\rm 1D}(R)$ are the one-dimensional potential
energy functions along $R$, with  $(\theta,\phi) = (0^\circ,0^\circ)$, 
and the ring modes set to the pyrrolyl equilibrium 
$\QQ = \gr{0}$; $U_{\rm ang}^\alpha(R,\theta, \phi)$ are the 
distance-dependent angular potentials vanishing for 
$(\theta,\phi) = (0^\circ,0^\circ)$ [the relation between the pair 
$(\theta,\phi)$ and the molecular configurations is exemplified in 
the caption to Fig.\ \ref{F01A}(b)]. Functions 
$\kappa_i^\alpha(R)$ are the $R$-dependent
gradients vanishing for all modes but $a_1$. Functions 
$\gamma_{ij}^\alpha(R)$ are 
the $R$-dependent Hessians with respect to normal modes. Both 
$\kappa_i^\alpha(R)$ and $\gamma_{ij}^\alpha(R)$ are evaluated
at $\QQ = \gr{0}$. Hessian matrices
${\bm \gamma}^\alpha$ are four-block diagonal: $\boldsymbol{\gamma}^\alpha =
\boldsymbol{\gamma}^\alpha_{a_1} \oplus
\boldsymbol{\gamma}^\alpha_{a_2} \oplus
\boldsymbol{\gamma}^\alpha_{b_1} \oplus
\boldsymbol{\gamma}^\alpha_{b_2}$.

The off-diagonal diabatic couplings $V^{\alpha \beta}$ are given by
\begin{eqnarray}
V^{\alpha \beta}(R,\theta,\phi,\QQ) & = &
U_R^{\alpha\beta}(\RR) + W_Q^{\alpha\beta}(\QQ|R) \nonumber \\ 
&= & U^{\alpha
\beta}_\mathrm{ang}(R,\theta,\phi) + {\sum_i}^{\Gamma_\alpha \times
\Gamma_\beta} \lambda^{\alpha \beta}_i(R) Q_{\Gamma_\alpha \times
\Gamma_\beta}(i) \ ,
\label{V_coup}
\end{eqnarray}
Functions $U^{\alpha \beta}_\mathrm{ang}(R,\theta,\phi)$
are constructed on the 3D coordinate grid by first applying the 
regularized diabatization procedure of K\"oppel et al.\cite{KGM01} 
to the raw adiabatic ab initio energies and next interpolating between
grid points using cubic splines. 
While $U^{\alpha \beta}_\mathrm{ang}(R,\theta,\phi)$
do not have an analytical representation,  near the CIs with $\widetilde{X}$
the construction
algorithm ensures that they follow the lowest allowed orders
in the symmetry-adapted spherical harmonics, namely
$U^{X
A_2}_\mathrm{ang} \sim \sin^2 \theta \sin(2\phi)$ and $U^{X
B_1}_\mathrm{ang} \sim \sin \theta \cos \phi$.
The $\QQ$-dependent couplings $W_Q^{\alpha\beta}(\QQ|R)$
are linear in the 
ring modes: The $X/A_2$
and $X/B_1$ coupling terms are promoted by the vibrational  modes of $a_2$ and
$b_1$ symmetry, respectively. The
matrix element $W^{A_2 B_1}$ between the states $A_2/B_1$ is not included
in the quantum mechanical calculations.

The quasi-diabatic representation used in the Hamiltonian of Eq.\ (\ref{Ham})
is local, i.e. a given off-diagonal matrix 
element $V^{\alpha\beta}$
is non-zero only in a (broad) vicinity of the CI
between states $\alpha$ and $\beta$,
where a non-vanishing transition probability between 
quasi-diabatic states is expected.
For the second term in Eq.\ (\ref{V_coup}), this is achieved
by using the following functional form for the coupling strength
$\lambda_i^{\alpha \beta}$:
\begin{equation} 
\lambda_i^{\alpha \beta}(R) =
\lambda_{\mathrm{CI},i}^{\alpha \beta} \exp \left(- \left| \frac{R -
R_\mathrm{CI}^{\alpha \beta}}{\Delta} \right|^n \right) \ ,
\label{loclam}
\end{equation}
where $R_\mathrm{CI}^{\alpha \beta}$ is the position of the CI 
between $\alpha/\beta = X/A_2$ or $X/B_1$. The parameters
$\lambda_{\mathrm{CI},i}^{\alpha \beta}$, $\Delta$ and $n$ are tuned
\lq by eye' in order to obtain smooth diabatic
Hessians for the coupled states. Similar attenuation functions are
applied to the mixing angles of the regularized adiabatic-to-diabatic
transformation in $(R,\theta,\phi)$. Examples of the local diabatic
coupling functions are given in Fig.\ \ref{F01}(b).

\subsection{Ab initio parameterization of the molecular Hamiltonian}
\label{param}

\subsubsection{Quantum chemical calculations}
\label{qchem}

The matrix elements of the molecular Hamiltonian,
Eqs.\ (\ref{Ham}), (\ref{V_op}), and (\ref{V_coup}), are constructed from
the ab initio energies obtained using the  
electronic structure calculations performed with the
aug-cc-pVTZ (AVTZ) basis set of Dunning.\cite{D89} The basis set is 
further augmented with the
diffuse $s$ and $p$ functions added
to the N and H atoms of the dissociating bond (one set of $s$
and $p$ functions for N and two sets for
H). The exponents of these functions are derived in an even tempered
manner from the most diffuse $s$ and $p$ functions
of the AVTZ basis by dividing
the exponents successively by a factor of 3.0.\cite{VLMSD05}
This extension is necessary to correctly describe
the Rydberg character of the states $1^1\!A_2(\pi\sigma^*)$ 
and $1^1\!B_1(\pi\sigma^*)$,  and the resulting
basis set is referred to as AVTZ+. 

Most calculations are performed at the CASPT2 level of theory. 
The reference wavefunctions are obtained from the
state-averaged CASSCF calculations including the states 
$\widetilde{X}^1A_1$, $1^1\!A_2(\pi\sigma^*)$ 
and $1^1\!B_1(\pi\sigma^*)$. The active space 
comprises five $\pi$ valence molecular orbitals, three of $b_1$ and
two of $a_2$ symmetry, the $9a_1(\sigma)$ and the $10a_1(3s/\sigma^*)$
orbitals.\cite{NOTE-PYR01A-01} This setting for the active space, consisting 
of eight electrons in seven orbitals, will be denoted as $(8_e,7_o)$. 
The choice of this relatively small active space guarantees that the 
ab initio calculations are running smoothly for small as well as asymptotic
interfragment distances $R$. 
Electronic structure calculations are performed using the highest possible
symmetry. The matrix elements of the potential
matrix  $\MatVar{{\bf U}}_R(\RR)$ are calculated 
as functions of the disappearing modes using
the symmetry group $C_1$.  
The blocks of the $\QQ$-dependent  Hessian matrices with symmetries
$a_1, \ a_2, \ b_1 \ \mbox{and} \ b_2$ are calculated 
separately using $C_{2v}$, $C_2$, $C_s$ and $C_s^\prime$ symmetries,
respectively.
Coordinate dependent TDM functions,
necessary to properly describe the optical excitation of the $\pi\sigma^*$
states from the ground electronic state, are calculated at the CASSCF
level. In addition to the main batches, several
sets of benchmark calculations were performed with the AVTZ+ basis set 
using  the
multi-reference configuration interaction (with Davidson correction 
applied,\cite{LD74} MRCI+Q) and the equation of motion coupled cluster
(EOM-CCSD) methods. In all calculations, MOLPRO
suite of programs was used.\cite{MOLPRO-FULL}

The first step in the construction of matrix elements is the calculation of
the minimum energy path (MEP) for hydrogen detachment from the N---H group 
in the first excited state $1^1\!A_2(\pi\sigma^*)$. The MEP is calculated
along the Jacobi coordinate $R$; the grid $\{R_i\}$ consists of  28
points lying between $R_{\rm min} = 3.5\,a_0$ and
$R_{\rm max} = 8.3\,a_0$. 
Along this path, the molecule is constrained to
$C_{2v}$ geometries with $(\theta,\phi) = (0^\circ,0^\circ)$. 
In the subsequent steps, the
full dimensional quasi-diabatic representation is built using this MEP
as a reference. This is 
conceptually similar to the reaction surface Hamiltonian\cite{MHA80,CM84} 
with the difference that a single set
of normal modes is used along the MEP. Finally, 
in order to simulate the 
excitation process, the geometry of pyrrole in the ground electronic
state is also optimized and used as a reference for the calculation of the
Herzberg-Teller TDM. 
The optimized structures of pyrrole and pyrrolyl are 
compared in Fig. \ref{F01A}. 
The main structural changes in going from the parent to the radical 
are the increase of the $\mathrm{C-C^\prime}$ bond  
(by $0.16 \ a_0$) and the decrease of the $\mathrm{C^\prime - C^\prime}$ 
bond (by $0.14 \ a_0$).

\subsubsection{3D diabatic potentials for the disappearing modes}
\label{modes3D}

The CASPT2 energies of the lowest three
electronic states are calculated on 
a three-dimensional grid $(R_i,\theta_j,\phi_k)$, with the nodes $R_i$ being
grid points on  the MEP. The grid points $\theta_j$ in the polar
angle cover the range $[0^\circ,90^\circ]$ with a step of 5$^\circ$; for a set
of $R_i$ values, calculations over the  
range $[0^\circ,180^\circ]$ are performed, and the  
remaining missing energies for $\theta > 90^\circ$ are extrapolated.
The grid in the azimuthal angle $\phi$ ranges from $0^\circ$ 
to $90^\circ$ with a step of
15$^\circ$; energies for larger $\phi$ are reconstructed using $C_{2v}$
symmetry of the pyrrolyl ring. The 1D cuts along $\theta$ and $\phi$ are
exemplified in Fig.\ \ref{F01}(c,d). In all three states, the geometries
with $(\theta,\phi) = (0^\circ,0^\circ)$ correspond to a minimum. In the
excited states, this minimum is local and the hydrogen atom lying in the
plane of the ring has to overcome
a potential barrier in order to reach the out-of-plane
geometries with $\theta > 20^\circ$. Thus the calculation predicts
that pyrrole 
in the excited $\pi\sigma^*$ states remains  near-planar 
in the initial stages of dissociation. 

Quasi-diabatic representation for the calculated states is constructed
in two steps. In the first step, the states $\widetilde{X}$ and 
$1^1\!A_2(\pi\sigma^*)$ are
diabatized. The angular modes do not couple them 
at either  $\phi = 0^\circ$ or
$90^\circ$, and the impact of $\theta$ and $\phi$
on the non-adiabatic dynamics in the $X/A_2$ pair 
is exceedingly weak. For this reason, the states $\widetilde{X}$ and 
$1^1\!A_2(\pi\sigma^*)$ 
can be diabatized by a relabeling the adiabatic energies. 
The tiny
coupling matrix element between these states for $\phi \ne 0, 90^\circ$, 
is modeled  as 
\begin{equation} 
U^{X A_2}_R(R,\theta,\phi)
= \lambda^{XA_2}_{\rm ang}(R) \sin^2 \theta\sin(2\phi) \, ,
\label{uxa2}
\end{equation}
with the $R$-dependent 
function $\lambda^{XA_2}_{\rm ang}(R) =
c\left[1-{\rm atan}\left((R-R_0)/\Delta R\right)\right]$. The 
parameters $c$, $R_0$, and $\Delta R$ 
are chosen \lq by eye' in order to give smooth diabatic curves
for a full range of $\phi$. 

In the second step, the states $\widetilde{X}$ and $1^1\!B_1(\pi\sigma^*)$ 
are transformed to the quasi-diabatic representation. The orthogonal 
transformation between the diagonal matrix of the adiabatic energies of 
$\widetilde{X}$ and $1^1\!B_1(\pi\sigma^*)$ 
and the non-diagonal diabatic potential matrix 
is effected by the $2\times 2$ adiabatic-to-diabatic 
transformation matrix $\MatVar{{\bf S}}(\RR)$,
\begin{equation}
  \MatVar{{\bf S}}(\RR) = 
\left(
\begin{array}{lr}
\cos\Theta & -\sin\Theta\\
\sin\Theta & \cos\Theta
\end{array}
\right)
\label{adt1}
\end{equation}
where $\Theta$ is the coordinate dependent mixing angle constructed 
with the regularized diabatic state method of K\"oppel et
al.\cite{KGM01} To this end, two potential energy differences between adiabatic
states are used. One, denoted $\Delta_1$, is evaluated in  
the $C_{2v}$ symmetric subspace $(R,\theta,\phi) = (R,0,0)$ 
along the Jacobi coordinate $R$. 
Another, denoted $\Delta_2$, is evaluated  
for the $C_1$ subspace $(R_\mathrm{CI}^{XB_1},\theta,\phi)$ with the distance
$R$ fixed at the CI degeneracy point. The mixing angle is defined as
\begin{equation}
\Theta = \frac{1}{2} \arctan\frac{\Delta_2}{\Delta_1} \, .
\label{adt2}
\end{equation}
By construction, 
the diabatic potential matrix elements in the
vicinity of the degeneracy point are linear in the displacements from 
$(R_\mathrm{CI}^{XB_1},0,0)$,  
i.e. $U^X,\, U^{B_1} \sim (R-R_\mathrm{CI}^{XB_1})$ and $U^{X
B_1} \sim \sin \theta \cos \phi$.
Away from the degeneracy point, the mixing angle $\Theta$
is attenuated with the
function of the form of Eq.\ (\ref{loclam}), so that the adiabatic-to-diabatic
transformation is localized to the intersection region. 
Finally, all diabatic matrix elements set on the
grid are interpolated using cubic splines. It is ensured that they
become independent of the disappearing angles 
$\theta$ and $\phi$ as $R$ reaches the asymptotic region $R > 7.0\,a_0$. 

1D cuts through the potential energy surfaces of the three diabatic states
are shown in Fig. \ref{F01}. 
The minimum of the $\widetilde{X}$ state [panel (a)]
is located at $R \approx 4.1 \ a_0$. In the excited states,  
local minima are found at $4.23 \ a_0$ for 
$1^1\!A_2(\pi\sigma^*)$ and at
$4.14 \ a_0$ for $1^1\!B_1(\pi\sigma^*)$.  
These minima are separated by $\le 0.10$\,eV high
barriers from the repulsive portions of the potential curves. 
The energy parameters of the ab
initio PESs are summarized in Tables \ref{T01} and \ref{T01A} 
which are discussed
in the next section.  

Two-dimensional (2D) contour plots of the splined diabatic potentials 
$\widetilde{X}$ and $1^1\!A_2(\pi\sigma^*)$ 
in the $(R,\theta)$ plane are shown in Fig.\ \ref{F02}; 
the contour plots for the state $1^1\!B_1(\pi\sigma^*)$ 
are discussed in Paper III.
In the ground electronic state,  
the coordinates $R$ and $\theta$ are 
strongly mixed in the FC zone, around $R = 4.15\,a_0$ and 
$(\theta,\phi) = (0^\circ,0^\circ)$: Contour lines around 
the potential minimum have a characteristic \lq banana' shape indicating
that the equilibrium pyrrolyl---H distance shrinks as the H-atom moves
out of the plane of the pyrrolyl ring. In the potential $U^X_R(R,\theta,\phi)$,
the quantum mechanical (anharmonic) out-of-plane bending
frequency of the H-atom (along $\theta$, $\phi = 0^\circ$) is of the
order of  $406$\,cm$^{-1}$ which is about three times  
lower than the anharmonic 
in-plane bending frequency of $\sim 1281$\,cm$^{-1}$
(along $\theta$, $\phi = 90^\circ$).
In the state $1^1\!A_2(\pi\sigma^*)$, $R$ and $\theta$ are also mixed, and the
height of the dissociation barrier around $R \approx 4.7\,a_0$ 
depends on $\theta$. The barrier of $0.09$\,eV (cf. Table \ref{T01}) 
along the straight dissociation path with $\theta = 0$
is always the lowest, 
which is yet another indication that almost no torque  along $\theta$ 
is exerted 
on the H-atom  in the initial stage of the photodissociation 
in the state $1^1\!A_2(\pi\sigma^*)$.

\subsubsection{21D diabatic potentials for the non-disappearing modes}
\label{modes21D}

The parameters of the $\QQ$-dependent part of the diabatic 
Hamiltonian matrix are calculated
as first and second derivatives with respect to deviations from the
MEP in the state  $1^1\!A_2(\pi\sigma^*)$. The $C_{2v}$ symmetry is preserved
along the MEP, and the raw ab initio energies and their first
derivatives are diabatized by a trivial relabeling of the 
calculated points. Most ab initio 
second derivatives (i.e., elements of the Hessian matrices) are
also in this \lq trivially diabatic' form. 
Exceptions are Hessian blocks involving
symmetry breaking coupling modes and diverging near the CIs; diabatization is
required in order to remove the resulting singularities and to fix 
the diabatic coupling strengths 
$\lambda_i^{\alpha\beta}$. The Hamiltonian matrix is constructed in the 
following sequence of ab initio calculations and transformations: 

\begin{enumerate}
\item {\it Geometry optimization and the normal
mode analysis for the ground electronic state $\widetilde{X}^2\!A_2$ of 
pyrrolyl}.  In this step, the dimensionless normal coordinates
$\QQ$ of pyrrolyl are defined. They are related to the
Cartesian coordinates $\mathbf{X}$ of the atoms via the rectangular 
transformation matrix $\MatVar{{\bf L}}$ with elements
$L_{ir}$:
\begin{equation}
Q_i = \sum_{r=1}^{27} \sqrt{\frac{\omega_i M_r}{\hbar}} F_{ir} X_r = 
\sum_{r=1}^{27} L_{ir} X_r \ , i = 1, 21 \, ,
\end{equation}
where $M_r$ is the mass of the atom associated with the coordinate
$X_r$, $\omega_i$
is the frequency of the normal mode $Q_i$, and $\{F_{ir}\}$ is the
matrix of eigenvectors of the mass-weighted Cartesian Hessian from which 
the rows corresponding to the rigid pyrrolyl translations
and rotations are removed. 
The values of normal modes along the MEP are denoted
$\QQ = \QQ_\mini(R)$. 

\item {\it Calculation of the 
Cartesian gradient vectors $\{\bar{g}_i^\alpha\}$ and Hessian matrices
$\{\bar{\gamma}_{\Gamma,ij}^\alpha\}$.} They are calculated for  
$\alpha = X$, $A_2$, and $B_1$ along the MEP. The gradients and Hessians with
respect to normal modes (without overbar) are given by
\begin{eqnarray} 
g^\alpha_i & = & \sum_r
\left(\MatVar{{\bf L}}^{-1}\right)_{ri} \bar{g}^\alpha_r \\
\gamma^\alpha_{\Gamma,ij} 
& = & \sum_{rs} \left(\MatVar{{\bf L}}^{-1}\right)_{ri}
\left(\MatVar{{\bf L}}^{-1} \right)_{sj} \bar{\gamma}^\alpha_{\Gamma,rs} \ .
\end{eqnarray} 
Non-vanishing 
gradients point
along the normal modes of $a_1$ symmetry.

\item {\it Reconstruction of gradients at $\QQ = \gr{0}$. }
The gradients $\gr{g}^\alpha(R)$, computed at $\QQ = \QQ_\mini(R)$ in the 
previous step,
vanish for $\alpha = A_2$ and differ from zero for $\alpha = X, B_2$. 
Together with the Hessian matrices 
$\MatVar{\bm{\gamma}}^\alpha_{\,\,\Gamma}(R)$, they
are used to reconstruct the gradients $\boldsymbol{\kappa}^\alpha(R)$
at
$\QQ = \gr{0}$ which enter Eq.\ (\ref{V_op}) for the diagonal elements of the
Hamiltonian:
\begin{equation} 
\boldsymbol{\kappa}^\alpha(R) = \gr{g}^\alpha(R) -
\MatVar{\boldsymbol{\gamma}}_{\,\,\Gamma}^\alpha(R) \QQ_\mini(R) \, .
\label{Reconstruction1}
\end{equation}
The function $U^\alpha_{\rm 1D}(R)$ along the $\QQ = \gr{0}$ cut
[cf. Eq.\ (\ref{V_op})] is given by 
\begin{equation} 
U^\alpha_{\rm 1D}(R) = U^\alpha_{\rm Relax}(R) -
\QQ_\mini^T(R) \gr{g}^\alpha(R) \ .
\label{Reconstruction2}
\end{equation}
Here $U^\alpha_{\rm Relax}(R)$ are the 
potential profiles along the MEP
shown in Fig.\ \ref{F01}(a,b). The gradients 
$\boldsymbol{\kappa}^\alpha(R)$ and $\gr{g}^\alpha(R)$, as well as the
potentials $U^\alpha_{\rm 1D}(R)$ and $U^\alpha_{\rm Relax}(R)$, refer to
the high symmetry $C_{2v}$ subspace and are \lq trivially diabatic'. They are
smooth functions of $R$ near the CIs. 

\item {\it Evaluation of the off-diagonal diabatic matrix elements.}
The matrix elements $W_Q^{X A_2}$ and $W_Q^{X B_1}$ 
in Eq.\ (\ref{V_coup}) are linear functions
of the coupling modes at the CIs; for 
$W_Q^{X A_2}$, these are 
the normal modes $Q_{a_2}$ of $a_2$ symmetry, for  $W_Q^{X B_1}$  ---
the normal modes $Q_{b_1}$ of $b_1$ symmetry. 
The corresponding proportionality coefficients, 
$\lambda^{XA_2}_i(R)$ and $\lambda^{XB_1}_i(R)$ [cf. Eq.\ (\ref{loclam})],
are found using a property-based diabatization described in 
detail in Appendix \ref{appb}. The chosen properties
are the $a_2$ and $b_1$ symmetry blocks of the 
adiabatic Hessian matrices. While other symmetry blocks 
are \lq trivially diabatic' and vary smoothly with $R$, the matrix elements in
the blocks 
$\MatVar{\boldsymbol{\gamma}}_{\,\,a_2}^X(R)$ and 
$\MatVar{\boldsymbol{\gamma}}_{\,\,a_2}^{A_2}(R)$
diverge as $R$ approaches the $X/A_2$ CI; 
similarly, the matrix elements in the blocks 
$\MatVar{\boldsymbol{\gamma}}_{\,\,b_1}^X(R)$, and
$\MatVar{\boldsymbol{\gamma}}_{\,\,b_1}^{B_1}(R)$ 
diverge near the $X/B_1$ crossing.
The singularities, arising in these symmetry blocks because the 
adiabatic energies along the coupling modes are cusped at the intersections, 
are removed in the quasi-diabatic representation. As shown in 
Appendix \ref{appb}, the smooth diabatic Hessian blocks 
$\{ \left.\gamma^\alpha_{\Gamma,ij}(R)\right|_{\rm dia} \}$, used in
Eq. (\ref{V_op}), are related to the ab
initio Hessian blocks
$\{ \left.\gamma^\alpha_{\Gamma,ij}(R)\right|_{\rm adia} \}$ via:
\begin{eqnarray}
\left.\gamma^{X}_{\Gamma,ij}(R)\right|_{\rm dia} 
& = &  
\left.\gamma^{X}_{\Gamma,ij}(R)\right|_{\rm adia} + 2\frac{\lambda_i^{X A_2}(R) \lambda_j^{X A_2}(R)}{U_{\rm Relax}^{X}(R) - U_{\rm Relax}^{A_2}(R)}, \ \  \Gamma,\Gamma_i, \Gamma_j = a_2 \nonumber \\
\left.\gamma^{X}_{\Gamma,ij}(R)\right|_{\rm dia} 
& = &   
\left.\gamma^{X}_{\Gamma,ij}(R)\right|_{\rm adia} + 2\frac{\lambda_i^{X B_1}(R) \lambda_j^{X B_1}(R)}{U_{\rm Relax}^{X}(R) - U_{\rm Relax}^{B_1}(R)}, \ \  \Gamma,\Gamma_i, \Gamma_j = b_1 \nonumber \\
\left.\gamma^{A_2}_{\Gamma,ij}(R)\right|_{\rm dia} 
& = &  
\left.\gamma^{A_2}_{\Gamma,ij}(R)\right|_{\rm adia} - 2\frac{\lambda_i^{X A_2}(R) \lambda_j^{X A_2}(R)}{U_{\rm Relax}^{X}(R) - U_{\rm Relax}^{A_2}(R)}, \ \Gamma,\Gamma_i, \Gamma_j = a_2 \nonumber \\
\left.\gamma^{B_1}_{\Gamma,ij}(R)\right|_{\rm dia} 
& = &  
\left.\gamma^{B_1}_{\Gamma,ij}(R)\right|_{\rm adia} - 2\frac{\lambda_i^{X B_1}(R) \lambda_j^{X B_1}(R)}{U_{\rm Relax}^{X}(R) - U_{\rm Relax}^{B_1}(R)}, \ \   \Gamma,\Gamma_i, \Gamma_j = b_1  \ . \label{AdToDiaHess}
\end{eqnarray}

When the energy gaps 
$\Delta U_{\rm Relax} = U_{\rm Relax}^{X}(R) - U_{\rm Relax}^{\alpha}(R)$  
are large, the differences between the adiabatic and the diabatic
Hessians are negligible. In the vicinity of the state intersection, 
the coupling strengths
$\lambda^{XA_2}_{{\rm CI},i}$ and $\lambda^{XB_1}_{{\rm CI},i}$ 
in the singular 
terms $\lambda_i \lambda_j / \Delta U_{\rm Relax}$ in Eq. (\ref{AdToDiaHess})
are adjusted to 
compensate the divergence of the adiabatic Hessians, making matrix elements
in the
diabatic blocks smooth functions of $R$.  The local
character of the diabatic representation is enhanced by the $R$-dependence
of the coupling coefficients $\lambda_i$ in Eq.\ (\ref{loclam}); 
parameters of the localizing exponential function are tuned \lq by eye'. 
Note that for the Hessian matrix in the state $\widetilde{X}$, the CI
$X/A_2$ affects exclusively the block $a_2$, while the CI $X/B_1$
affects exclusively the block $b_1$ (see Appendix \ref{appb}).  

\item {\it Spline interpolation on the $R$ grid.}

The diabatized functions $U^\alpha_{\rm 1D}(R)$, $\{
\kappa_i^\alpha(R) \}$, and $\{ \gamma_{ij}^\alpha(R) \}$,  
set on the discrete 
grid in $R$, are interpolated 
using cubic splines. In this way, fitting with Morse curves or
application of ad hoc switching functions, ubiquitous in the theoretical
studies of the photodissociation of aromatic molecules,\cite{VLMSD05,NW14} 
are avoided. 

\end{enumerate}

The quality of the resulting potential energy surfaces is illustrated in 
Tables \ref{T01}---\ref{T02}. 
The quantum mechanical excitation energies (band origins $T_0$), the potential
barrier heights $E^\ddagger$, and the  
dissociation thresholds $D_0$ for the $\pi\sigma^*$ states 
calculated using the CASPT2 method are given in 
Table \ref{T01}; harmonic zero-point energies for the ground and the
excited electronic states are used.  Comparison with the 
the experimental values in the same Table shows that 
the calculated $T_0$ and $D_0$ are underestimated by 
about 0.5\,eV. At the same time, the energy gaps between 
the states $1^1\!A_2(\pi\sigma^*)$ and $1^1\!B_1(\pi\sigma^*)$ 
in the FC zone and in the asymptotic region are reproduced to within 
0.05\,eV, i.e. an order of magnitude better. This indicates that 
the shapes of the 
calculated potential energy surfaces are qualitatively correct. 
The 
potential barrier along the MEP, $E^{\ddagger}_{\rm MEP}$, 
located at $R_b = 4.6\,a_0$ is about 0.09\,eV
for the state $1^1A_2(\pi\sigma^*)$, and 0.08\,eV
for the state $1^1B_1(\pi\sigma^*)$. Both energies, measured relative to the
local minima, are somewhat lower
than previously assumed. For example, Domcke and co-workers found for the 
states $1^1\!A_2$ and $1^1\!B_1$ barrier heights of 0.40\,eV 
and 0.26\,eV, respectively, for ring geometry fixed to pyrrole equilibrium.\cite{VLMSD05} These values compare well with the barrier heights in the
$\pi\sigma^*$ states constructed by Neville and Worth (0.40\,eV and 0.30\,eV, 
respectively).\cite{NW14}
The corresponding barriers in our potentials are 
0.21\,eV [for $1^1A_2(\pi\sigma^*)$]
and 0.05\,eV  [for $1^1B_1(\pi\sigma^*)$].
 One might expect, however, that the values
$E^{\ddagger}_{\rm MEP}$, for which a comparison is not available, are more dynamically relevant than 
$E^{\ddagger}_{\rm pyr}$.

In order to rationalize the origin of the detected deviations 
and to put them into perspective, we performed a series of
additional ab initio calculations of
the vertical excitation energies $T_v$ and the classical dissociation 
thresholds $D_e$. These are defined as differences between the 
ab initio energies without zero-point energy corrections.
All calculations use the same AVTZ+ basis set. Two post-CASSCF
methods, CASPT2 and MRCI, are applied with several active spaces and compared
with the  EOM-CCSD method. 
The results are summarized in Table \ref{T01A};  
previously published data are also presented for comparison. 

The vertical excitation energies lie between 4.45\,eV and 5.33\,eV
for the  state $1^1\!A_2(\pi\sigma^*)$,
and between  5.03\,eV and 6.12\,eV  for the state $1^1\!B_1(\pi\sigma^*)$. 
As expected,\cite{FVSELK11} the CASPT2 calculations predict lower 
$T_v$ values than either the 
MRCI calculations with Davidson correction, or the EOM-CCSD or ADC(2)
methods. Exceptions are the CASPT2 energies
calculated in Ref.\ \onlinecite{NW14}: For example, 
$T_v$ for the state $1^1\!A_2(\pi\sigma^*)$
is almost 0.8\,eV higher than the CASPT2 and MRCI values obtained 
in this work. The energies $T_v$ calculated using the CASSCF based methods
tend to grow with growing active space. This, however, is also a propensity
rather than a rule:  The MRCI
calculation of Ref.\ \onlinecite{FVSELK11}, performed with a modest active
space,  gives the largest value of $T_v = 5.33$\,eV 
for the state $1^1\!A_2(\pi\sigma^*)$.  

The lowest dissociation threshold $D_e$, diabatically correlated with
$1^1\!A_2(\pi\sigma^*)$, is located between 3.17\,eV and 4.00\,eV. 
$D_e$ is a surprisingly robust quantity, and changes little with 
changing method or changing active space. Exception is again the CASPT2
calculation of Ref.\ \onlinecite{NW14}, which predicts a strikingly low
$D_e$ value. The spread in $D_e$ values is larger for the  next
threshold diabatically correlating with
$1^1\!B_1(\pi\sigma^*)$: The calculated $D_e$ values lie between 4.09\,eV and
5.11\,eV and vary irregularly with changing active space. 

The EOM-CCSD and ADC(2) methods are consistent in predicting 
$T_v \sim 5$\,eV for
the  state $1^1\!A_2(\pi\sigma^*)$ and 
$T_v \sim 6$\,eV for
the  state $1^1\!B_1(\pi\sigma^*)$. These methods are known 
to cope easily with the mixed Rydberg/valence character of
the electronic wave functions and to deliver high
quality electronic energies for heteroaromatic 
molecules.\cite{TS97} Unfortunately, they are not useful in the 
dissociation region electronic structure in which becomes explicitly
multiconfigurational. The ability to calculate global potential energy surfaces
across the FC zone and into the asymptotic dissociation channels is 
instrumental for our photodissociation study. 
 The CASPT2
calculations of the PESs in this work represent a trade off between the 
accuracy, feasibility, and stability of hundreds of underlying
batch runs.

Harmonic 
vibrational frequencies calculated for the minimum of the pyrrolyl ground 
state $\widetilde{X}^2\!A_2$, and for the optimized minima of pyrrole
in the state $1^1\!A_2(\pi\sigma^*)$ and in the ground state 
$\widetilde{X}^1\!A_1(\pi\pi)$
are summarized in  Table \ref{T02}. The normal modes $Q_{\Gamma}(i)$  
in the pyrrolyl minimum are enumerated for each symmetry block
in order of increasing frequency. The frequencies of pyrrole 
are ordered differently: They are listed according to pyrrolyl normal
modes with 
which they correlate. This does not always correspond to an increasing
frequency order, because of the Duschinsky mixing which 
especially affects modes with similar frequencies. 
Three normal modes of pyrrole, belonging to the irreps 
$a_1$, $b_1$, and $b_2$, 
\lq disappear' upon dissociation and correlate with the 
translation and rotations of the fragments. Their Jacobi labels $R$, 
$\theta$, and $\phi$ are included in Table \ref{T02} but the 
corresponding zero frequencies are omitted for pyrrolyl. 

The infrared spectrum of pyrrole was studied 
experimentally.\cite{MLH01,CPCESGLS03} Measured
vibrational frequencies are shown parenthetically in Table \ref{T02}, 
providing another assessment of the ab initio quality.
The CASPT2 calculations overestimate the experimental
 frequencies by less than  
$100$\,cm$^{-1}$ (for frequencies below $\sim 1500$\,cm$^{-1}$); the deviation
grows to about 
$\sim 300$\,cm$^{-1}$ for frequencies above $\sim 3000$\,cm$^{-1}$, but remains
within 10\%.  Note
that the theoretical frequencies in Table \ref{T02} are harmonic and no
scaling factors have been applied; deviations are
therefore within the limit expected for the anharmonicity corrections.

\begin{figure}[h!]
\caption{Contour plots of the states $\widetilde{X}$ 
(blue, bottom) and $1^1\!A_2(\pi\sigma^*)$
(red, top) in the $(R,Q_{a_1}(1))$ plane. 
All other normal modes are kept fixed at the pyrrolyl equilibrium, 
$\QQ = \boldsymbol{0}$. Energies marked on the contour lines are in eV.
The energy zero is at the minimum of the $\widetilde{X}$ state of pyrrole. 
The black dot is the FC point, and the solid line is the MEP in the 
state $1^1\!A_2(\pi\sigma^*)$ state.} 
\label{F03A}
\bigskip
\includegraphics[scale=0.73]{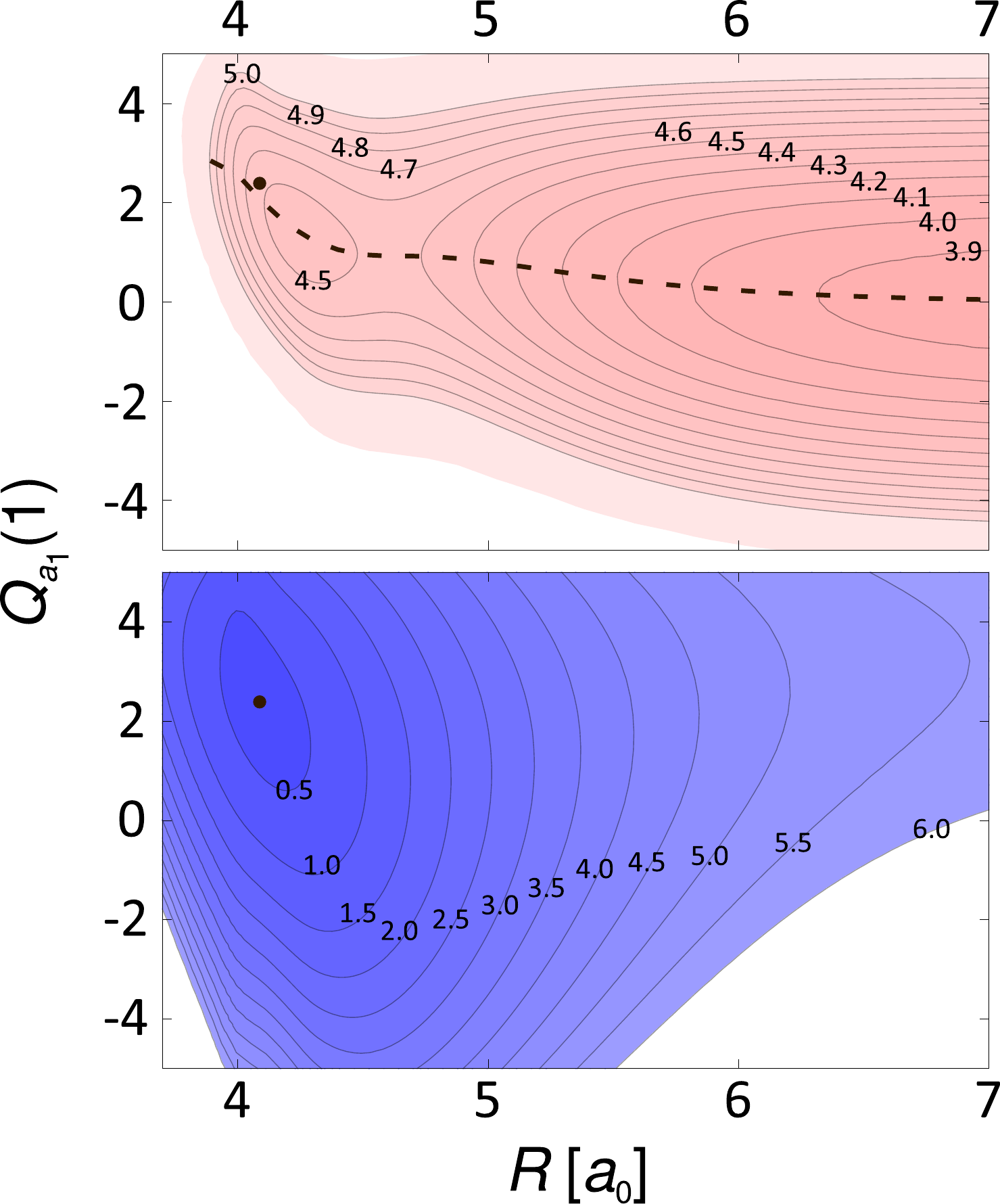}
\end{figure}

\begin{figure}[h!]
\caption{Contour plots of the states $\widetilde{X}$ 
(blue, middle) and $1^1\!A_2(\pi\sigma^*)$
(red, top) in the $(R,Q_{a_2}(2))$ plane. The diabatic coupling surface is  
shown in the bottom panel (green). 
All other normal modes are kept fixed at the pyrrolyl equilibrium, 
$\QQ = \boldsymbol{0}$. Energies marked on the contour lines are in eV.
The energy zero is at the minimum of the $\widetilde{X}$ state of pyrrole. 
The black dot is the FC point, and the solid line is the MEP in the 
state $1^1\!A_2(\pi\sigma^*)$.} 
\label{F03B}
\bigskip
\includegraphics[scale=0.73]{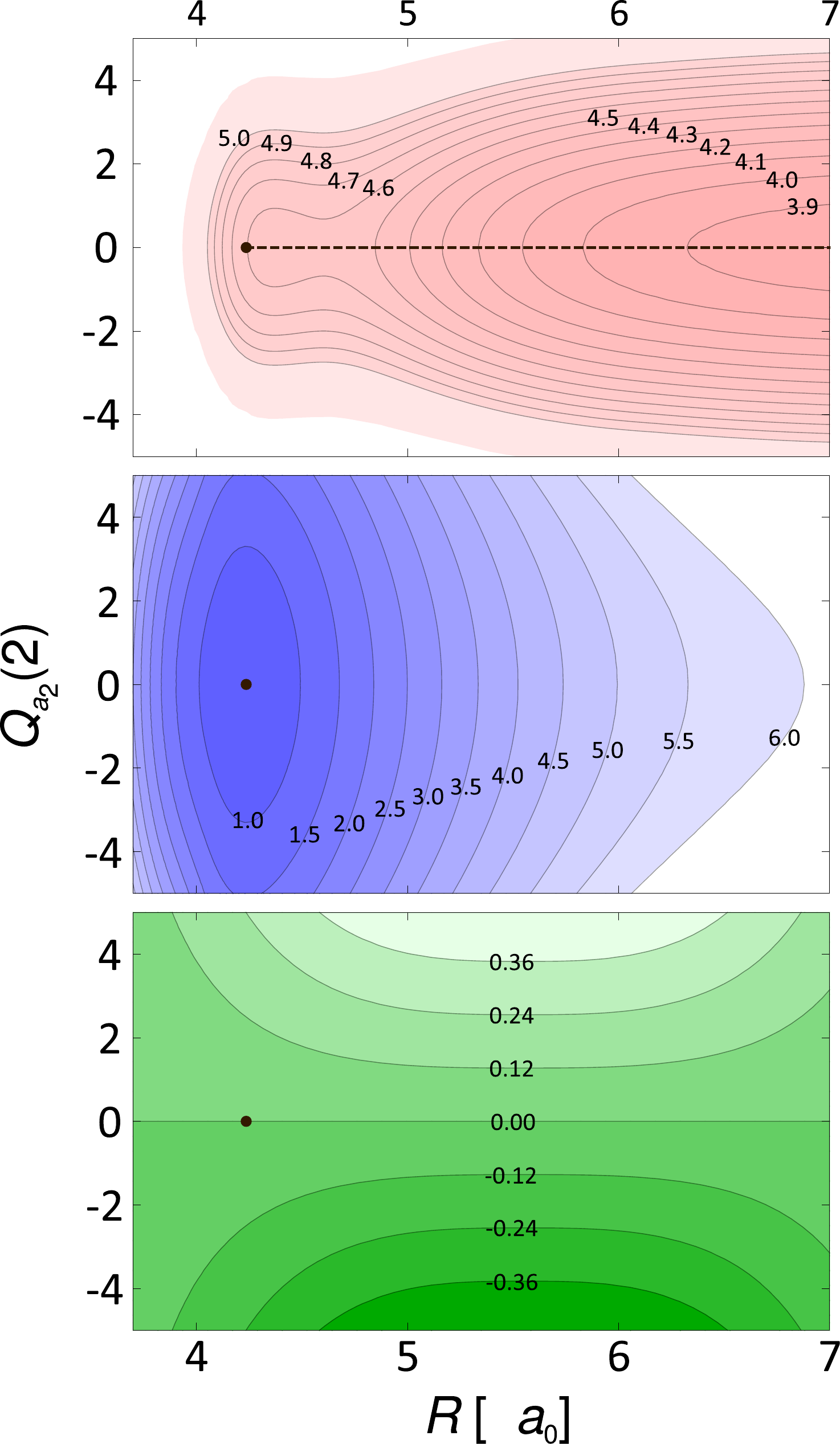}
\end{figure}

Contour maps of the ab initio PESs of the states $\widetilde{X}$ and 
$1^1\!A_2$  are shown in Fig.\ 
\ref{F03A} in which the dependence of the electronic energies  on 
the totally symmetric mode $Q_{a_1}(1)$ and the Jacobi
coordinate $R$ is illustrated; 
the remaining modes are kept fixed in the pyrrolyl minimum,
$\QQ = \gr{0}$. The included mode has the
largest displacement between the parent geometry in $\widetilde{X}$ 
($Q_{a_1}(1) = 2.24$)
and the asymptotic fragment equilibrium ($Q_{a_1}(1) = 0$). 
As a result, the MEPs 
in the $(R,Q_{a_1}(1))$ plane are curved in both electronic states.
The FC point $Q_{a_1}(1) = 2.24$ lies on the MEP of the ground electronic
state. In the excited electronic state, this point is displaced from
the MEP (see Fig. \ref{F03A}), 
and the dynamics along $R$ and $Q_{a_1}(1)$ are expected to be 
correlated during dissociation. 
In contrast, the anharmonic coupling between the totally symmetric
coordinate $R$
and the non-totally symmetric modes is substantially weaker in the 
constructed PESs. This is 
illustrated in 
Fig. \ref{F03B} for the mode  $Q_{a_2}(2)$. 
Upon vertical excitation, the displacements of the non-totally symmetric
modes from the $C_{2v}$ FC geometry vanish. The diabatic potentials 
are stationary with respect to the 
non-totally symmetric distortions, and the Hessian blocks, evaluated at 
 $\QQ^\Gamma = \gr{0}$ for $\Gamma \ne a_1$, are positive definite. 
The main effect of the intrastate coupling between $R$ and 
$Q_{a_2}(2)$ is the change of the vibrational frequency of the ring deformation
mode as pyrrole dissociates, and the average forces, acting on the 
diabatically dissociating
wave packet along the non-totally symmetric modes, are weak. 
Note that the off-diagonal diabatic coupling
element, shown in the bottom panel of Fig.\ \ref{F03B}, is another source
of coupling between vibrational 
modes, albeit in different electronic states.\cite{NOTE-PYR01A-02}

\subsection{Ab initio transition dipole moment functions}
\label{TDM}

The ab initio TDMs with $\widetilde{X}$ are
computed with MOLPRO using the CASSCF method. 
The molecular axes $(x,y,z)$
in these calculations are chosen as shown in Fig.\ \ref{F01A}(a). 
Only the TDMs for the transition 
$1^1\!A_2(\pi\sigma^*) \leftarrow \widetilde{X}$  are explicitly considered 
here; 
TDMs for the transition $1^1\!B_1(\pi\sigma^*) \leftarrow \widetilde{X}$  
are discussed in paper III.

The vector functions $\bmu^{A_2}(\RR,\QQ)$ are constructed using the 
Herzberg-Teller expansion, linear in deviations from the FC 
geometry.\cite{BUNKERJENSEN06} 
The TDMs are sums of the $\RR$- and $\QQ$-dependent terms: 
\begin{equation}
\bmu^{A_2}(\RR,\QQ) \approx \bmu_R^{A_2}(\RR) + \bmu_Q^{A_2}(\QQ) \, .
\label{tdm0}
\end{equation}
The symmetry properties of $\bmu^{A_2}$ are crucial
for calculating and understanding the absorption spectra and the 
photofragment distributions.  

\begin{figure}[h!]
\caption{
The TDM components $\mu_x$ (a) and $\mu_y$ (b) for the transition
$1^1\!A_2 \leftarrow \widetilde{X}$  as functions of the
polar angle $\theta$ for a set of listed Jacobi distances $R$.
Panel (c) shows the component $\mu_z$ of the same
TDM  as a function of the normal mode $Q_{a_2}(3)$.  
Dashed lines are the coordinate dependences used in the calculations of
the absorption spectra. Gray solid lines show the Gaussian profiles of the
initial vibrational state in $\widetilde{X}$. 
} 
\label{F03C}
\bigskip
\includegraphics[scale=0.3]{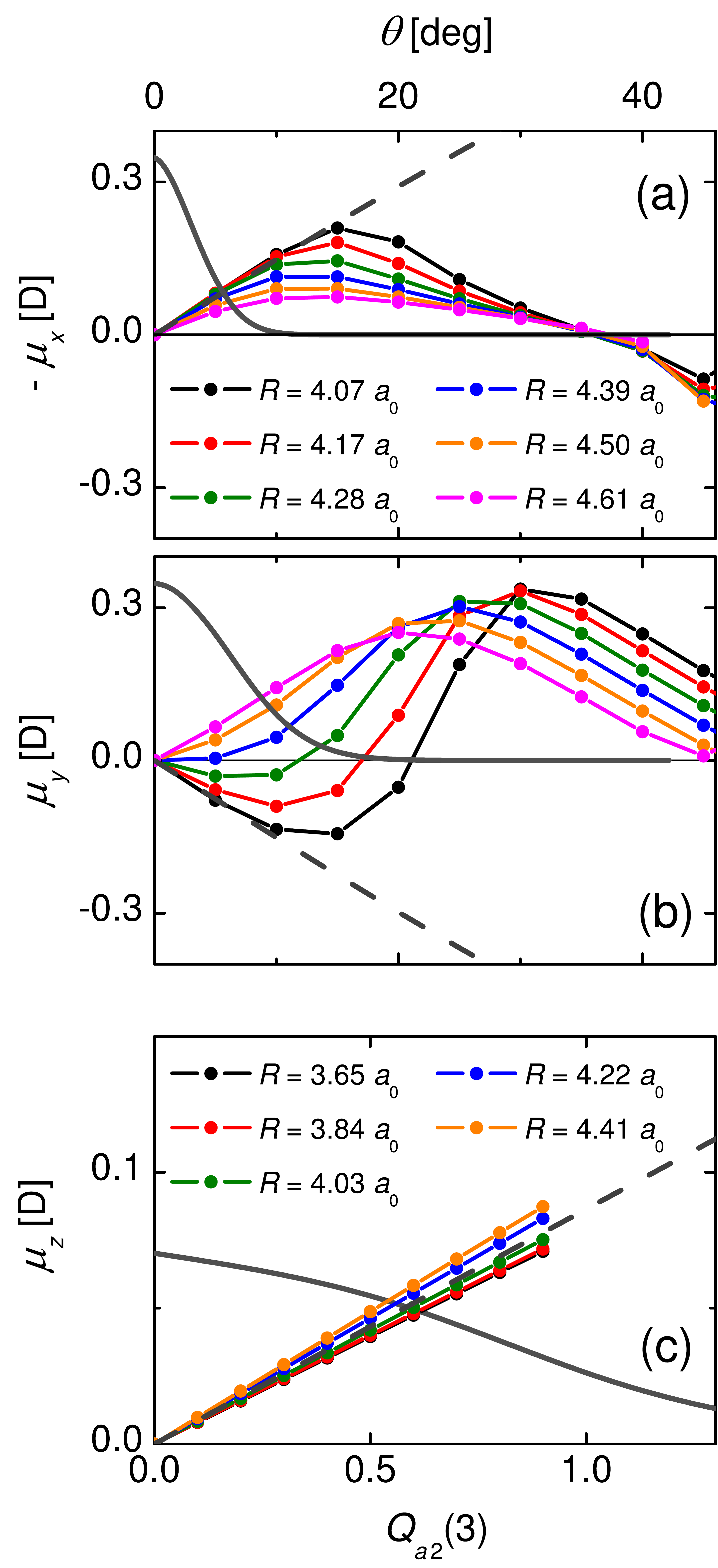}
\end{figure}

The transition $1^1\!A_2(\pi\sigma^*) \leftarrow \widetilde{X}$ 
is forbidden by symmetry at
$C_{2v}$ geometries. It becomes vibronically allowed via the TDM components
$\mu_x$, $\mu_y$, or $\mu_z$ if pyrrole  undergoes distortions of
$b_2$, $b_1$ and $a_2$ symmetry, respectively;  the $b_1$ and
$b_2$ distortions include displacements along the polar angle $\theta$ with
$\phi = 0$ (out-of-plane) and $\phi = 90^\circ$ (in-plane). The lowest order
Herzberg-Teller expansion around the FC point,  
compatible with Eq.\ (\ref{tdm0}), reads as
\begin{subequations}
\label{muHT}
\begin{align}
\mu^{A_2}_x(\RR,\QQ) & = & \mu^{A_2}_{x,\theta}(R_{\rm FC}) 
\sin(\theta) \sin(\phi) + 
\sum_{i=1}^7 \mu^{A_2}_{x,i}(R_{\rm FC})  Q_{b_2}(i) \  ,  
\label{mu_x_A2} \\
\mu^{A_2}_y(\RR,\QQ) & = &  \mu^{A_2}_{y,\theta}(R_{\rm FC})  
\sin(\theta) \cos(\phi) + \sum_{i=1}^3 \mu^{A_2}_{y,i}(R_{\rm FC})  
Q_{b_1}(i) \ ,  
\label{mu_y_A2} \\
\mu^{A_2}_z(\RR,\QQ) & = &  \sum_{i=1}^3 \mu^{A_2}_{z,i}(R_{\rm FC})  
Q_{a_2}(i) \, .  
\label{mu_z_A2} 
\end{align}
\end{subequations}
The real spherical harmonics $p_x$ and $p_y$ are chosen as the angular basis
functions. The coefficients in the Herzberg-Teller expansion 
are essentially 
derivatives of the TDM components with respect to normal coordinates. Their
values, calculated at the FC point $R_{\rm FC}$, 
are given in Table \ref{T02}. Most coefficients, non-vanishing by symmetry,
 are nevertheless small. 
The modes which significantly mediate the
excitation of the state $1^1\!A_2(\pi\sigma^*)$ are the bending
vibration along $\theta$ (transitions via $\mu_{x}$ and $\mu_{y}$), as well as 
the ring deformation  
$Q_{a2}(3)$ of $a_2$ symmetry (transition via $\mu_{z}$). The ab initio
TDMs along these coordinates are shown in Fig.\ \ref{F03C}. For large 
deviations from the FC point, the TDMs are complicated functions of 
displacements. However, the parent 
wave function of the ground vibrational state in $\widetilde{X}$
is localized around the FC point;  the shape of this wave function along
the Jacobi angle $\theta$ and the normal mode $Q_{a2}(3)$ 
is also illustrated in Fig.\ \ref{F03C}. 
Within the width of the initial wave function,   
the TDMs are linear and the expansions of Eq.\ (\ref{muHT}) are valid.  
It is also clear from Fig.\ \ref{F03C} that the Herzberg-Teller coefficients
depend on the interfragment distance $R$: Even the sign of the coefficient
can change as $R$ is varied in a broad vicinity of $R_{\rm FC}$ [this happens,
for example, for the component $\mu_y$ in panel (b)]. 
In the expansion of Eq.\ (\ref{muHT}) 
and in the calculation of the absorption spectra, this dependence is neglected
and the expansion coefficients are fixed to values corresponding to
 the dashed lines in  Fig.\ \ref{F03C}. In paper II, 
the comparison of the theoretical
photofragment distributions with experiment requires the $R$-dependence of
the TDMs to be incorporated into the calculations.

Strictly speaking, 
the Herzberg-Teller expansion in Eqs.\ (\ref{tdm0}) and (\ref{muHT}) is 
applicable to the adiabatic rather than diabatic states. In the globally  
diabatic representations, the TDMs are often taken coordinate independent,
with different diabats distinguished as \lq dark' or \lq bright' states;
such scheme was employed by Neville and Worth.\cite{NW14} Its
consistent implementation requires, however,
that all important intensity lending bright electronic states are
included in the calculation. The $\pi\pi^*$ states, carrying the oscillator
strength at the FC geometry, are missing in our calculations which are
restricted to the lowest $\pi\sigma^*$ states only.  Additionally, the CI
in the pair
$1^1\!A_2(\pi\sigma^*)/\widetilde{X}$ is located outside the FC zone and is
diabatized locally: The adiabatic-to-diabatic transformation matrix
[see e.g. Eq.\ (\ref{adt1})] smoothly goes over into a unit matrix 
as the interfragment distance $R$ moves outside a $1\,a_0$-wide
strip around the intersection. In the locally diabatic representation, the
adiabatic and the diabatic states in the FC zone coincide, and the 
Herzberg-Teller expansion, with the coefficients 
obtained directly from the electronic structure calculations, is justified. 

\section{Calculations of the absorption spectra}
\label{qm}

\subsection{Quantum mechanical calculations}
\label{qm1}

The linear absorption spectrum for the transition 
$1^1\!A_2(\pi\sigma^*) \leftarrow \widetilde{X}^1\!A_1(\pi\pi)$ is calculated
quantum mechanically using the molecular Hamiltonian of Eq.\ (\ref{Ham}). 
The ground vibrational state $\Psi_0(\RR,\QQ)$ 
of the Hamiltonian $\hat{T} + V^{X}$  is taken as
 the initial state of the parent molecule. The wave function 
$\Psi_0(\RR,\QQ)$ is strongly localized near 
$R = R_{\rm FC} \approx 4.15\,a_0$ [see Fig.\ \ref{F01}(a)] where 
the off-diagonal 
diabatic coupling matrix elements $V^{\alpha \beta}$ are negligible,
and the locally diabatic potential $ V^{X}$ is very close to the
adiabatic ground electronic state. 
The molecular state immediately after photoexcitation is given by 
\begin{equation}
\Phi_\epsilon(t=0) = \left( \bmu^{A_2}(\RR,\QQ)\cdot 
\hat{\boldsymbol{\epsilon}}\right) \Psi_0(\RR,\QQ) \, ,  
\label{InitState2}
\end{equation}
where $\hat{\boldsymbol{\epsilon}}$ denotes the polarization
vector of the electric field of the incident light. 
Initially, only the state $1^1A_2(\pi\sigma^*)$ is populated. 
The absorption spectra of the isolated state  $1^1A_2(\pi\sigma^*)$  and of
the coupled pair $\widetilde{X}/1^1A_2(\pi\sigma^*)$ are calculated 
using the MCTDH program package.\cite{BJWM00} 
First, the autocorrelation functions
for a given polarization direction $\epsilon = x, y$ or $z$, 
\begin{equation}
S_\epsilon(t) = \langle \Phi_\epsilon(0)| \exp 
\left(-i\hat{H} t \right) | \Phi_\epsilon(0) \rangle\, ,
\label{auto}
\end{equation}
are evaluated via the propagation on a discrete time grid. Next, the 
absorption spectra are 
calculated as Fourier transforms of $S_\epsilon(t)$:  
\begin{equation}
\sigma_\epsilon(E_{\rm ph}) =  
\frac{E_{\rm ph}}{2\epsilon_0 c} \int_{-\infty}^\infty 
S_\epsilon(t) {\rm e}^{i E_{\rm ph} t} {\rm d}t \, .   
\label{spectrum1}
\end{equation}
The photon energy $E_{\rm ph}$ is measured relative to the 
energy $E_0$ of the state $\Psi_0(\RR,\QQ)$. 
Averaging over the orientations of
the electric field gives the total absorption spectrum
\begin{equation}
\sigma_{\rm tot}(\omega) = 
\frac{1}{3} \sum_{\epsilon=x,y,z} \sigma_\epsilon(E_{\rm ph})
\label{spectrum2}
\end{equation}
An overview  of the quantum mechanical calculations discussed in this work
is given in Table \ref{T06}.  The calculations differ in the 
number of included electronic states (one or two), and in the 
number of dynamically active degrees of freedom, ranging from 6 to 15; 
the remaining degrees of freedom are kept fixed.

The diffuse bands in the absorption spectra are analyzed using the 
low storage filter diagonalization\cite{MT97A,M01A,CG97} applied to 
the autocorrelation functions $S_\epsilon(t)$. For long time signals,
this method allows one to decompose the spectrum into a set of resonance
states [eigenstates of the Hamiltonian of Eq.\ (\ref{Ham})]
with energies $E_l$ and widths $\Gamma_l$. Resonance states provide a direct
connection to the time resolved experiments. Indeed, the widths of
the intense (non-overlapping) resonances in the spectra are related
to the state-specific lifetimes via
\begin{equation}
\tau_{\rm res}(l) = \hbar/\Gamma_l \, .
\label{life1}
\end{equation}
Lifetimes $\tau_{\rm res}(l)$ 
provide theoretical counterparts to the  measured dissociation times
$\tau_d$. 
We also implement an  
alternative method 
to estimate the dissociation time scales associated with the 
spectra $\sigma_\epsilon(E_{\rm ph})$ based the 
time dependence of the population in the inner region of the potential
(i.e., the survival probability),  
\begin{equation}
P_\epsilon^\mathrm{FC}(t) = 
\frac{\left\bra \Phi_\epsilon(t) \left| \Theta(R_b - R) \right| 
\Phi_\epsilon(t) \right\ket}{\left\bra \Phi_\epsilon(0) \left| 
\Phi_\epsilon(0) \right. \right\ket}\, .
\label{fit1}
\end{equation}
Here $\Theta(x)$ is the Heaviside step function, and $R_b$ is the outer
boundary of the region in which H-atom and pyrrolyl are strongly interacting. 
For the $A_2$ state, this boundary is defined as a position of the barrier
top at $R_b = 4.6\,a_0$. 
The dissociation lifetimes are determined from the 
fit of the functions  $P_\epsilon^\mathrm{FC}(t)$ to an empirical expression
\begin{equation}
P_\epsilon^\mathrm{FC}(t) \approx
 \Theta(T_0 - t) + 
\Theta(t - T_0) \left[ a \exp\left(- \frac{(t - T_0)^2}{T_1^2} \right) + 
(1 - a)  \exp\left(- \frac{t - t_0}{T_2}\right) \right] \, ,
\label{fit2}
\end{equation}
depending on the parameters $T_0$, $T_1$, $T_2$, and $a$. The (very short)
transient
time $T_0$ signifies the time it takes the excited molecule to reach the 
boundary of the interaction region. The Gaussian term describes fast
direct dissociation with the time constant $T_1$, while the exponential term
accounts for the fraction of molecules trapped in resonance
states with the (average) lifetime $T_2$. The results of this analysis are
discussed in Sect.\ \ref{res_abs_2} and \ref{res_abs_3}.

\subsection{Absorption spectrum as a convolution}
\label{FCconv}

The quantum mechanical calculations described in the preceding section  
can be considerably simplified because the dissociation
dynamics in the repulsive $\pi\sigma^*$ states is mainly direct, 
and already the initial stages of the time evolution in the
excited state reveal the shape of the absorption spectrum. 
The N---H stretching frequency in
the ground electronic state is large, $\sim 3915$\,cm$^{-1}$, and the 
wave function $\Psi_0(\RR,\QQ)$ in Fig.\ \ref{F01}(a) is localized around 
$R_{\rm FC}$. This has two consequences. 
First, $\Psi_0(\RR,\QQ)$ is
accurately approximated by a product of an $\RR$- and a $\QQ$-dependent factor,
\begin{equation}
\Psi_0(\RR,\QQ) \approx \Psi_R(\RR) \Psi_Q(\QQ) \, .  
\label{InitState}
\end{equation}
Indeed, the Hessian matrix near $R_{\rm FC}$ is approximately
block diagonal, and the couplings between three 
coordinates of the dissociating H-atom on the one hand 
and the coordinates of the pyrrolyl moiety on the other hand are 
small. In the Herzberg-Teller approximation of Eq.\ (\ref{tdm0}), 
the photoexcited state $\Phi_\epsilon(0)$ 
either is in the same product form, 
\begin{equation}
\Phi_\epsilon(0) \approx F_R(\RR)f_Q(\QQ) \, ,
\label{InitState3}
\end{equation}
(if one of the TDM components $\bmu^{A_2}_R$ or 
$\bmu^{A_2}_Q$ in Eq.\ (\ref{tdm0}) vanishes)
or is a  sum of several such product terms 
(if both components $\bmu^{A_2}_R$ and 
$\bmu^{A_2}_Q$ are nonzero).  
Second, the diabatic potential matrix ${\bf W}_Q(\QQ|R)$ in Eq.\ (\ref{V_gen})
and, in particular, the functions 
$\kappa_i^\alpha(R)$ and $\gamma_{ij}^\alpha(R)$ in Eq.\ (\ref{V_op}), 
vary slowly with $R$ and can be fixed to their
values at a near-equilibrium distance $R \approx R_{\rm FC}$. Thus, 
the dynamics  
in the FC zone is approximately described by the Hamiltonian
\begin{eqnarray}
\hat{\MatVar{{\bf H}}}_0 & = & \hat{\MatVar{{\bf H}}}_R(\RR) + 
\hat{\MatVar{{\bf H}}}_Q(\QQ|R_{\rm FC}) \nonumber \\
\hat{\MatVar{{\bf H}}}_R & = & \hat{T}_R\MatVar{{\bf 1}} + 
\MatVar{{\bf U}}_R(\RR) \nonumber \\
\hat{\MatVar{{\bf H}}}_Q & = & 
\hat{T}_Q\MatVar{{\bf 1}} + \MatVar{{\bf W}}_Q(\QQ|R_{\rm FC}) \, ,
\label{Ham1}
\end{eqnarray}
which is a sum of two operators depending 
on $\RR$ and $\QQ$, respectively. The operators 
$\hat{\MatVar{{\bf H}}}_R$ and $\hat{\MatVar{{\bf H}}}_Q$ 
commute because in the
local quasi-diabatic representation the off-diagonal diabatic 
matrix elements vanish by construction. As a
result, the vibrational motion of the ring is 
decoupled from the dissociative dynamics along $\RR$. 
This separable approximation is valid for any number of locally diabatic
electronic states. 

The separability of the dissociative dynamics in the $\RR$-space 
and the vibrational dynamics in the $\QQ$-space  allows one
to express the total absorption spectrum as a convolution of the spectra
originating from the two spaces. This is demonstrated below for  
the dissociation taking place in the single 
state $1^1\!A_2(\pi\sigma^*)$ photoexcited via 
the TDM $\mu_z^{A_2}$ depending on a single $a_2$-symmetric 
mode. TDMs obeying the general 
Herzberg-Teller expansion  are treated in Appendix\ \ref{appa}. 

The autocorrelation function $S(t)$  of the initial state of 
Eq.\ (\ref{InitState3}) under the Hamiltonian of Eq.\ (\ref{Ham1})
is given by a product of the autocorrelation functions of the two subsystems:
\begin{equation}
S(t) \approx \langle F_R| \exp 
\left(-i\hat{H}_R t \right) | F_R \rangle_R\,\cdot\,
\langle f_Q| \exp 
\left(-i\hat{H}_Q t \right) | f_Q \rangle_Q \equiv s_R(t)s_Q(t)
\, ,
\label{auto1}
\end{equation}  
where the spatial integration variables are explicitly 
indicated for each set of angular brackets. 
Next,
the Fourier integral over $S(t)$ [i.e., the spectrum $\sigma(E_{\rm ph})$]
is transformed into a convolution of the Fourier
integrals over the functions $s_R(t)$ and $s_Q(t)$ via a convolution theorem
(introduce an integration over
the second time variable
$\delta(t-\tau)\,{\rm d}\tau$, replace the $\delta$-function
with an integral
$\exp\left[i(t-\tau)\omega\right]\,{\rm d}\omega$, and isolate the individual
Fourier integrals). Using the \lq spectral functions' without energy 
prefactors, 
\begin{eqnarray}
\bar{\sigma}_R(E) & = & \int_{-\infty}^\infty
s_R(t) {\rm e}^{i E t} {\rm d}t \, ,\nonumber \\
\bar{\sigma}_Q(E) & = & \int_{-\infty}^\infty
s_Q(t) {\rm e}^{i E t} {\rm d}t \, ,
\label{convfact}
\end{eqnarray}
the absorption cross section can be written as 
\begin{equation}
\sigma(E_{\rm ph}) =
\frac{E_{\rm ph}}{2\epsilon_0 c} \int_{-\infty}^\infty
\bar{\sigma}_R(E_{\rm ph}-\omega)\bar{\sigma}_Q(\omega)
{\rm d}\omega \, .
\label{spectrum3}
\end{equation}
In the $\RR$-space, the motion of the wave packet is 
(directly or indirectly) dissociative, while the motion 
in the quadratic potentials of the $\QQ$-space is bound. Thus, the 
absorption spectrum in Eq.\ (\ref{spectrum3}) 
consists of a series of 
excitations of the pyrrolyl ring broadened by the dissociation of the 
hydrogen atom. 

In the practical applications in this paper, the 
two convolution factors $\bar{\sigma}_R$ and $\bar{\sigma}_Q$ 
are calculated as follows. For the $\RR$-space, the initial 
3D wavefunction $\Psi_R(\RR)$ 
is defined as the ground vibrational state of the Hamiltonian for the
electronic state $\widetilde{X}$, with all ring modes $\QQ$ fixed to their values
at the $\widetilde{X}$ minimum. Next, $\Psi_R(\RR)$ is multiplied by the 
appropriate TDM functions (e.g. $\mu_x$ and $\mu_y$), and the resulting 
functions $F_R(\RR)$ are independently propagated with the
Hamiltonian $\hat{H}_R = \hat{T}_R + U_R(\RR)$, with the 
function $U_R(\RR)$ taken along the MEP in the state 
$1^1\!A_2(\pi\sigma^*)$. This gives the factors 
$\bar{\sigma}_R$ for each polarization. In a similar fashion, 
the bound vibrational  spectrum $\bar{\sigma}_Q$ is  calculated 
using the Hamiltonians 
$\hat{H}_Q = \hat{T}_Q + W^\alpha(\QQ|R_\mathrm{FC})$ 
of the states $\alpha = X$ [giving rise to the 
initial states $\Psi_Q(\QQ)$ and $f_Q(\QQ)$] and  
$\alpha=A_2$ (giving the final spectrum). 
As shown in Appendix \ref{appa}, more convolution terms 
are needed to approximate the absorption spectrum 
if the transition is 
induced by a TDM in a more general form. 
The accuracy of the approximation  is discussed in Sect.\ 
\ref{res_abs}. 

The convolution approach to diffuse 
absorption spectra can be considered as
an extension of the familiar FC computations of bound$-$bound 
transitions\cite{SGK00,BKPMKTS10,PMD06,CB12,GKT13} to the case
of dissociating systems. This approach considerably simplifies the assignment
of the diffuse spectral bands and, moreover, has several clear computational 
advantages. Indeed, the spectrum 
$\bar{\sigma}_Q(E)$ is given by the FC overlap integrals,  
\begin{equation}
\bar{\sigma}_Q(E) = \sum_{\bm m} 
\left|\langle \varphi_{\bm m}(\QQ)|f_Q(\QQ)\rangle\right|^2
\delta(E-E_{\bm m})
\, ,
\label{spectrum4}
\end{equation}
between the eigenfunctions $\varphi_{\bm m}(\QQ)$ (with energies
$E_{\bm m}$) of the non-disappearing modes in the FC zone
and the initial state $f_Q$.  
The harmonic stick spectrum $\bar{\sigma}_Q(E)$ 
can be efficiently calculated analytically
using the techniques developed for the FC factors in polyatomic
molecules,\cite{DMM77,TP01,CB12} and the only required 
ab initio input are the 
Hessians in the ground and the excited electronic states 
at a single point $R \approx R_{\rm FC}$. The 
main computational effort goes into the construction of the
potential matrix $\MatVar{{\bf U}}_R$ on 
the 3D grid of the disappearing modes
$(R,\theta,\phi)$, needed 
to evaluate the direct dissociation factor $\bar{\sigma}_R(E)$ --- 
but the number of grid points does not depend on the size of the molecule.

The convolution calculations are further simplified if, as in 
thiophenol,\cite{ADDKNO08} the 
$\pi\sigma^*$ state is purely repulsive. In this case, the structureless
spectrum $\bar{\sigma}_R(E)$  can be accurately reconstructed
using the reflection principle\cite{CS83A} which only requires 
the gradient of the 3D potential at the FC point. In the most optimistic
scenario, a convolution
calculation of the diffuse absorption spectrum becomes 
purely analytical, while all ab initio calculations 
refer to a single molecular
geometry near the FC point.


\section{Results}
\label{res_abs}

Absorption spectra for the $1^1\!A_2(\pi\sigma^*) \leftarrow \widetilde{X}$ 
transition are calculated using the Hamiltonian of Eq. (\ref{Ham}). 
Three calculations are discussed below, in which the following degrees of 
freedom are included: (i) $R,\theta,\phi,Q_{b_1}(1,2,3)$; 
(ii) $R,\theta,\phi,Q_{a_1}(1,2,3,4,5,6,7,8)$; and 
(iii) 
$R,\theta,\phi,Q_{a_1}(1,2,5),Q_{a_2}(1,2,3),Q_{b_1}(1,2,3),Q_{b_2}(1,3,5)$. 

Calculations (i) and (ii) are performed for the isolated 
electronic state $1^1\!A_2$ 
and highlight the specific absorption features due to the
non-totally symmetric (irrep $b_1$) and the totally symmetric (irrep $a_1$) 
modes, as well as the accuracy of the convolution approximation for them. 
All displacements for the  $b_1$ modes vanish by symmetry for pyrrole and
pyrrolyl regardless of the electronic state. 
The $a_1$ modes in FC region of the $A_2$ state are displaced relative to the 
equilibrium geometries of either pyrrole or
pyrrolyl. It is justified to consider the
single state dynamics for these coordinates: 
The coupling $U^{XA_2}_\mathrm{ang}(R,\theta,\phi)$ with the ground electronic
state is small everywhere, while the important coupling modes of $a_2$ symmetry
are not included. 
The MCTDH settings for these calculations are 
summarized in Table \ref{T07}. 
For each combined mode, the number $n_i$ of single particle functions (SPF) 
is chosen in order to have the population below $10^{-3}$ for the least  
populated natural orbital. 
The convergence with respect to the discrete variable representation (DVR) 
grid size ($N_i$) was ensured by transforming the single-particle functions 
into the finite basis representation (FBR) and checking that the FBR function with the highest quantum number has a population below $10^{-6}$. For the R coordinate, the grid spacing is $0.1\,a_0$, and a $3\,a_0$-wide complex absorbing potential $W = i \lambda\, \Theta(R - R_{\rm CAP})\, (R - R_{\rm CAP})^n$ is used, with $\Theta(x)$ the Heaviside step function, $R_{\rm CAP} = 10\,a_0$, 
$\lambda = 0.003\,{\rm hartree}/a_0^2$, $n = 2$.

The calculation (iii) is performed for the coupled pair $\widetilde{X}/A_2$. It
includes all modes of $a_2$ and $b_1$ symmetry, three $a_1$ modes with the 
largest displacement between the minima of pyrrole and pyrrolyl, as well
as three $b_2$ modes along which the Herzberg-Teller coefficients of the
TDMs are the largest.  This calculation accounts for 
the impact of the CI on the photodissociation dynamics and 
provides a realistic long wavelenegth spectrum of pyrrole 
comparable to the full-dimensional limit. 
In this calculation, smaller DVR grids for the disappearing modes are used
in order to speed up the computations and save memory (see Table \ref{T07}). We verified that these grid sizes were adequate for the 6D and the 11D calculations. Since the coordinates $R$ and $Q_{a_1}(1)$ are highly correlated, they were treated as a single combined mode. 
The dimension of the SPF basis ensures that all natural orbitals with populations above $10^{-3}$ are included. The convergence with respect to the number of SPF is fast in the $A_2$ state (the natural populations decrease exponentially) and slow in the $\widetilde{X}$ state. Therefore, a larger number of configurations would be necessary to fully converge the wave packet associated with the $\widetilde{X}$ state. However, the 
population transfer from $A_2$ to  $\widetilde{X}$ 
is under 
$10\%$ and increasing the number of configurations is expected to have minor
effect on the absorption spectra and the product state distributions.

The $A_2 \leftarrow \widetilde{X}$ transition is electric dipole forbidden, and
the absorption cross sections are small, of the order of 
$10^{-20} \ \mbox{cm}^{-1}$. The absorption corresponding to this transition
is overlayed by the intense band of the neighboring $\pi\pi^*$ 
states,\cite{RWYCYUS13} and the spectrum of the state $1^1\!A_2(\pi\sigma^*)$ 
has never been measured. The 
experimental characterization of the photodissociation in this state 
is more advanced in time domain.\cite{WNSSAWS15,LRHR04,RWYCYUS13} 
Below, the absorption spectra are discussed together with the 
autocorrelation functions, which contain information on 
the dissociation lifetimes. A summary of the experimental and the calculated 
lifetimes is given in Table \ref{T00}.

\subsection{6D absorption spectrum: Coordinates $R,\theta,\phi,Q_{b_1}(1,2,3)$}
\label{res_abs_1}

The isolated state $1^1\!A_2(\pi\sigma^*)$ in this calculation
is excited via the TDMs $\mu_x$ and $\mu_y$ given in Table \ref{T06}. 
The initial wave functions $\Phi_x(0)$ and $\Phi_y(0)$ 
have $b_2$ and $b_1$ symmetry, respectively. 
The TDM $\mu_y$ and the initial wave function 
$\Phi_y(0)$ consist of two components, one promoted by 
the out-of-plane bending angle $\theta$, and the other
by the  ring modes $Q_{b_1}$. The excluded coordinates are set equal to 
$\QQ_\mini(R)$, i.e. they adiabatically follow the MEP in the state
$1^1\!A_2(\pi\sigma^*)$ described in Sect. \ref{qchem}.

\begin{figure}[h!]
\caption{(a) Quantum mechanical 
absorption spectra for the isolated state $^1\!A_2$  
in the 6D calculation. 
The profiles $\sigma_x$ (red),  $\sigma_y$ (green), and the total
absorption spectrum $\sigma_{\rm tot}$ (black) are shown. 
(b) The same spectra as in (a) evaluated using the convolution approximation.
(c) The vibrational spectrum $\bar{\sigma}_Q$ used in the convolution. 
(d) The autocorrelation functions $|S_x|$ (red) and $|S_y|$ (green) versus 
time.
(e) The populations in the FC region $P^{\rm FC}_x(t)$ (red) and 
$P^{\rm FC}_y(t)$ (green) as functions of time. 
}
\label{F01res}
\includegraphics[scale=0.33]{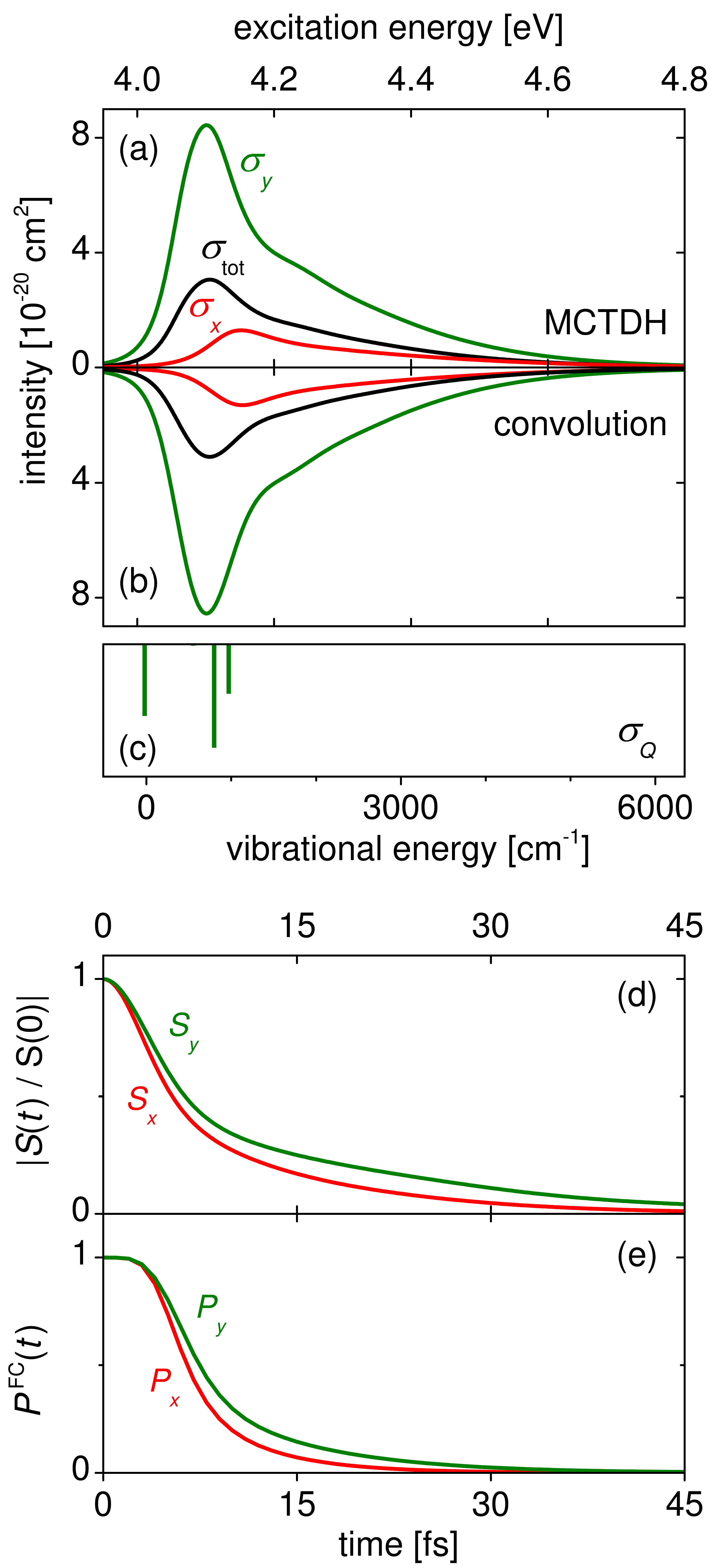}
\end{figure}

The spectra for the two polarizations, as well as 
the total absorption spectrum $(\sigma_x + \sigma_y)/3$,  
are shown in Fig.\ \ref{F01res}(a). The total absorption maximum of about
$ 3 \cdot 10^{-20} \mbox{ cm}^2$ is reached at $E_\mathrm{max} = 4.11$\,eV, 
close to the vertical excitation energy of $T_v = 4.19$\,eV. Note that 
this $T_v$ is smaller than the value of 4.80\,eV given in Table \ref{T01A}. 
The reason is the choice of the fixed values of the totally symmetric $a_1$
modes in this calculation: In the excitation zone, $\QQ = \QQ_\mini(R)$
corresponds to the minimum of the state $1^1\!A_2(\pi\sigma^*)$, but not of
the state $\widetilde{X}$ which is elevated by $\sim 0.6$\,eV. 
All spectra in Fig.\ \ref{F01res}(a) are composed of the main 
peak and the shoulder on the high energy side. The main peaks lie at  
$E_\mathrm{max} = 4.15$\,eV (for $\sigma_x$) and 
$E_\mathrm{max} = 4.10$\,eV (for $\sigma_y$), while the  
full widths at half maximum (FWHMs) are about 0.15\,eV. The broad maxima
in both components correspond to quasi-bound resonances
supported by the local minimum of the potential  $V^{A_2}$. The 
FWHM is determined by two factors, the number of the excited 
(mostly short lived) resonances and their linewidths. While
the vibrational assignments are addressed below, one aspect of the
excitation pattern follows directly from the molecular geometry. 
The N---H bond in the $1^1\!A_2(\pi\sigma^*)$ 
is elongated by $0.12\,a_0$ relative to the state $\widetilde{X}$, and the 
local minimum of $V^{A_2}$ is shifted to larger $R$ distances. 
Correspondingly, the  main peaks and the
shoulders are built on zero quanta and one quantum 
of N---H stretch, respectively. 

The maximum absorption for $\sigma_y$ is more than 8 times stronger
than for $\sigma_x$, even though the Herzberg-Teller coefficients for the 
TDMs $\mu_x$ along the in-plane bending and $\mu_y$ along the out-of-plane
bending modes are comparable (see Table \ref{T02}). The intensity of 
$\sigma_y$ is large due to a subtle enhancement effect: The 
out-of-plane bending frequency is $\sim 3$ 
times smaller than the in-plane one, and the range of $\theta$ covered by 
the parent wave function $\Psi_0(\RR,\QQ_{b_1})$ along the out-of-plane
direction is about $\sim \sqrt{3}$ broader than along the in-plane direction.
Consequently, the integrated TDM, sampled over an enhanced $\theta$ range, is
larger by a factor of $\sim 3$, and the calculated intensity 
for $\sigma_y$ is amplified by  $\sim 3^2$.  

Using the low storage filter diagonalization,
the main absorption peaks
can be uniquely assigned to intense resonance states located at 
$E_\mathrm{res} = 4.14$\,eV (for $\sigma_x$) and at $E_\mathrm{res} = 4.07$\,eV
and 4.11\,eV (for $\sigma_y$). These energies correspond to one quantum
excitations of the in-plane H-atom bend, the out-of-plane H-atom 
bend, and the ring mode $Q_{b_1}(3)$, respectively  
(corresponding frequencies are 
$0.13$\,eV, $0.08$\,eV, and $0.12$\,eV, see Table\ \ref{T02}). 
These modes have large Herzberg-Teller coefficients in Table \ref{T02},
and their excitation is expected. Several 
short lived resonance states are found 
at energies correlating with the spectral shoulders and assigned to overtone
excitations of the out-of-plane H-atom bend (for $\sigma_y$) and the
combination bands involving one quantum of N---H stretch (for $\sigma_x$
and $\sigma_y$).

The lifetimes $\tau_{\rm res}$ 
of the resonances corresponding to the main peaks provide 
estimations of the quantum mechanical dissociation time scale. 
We find 
5.6\,fs for $\sigma_x$ and 8\,fs for $\sigma_y$. These lifetimes are
of the order of the shortest dissociation time $\tau_d = 12$\,fs measured in
 Ref.\ \onlinecite{WNSSAWS15} (see Table \ref{T00}).  
The calculated $\tau_{\rm res}$ is short indicating that the effective 
potential barrier in the FC region is not high enough to support 
tunneling regime. In Ref.\ \onlinecite{WNSSAWS15}, this time scale was
associated with the direct dissociation in the state $1^1\!A_2(\pi\sigma^*)$, 
too. 

Dissociation time scale can also be assessed using the 
autocorrelation functions $S_\epsilon(t)$ and the population in the inner
potential region $P_\epsilon^\mathrm{FC}(t)$. They are depicted in
Fig. \ref{F01res}(d,e). The autocorrelation functions for both polarizations 
decay monotonically without visible recurrences, and the exponential
time constants of 7.5\,fs
(for $S_x$) and 9.5\,fs (for $S_y$) are
consistent with the resonance lifetimes. 
The populations in the FC zone [Eq.\ (\ref{fit2})] have a very
short \lq induction' plateau $T_0$, while the time scales for the direct
and indirect contributions are almost equal: $T_1 \approx T_2 
\approx 5.2$\,fs (for $P_x^\mathrm{FC}$) and $T_1 \approx T_2 
\approx 6.5$\,fs (for $P_y^\mathrm{FC}$). This is yet another indication that 
tunneling contribution is small. 
In order to verify this conclusion, we estimate
the isotope effect of the reaction,
\begin{equation}
\label{kie1}
KIE = \frac{\tau(D)}{\tau(H)} \, ,
\end{equation}
by substituting D for H in the N---H group (\lq pyrrole-$d_1$')
and repeating the quantum mechanical calculations. Based on the 
resonance lifetimes as well as the survival probabilities, a modest isotope 
effect of about 2.5---3.5 is predicted. This confirms 
that tunneling does not significantly affect dissociation in the state
$1^1\!A_2(\pi\sigma^*)$, and the lowest barrier $E^\ddagger_{\rm MEP}$
in Table \ref{T01} is dynamically relevant. Roberts et al. 
studied the isotope effect in the photodissociation of pyrrole and found
that the strength of the effect depends on the measured dissociation 
lifeimes: For
$\tau_d = 46$\,fs, they found $KIE \sim 2.9$ which is close to our result; 
for a much longer lifetime of 126\,fs, the measured $KIE$ is about 
11.\cite{RWYCYUS13} 

Figure \ref{F01res}(b) shows the spectra calculated using the 
convolution approximation of Eq. (\ref{spectrum3}) for each polarization
separately. The agreement with the MCTDH calculations is excellent
for all spectra. Comparison with calculations involving other pyrrolyl modes
and described below suggests that the convolution is especially effective for
the non-totally symmetric oscillators (like the $b_1$ modes) because their 
minima are not displaced in either $\widetilde{X}$ or $A_2$ states and they are
to a large extent decoupled from the H-atom coordinates $\RR$. As a result, 
the separable approximation of Eqs. (\ref{Ham1}) and (\ref{auto1}) is accurate.

The two convolution factors $\bar{\sigma}_R$ and $\bar{\sigma}_Q$ 
[Eq. (\ref{convfact})] can be distinguished in the spectra. In the excitation
via $\mu_x$, the initial vibrational wavefunction $f_Q(\QQ_{b_1})$ overlaps 
only with the ground state of the pyrrolyl ring in the state 
$1^1\!A_2(\pi\sigma^*)$. The discrete spectrum $\bar{\sigma}_Q$ 
(not shown) consists of
a single peak, and the convoluted spectrum $\sigma_x$ in Fig.\ \ref{F01res}(b)
coincides with the dissociative factor $\bar{\sigma}_R(E)$. 
For the $y$-polarized transition, 
the harmonic FC spectrum $\bar{\sigma}_Q$ includes several lines 
shown in Fig.\ \ref{F01res}(c). 
The position of $\bar{\sigma}_Q$ on the photon energy scale 
is chosen so that the vibrational states of the ring have  
correct excitation energies relative to the quantum mechanical 
band origin evaluated in the convolution approximation. 
The TDM $\mu_y$
depends on  both the out-of-plane H bending and the $b_1$ modes (cf. Table
\ref{T06}). Consequently, three vibrational states make the largest 
contribution to $\bar{\sigma}_Q$: 
The ground vibrational state of the ring (accessed because the excitation 
resides on the out-of-plane H bending), as well as the ring 
states with one quantum on the modes 
$Q_{b_1}(2)$ or $Q_{b_1}(3)$. 
The Herzberg-Teller coefficient for the mode $Q_{b_1}(1)$ is small, and this
mode is not excited. As expected,
this assignment is consistent with one given for the 
resonance states. This is an illustration that the convolution 
method can be effectively used to automatically 
label vibrational peaks in electronic
spectra. The convolution approximation confirms that the shoulder of the 
spectrum $\sigma_y$ is due to the same ring states combined with the overtone
excitations of the disappearing modes. This spectral pattern --- a given
set of the ring quanta \lq translated' to higher energy absorption
bands via additional excitations of the N-H stretch or H bend --- is a 
characteristic feature of the absorption spectrum of pyrrole.

\subsection{11D absorption spectrum: Coordinates $R,\theta,\phi,Q_{a_1}(1,2,3,4,5,6,7,8)$}
\label{res_abs_2}
In this calculation, all $a_1$ ring modes are included and all
remaining non-totally symmetric pyrrolyl modes are fixed to the equilibrium
values of $\QQ = {\bm 0}$. As in the 6D case, 
the isolated state $1^1\!A_2(\pi\sigma^*)$ 
is excited via the TDMs $\mu_x$ and $\mu_y$ (Table \ref{T06}) giving the 
initial states $\Phi_x(0)$ and $\Phi_y(0)$ of 
$b_2$ and $b_1$ symmetry. Each TDM consists of only one term depending on
the angular coordinates $(\theta,\phi)$  and representing
the in-plane ($\mu_x$) and the out-of-plane ($\mu_y$) excitations of
the H-atom bending in the state $1^1\!A_2(\pi\sigma^*)$. 

\begin{figure}[h!]
\caption{\scriptsize (a) Quantum mechanical 
absorption spectra for the isolated $1^1\!A_2$ state of pyrrole 
in the 11D calculation. 
The profiles $\sigma_x$ (red),  $\sigma_y$ (green), and the total
absorption spectrum $\sigma_{\rm tot}$ (black) are shown. 
(b) The same spectra as in (a) evaluated using the convolution approximation.
(c) The \lq spectra' $\bar{\sigma}_Q$ (sticks) and $\bar{\sigma}_R$ 
(gray shade) used in the convolution. 
(d) The  
autocorrelation functions $|S_x|$ (red) and $|S_y|$ (green) versus time.
(e) The populations in the FC region $P^{\rm FC}_x(t)$ (red) and 
$P^{\rm FC}_y(t)$ (green) as functions of time. 
The vibrational bands, marked with letters A--G in (a) and (b), are assigned
in terms of the quantum numbers $n_{a_1}(i), \ i = 1,8$ of the $a_1$ modes
of the pyrrolyl ring. 
The states mostly contributing to each peak of $\sigma_y$ are: 
(A) $n_{a_1}(i)=0$, for $i=1,...,8$; 
(B) $(n_{a_1}(1) = 1)$, $(n_{a_1}(2) = 1)$, $(n_{a_1}(3) = 1)$; 
(C, D) ($n_{a_1}(5) = 1$); 
($n_{a_1}(2) = 2$), ($n_{a_1}(2) = 1, n_{a_1}(3) = 1$); 
(E)  ($n_{a_1}(2) = 1, n_{a_1}(5) = 1$), ($n_{a_1}(3) = 1, n_{a_1}(5) = 1$);
(F) ($n_{a_1}(2) = 2, n_{a_1}(3) = 1$); ($n_{a_1}(1) = 1, n_{a_1}(2) = 2$); 
(G)  ($n_{a_1}(2) = 1, n_{a_1}(3) = 1, n_{a_1}(5) = 1$), 
($n_{a_1}(2) = 2, n_{a_1}(5) = 1$).
The listed ring excitations are accompanied by the lowest allowed
(\lq obligatory') excitations of the disappearing modes.  
}
\label{F02res}
\includegraphics[scale=0.28]{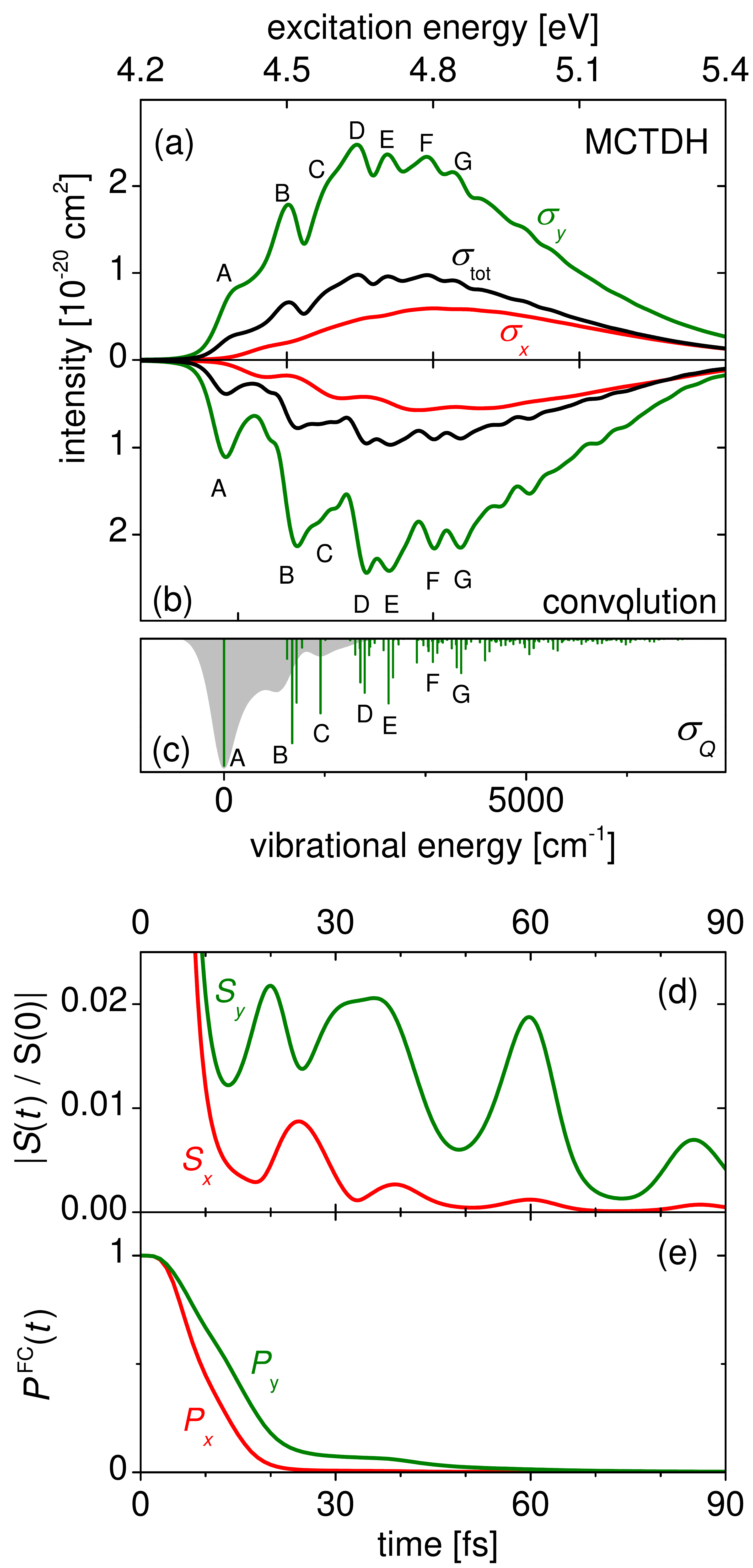}
\end{figure}

The total absorption spectrum $\sigma = (\sigma_x + \sigma_y)/3$ and the 
spectra  $\sigma_x$ and $\sigma_y$ are given in Fig. \ref{F02res}(a). 
Similarly to the 6D case, the absorption $\sigma_x$ (with the maximum of 
$6 \cdot 10^{-21} \ \mathrm{cm}^2$) is almost an order of magnitude weaker
than $\sigma_y$ (the maximum is $\sim 2.5 \cdot 10^{-20} \ \mathrm{cm}^2$).
For the total spectrum, 
the maximum intensity of $\sim 10^{-20} \ \mathrm{cm}^2$ is found at
$E_\mathrm{max} = 4.75$\,eV, near 
$T_v = 4.80$\,eV. The vertical excitation energy in this calculation is 
the same as for the full 21D Hamiltonian in Table  \ref{T01A}: With all
totally symmetric modes included, the potential well of
the 11D potential $V^X$ attains its global minimum, and the energy
separation with the
local minimum of $V^{A_2}$ is the largest.  The distinct features against 
the 6D case are the diffuse structures superimposed 
on the $\sim 0.6$\,eV broad background
in the spectrum $\sigma_y$ (more pronounced) and in the total absorption 
(less pronounced). In Fig. \ref{F02res}(a), 
they are labeled with letters A---G. Their assignment, clarified
below using the convolution method, reflects the 
principal geometry changes between the ground and the excited electronic 
state. The appropriate coordinates for such analysis are the 
normal modes $\widetilde{\QQ}_{a_1}$ of the pyrrolyl ring at the local 
minimum of the potential $V^{A_2}$. These modes are used in the following
discussion; they are numbered in order of 
their increasing frequency $\widetilde{\omega}$ (see Table \ref{T02}). 
The stronger the shift $\delta Q$ 
of a particular mode
between the minima of $V^X$ and $V^{A_2}$, the more overtones of 
this vibration are excited in the absorption spectrum.
The largest dimensionless
displacements $\delta Q$ are found for the modes 
$\widetilde{Q}_{a_1}(1)$ ($\widetilde{\omega} = 908\,{\rm cm^{-1}}$,
$\delta Q = 0.54$), $\widetilde{Q}_{a_1}(2)$ 
($\widetilde{\omega} = 1124\,{\rm cm^{-1}}$, $\delta Q = 1.32$), 
$\widetilde{Q}_{a_1}(3)$ ($\widetilde{\omega} = 1155\,{\rm cm^{-1}}$, 
$\delta Q = 1.03$), and 
$\widetilde{Q}_{a_1}(5)$ ($\widetilde{\omega} = 1584\,{\rm cm^{-1}}$, 
$\delta Q = 1.15$). 
Thus, several ring modes plus
the dissociation coordinate $R$ have appreciable shifts, and several distinct
vibrational progressions are expected in the 11D case.  
This is the reason why the overall width of the 11D spectra is about
four times the width of the 6D spectra. The remaining modes 
$\widetilde{Q}_{a_1}(4,6,7,8)$ are minimally shifted, 
and behave as \lq spectators'. 

\begin{figure}[h!]
\caption{Contour plots of the diabatic potentials of
the states $\widetilde{X}$ (blue, bottom) and $1^1\!A_2(\pi\sigma^*)$
(red, top) in the plane 
$(R,\theta)$. For each $R$, the normal modes are set to their values on 
the MEP  in the $1^1\!A_2$ state. 
The angle $\phi$ is set to 0\textdegree. 
Energies marked on the contour lines are in eV. The 
energy zero is at the global minimum of the 
$\widetilde{X}$ state of pyrrole.} 
\label{F02}
\bigskip
\includegraphics[scale=0.73]{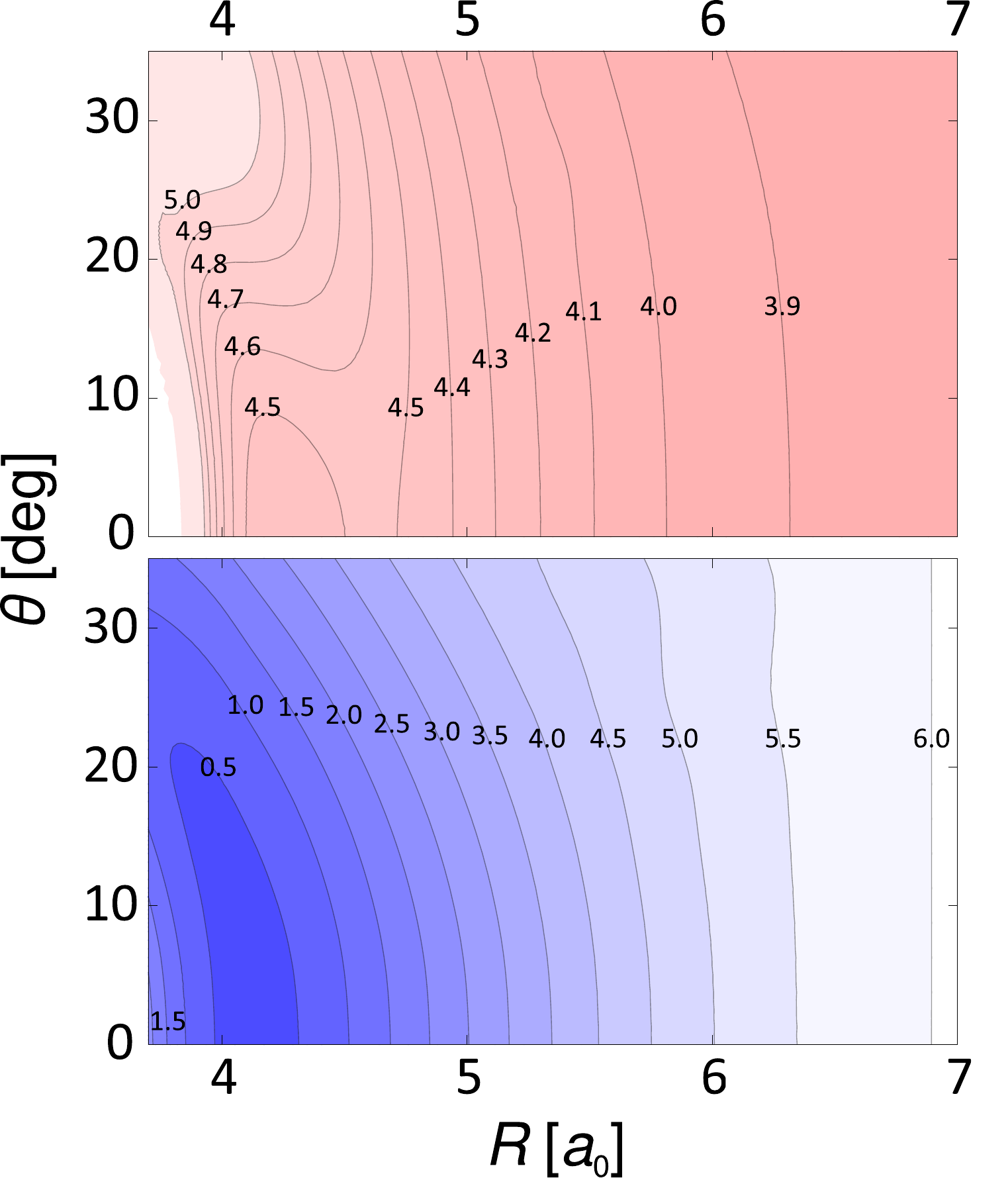}
\end{figure}

Note that although many vibrational states
are excited in both $\sigma_x$ and $\sigma_y$, 
the lines are considerably broadened by dissociation, and many
members of the vibrational progressions remain unresolved. The broadening is 
more pronounced in the spectrum $\sigma_x$ rendering it smooth 
especially if compared to the structured spectrum $\sigma_y$: 
Dissociation of the in-plane excited wave packet is considerably faster
than of the out-of-plane one. This is directly confirmed with the 
autocorrelation functions depicted in Fig.\ \ref{F02res}(d). 
Although both $S_x$ and $S_y$ rapidly decrease in the first several 
femtoseconds, the in-plane $S_x$ drops, relative to its value at $t=0$, 
by three orders of magnitude --- 5 times more
than $S_y$. The origin of this difference lies in the 
potential energy profiles of $V^{A_2}$ along the dissociation direction 
$R$ and along the polar angle $\theta$ for the H-atom moving in the 
state $1^1\!A_2(\pi\sigma^*)$ 
perpendicular to pyrrolyl plane ($\phi = 0^\circ$) and 
in the pyrrolyl plane ($\phi = 90^\circ$). The contour plot in Fig.\
\ref{F02}(a) is for $\phi = 0^\circ$. The local minimum near $R = 4.2\,a_0$, 
which is the deepest
for $\theta = 0^\circ$, persists for $\theta > 0^\circ$, too. 
 In fact, even H-atoms displaced by 
$\theta = 40^\circ$ above the ring still experience a barrier to
dissociation.
For the in-plane motion ($\phi = 90^\circ$; not shown in Fig.\ \ref{F02}), the
local minimum exists only for $\theta < 10^\circ$, and no barrier hinders
dissociation for larger in-plane displacements along $\theta$. As a result,
the dissociation is direct and fast for the in-plane excitations 
($\sigma_x$ in this case).

The dissociation time scale, prevailing near the absorption maximum, 
is established using the 
population in the inner
potential region $P_y^\mathrm{FC}(t)$ shown in Fig. \ref{F02res}(e). The 
plateau region $T_0$ is of the order of 1\,fs, the direct (Gaussian) 
decay proceeds with the time constant of $T_1 = 12.8$\,fs and accounts for 
80\% of dissociating molecules, while the indirect exponential 
lifetime is $T_2 = 23.8$\,fs and has a weighting factor of 20\%. These time
constants are also covered in the resonance spectrum obtained via 
filter diagonalization: Several intense resonance states are found in the 
vicinity of each diffuse structure in the spectrum $\sigma_y$, with the 
lifetimes ranging from 8\,fs (for $E_\mathrm{res} = 4.49$\,eV) to 
30\,fs (for $E_\mathrm{res} = 4.66$\,eV). 
The potential $V^{A_2}$ also supports
resonance states with lifetimes over 100\,fs. Such states are detected with
filter diagonalization at low excitation energies. However, their intensity
is very small in our calculations, less than 10$^{-4}$ of the intensity 
of the short
lived states, and they are not included in the analysis. 
The comparison with the 
experimental values is summarized in Table \ref{T00}.
The short time scale agrees with  $\tau_d = 12$\,fs measured\cite{WNSSAWS15}  
at $\lambda = 242$\,nm using the
time-resolved photoelectron spectroscopy sensitive to the population of the
excited state in the FC zone.
The longer time constant is close to the 
value of $19$\,fs from the same experiment, and also overlaps
with the dissociation times of $\tau_d = 39$\,fs 
at $\lambda = 249.5$\,nm (Ref.\ \onlinecite{KPNWF17}; the time
resolution of 3\,fs) and 46\,fs at $\lambda = 238$\,nm 
(Ref.\ \onlinecite{RWYCYUS13}; the time resolution of 30\,fs).

The isotope effect is estimated using Eq.\ (\ref{kie1}) 
in a separate quantum mechanical calculation of pyrrole-$d_1$. For the 
D-substituted molecule, the direct dissociation time $T_1$ increases only 
slightly, so that $KIE \approx 1.0$. The indirect time constant 
$T_2$ becomes 41.8\,fs with $KIE \approx 1.8$,  slightly smaller than in 6D.
Tunneling might indeed 
contribute to the time constants longer than 20\,fs, but this contribution
tends to decrease as more degrees of freedom are added.
Calculations of resonance states support this conclusion, and predict 
the isotope effect between 1.0 (for states with lifetimes of under 10\,fs)
to 2.5 (for lifetimes over 20\,fs), although in the 11D case it is not 
always possible to uniquely 
match  resonance states in pyrrole and in pyrrole-$d_1$. Again, the calculated
$KIE$ is in close agreement with the experimental estimate given in Ref.\
\onlinecite{RWYCYUS13} for similar lifetimes. 

The autocorrelation functions in Fig.\ \ref{F02res}(d) suggest vibrational
assignments 
in the time domain.  In $|S_y|$, the shortest recurrence time of 21 fs 
corresponds to the frequency of about 1600\,cm$^{-2}$ 
associated with the mode $\widetilde{Q}_{a_1}(5)$.
Another broad recurrence of nearly the same intensity is seen 
around 35\,fs. It is associated with low-frequency $a_1$ modes
$\widetilde{Q}_{a_1}(1,2,3)$, 
which span the frequency range 900--1150\,cm$^{-2}$. Based on this analysis
and assuming that the band A in the spectrum $\sigma_y$ 
carries zero quanta in the ring modes, we expect to find 
vibrational states (i) with one quantum on 
$\widetilde{Q}_{a_1}(5)$ (e.g. near the shoulder C at $\Eph = 4.60$\,eV) and
(ii) with one quantum on $\widetilde{Q}_{a_1}(1,2,3)$ (e.g. near the band B
peaking at $\Eph = 4.50$\,eV). This is in line with the 
discussion based on shifts $\delta Q$ of normal modes in the
initial and final electronic states. 

Fig. \ref{F02res}(b) depicts the spectra $\sigma_x$ and $\sigma_y$, 
calculated using the convolution method. 
The resulting absorption profiles agree well with the exact 
MCTDH calculation: 
The convoluted spectra are correctly positioned on the energy scale and the 
moments of the spectral envelope (width, asymmetry, etc.) are well reproduced.
The lower resolution of the $\sigma_x$ profile, compared to $\sigma_y$, 
is also correctly captured. On the other hand, the absorption bands 
in the convoluted spectrum $\sigma_y$ are slightly
more pronounced than in its MCTDH counterpart. This indicates
that the spectral broadening, 
represented by the factor $\bar{\sigma}_R$, 
is underestimated and the coupling between $\QQ$- and $\RR$-spaces
is only approximately taken into account. 
In fact, 
the  11D calculation including all $a_1$ vibrational modes represents a 
stringent test for the method, because anharmonic coupling with the 
totally symmetric dissociation coordinate $R$ is symmetry-allowed, and
for many $a_1$ modes this coupling is indeed strong.  

In these calculations, the 
convolution factors $\bar{\sigma}_R$ exemplified in Fig.\ \ref{F02res}(c) 
consist of 
the main peak and several high energy shoulders, i.e. 
they are similar to those 
discussed in the 6D case. Their FWHMs of 0.18\,eV (for $\mu_x$) and 
0.12\,eV (for $\mu_y$) also compare well with 6D. The 
spectrum $\bar{\sigma}_R$ is broader for the $x$-polarized 
case, indicating faster decay of the in-plane excitation,
in agreement with the analysis given above in terms of the potential energy
curves.  The total spectral width of 0.6\,eV 
is approximately four times larger than the width of $\bar{\sigma}_R$, so 
that most of the spectral broadening is due to the long 
$a_1$ progressions in the convolution factor $\bar{\sigma}_Q$ 
(FWHM of $\sim 0.4\,{\rm eV}$) shown in 
Fig.\ \ref{F02res}(c).  
In the calculation of $\bar{\sigma}_Q$, 
the value of the dissociation coordinate for the state 
 $1^1\!A_2$ is fixed at $R_\mathrm{FC}=4.23\,a_0$. 
The vibrational ground state of the ring in the state 
$\widetilde{X}$ is taken at $R_\mathrm{FC}=4.10\,a_0$.

The sequence of vibrational peaks in the convoluted spectrum $\sigma_y$ 
is similar to the one found in the exact MCTDH spectrum. 
The intense peaks in the exact and the convoluted spectra are matched by 
visual inspection, and the corresponding bands are 
labeled with the same letters in Figs. \ref{F02res}(a) and (b). 
The same bands can be found in the 
total spectrum $\sigma_{\rm tot}$, too, albeit  with smaller intensity. 
They can be unequivocally matched to the 
peaks in the harmonic convolution factor $\sigma_Q$ in Fig. \ref{F02res}(c). 
One 
advantage of the convolution method is that the constructed spectrum --- like
any other FC spectrum --- is 
automatically assigned in terms of the ring modes, 
because each vibrational contribution to
$\bar{\sigma}_Q$ [Eq. (\ref{spectrum4})] is known. In addition, the
dissociative factor $\bar{\sigma}_R$ giving the band shape of each spike in
$\bar{\sigma}_Q$ helps to identify the excitations of the disappearing modes
in the absorption bands. Guided by the shape of the spectrum
$\bar{\sigma}_R$, we distinguish to main groups of excitations contributing
to a given band, one due to the lowest allowed (\lq obligatory') excitation
of the disappearing modes and another involving overtone excitations in the
$\RR$-space. The first group is associated with the main peak in 
$\bar{\sigma}_R$, the second --- with the shoulders. The 
assignments for the bands A---G corresponding to the first group 
are given in the caption to Fig.\ \ref{F02res}. The resulting assignments
confirm the previous analyses based on the shifts $\delta Q$ and the 
autocorrelation function. Indeed, 
the most intense peaks involve excitations in the 
modes which are strongly displaced between the minima of $V^X$ and $V^{A_2}$,
namely $\widetilde{Q}_{a_1}(2)$, $\widetilde{Q}_{a_1}(5)$, which give rise to the most 
intense progressions, and $\widetilde{Q}_{a_1}(3)$, which gives rise to 
slightly weaker peaks. The 
vibrational couplings between $R$ and the $a_1$ modes are neglected in 
the convolution method based on the separable approximation, and
the vibrational frequencies do not perfectly coincide with those in 
the MCTDH calculation.  For example, 
the band B, involving one quantum excitations on the modes 
$\tilde{Q}_{a_1}(2)$ and $\tilde{Q}_{a_1}(3)$, appears to be shifted to 
higher energies by $\sim 120\,{\rm cm}^{-1}$;
the energy of the shoulder C, 
to which the mode $Q_{a_1}(5)$ contributes significantly,
is underestimated by $\sim 130\,{\rm cm}^{-1}$.

The second group of excitations involving overtones of the disappearing
modes can also be distinguished in many bands. One example is the normal 
vibrational ring state ${\bf 0}$ which, combined with the obligatory
out-of-plane bending excitation $n_\theta = 1$, provides the main assignment
of the band A in the spectrum $\sigma_y$. The same ring state, augmented with
two more bending quanta ($n_\theta = 3$), contributes to the band B and,
augmented with the NH stretch excitation ($n_R = 1$), to the band D. Further, 
the ring excitations contributing to the band B are also found in the band D
(with the $\RR$-space excitation $n_\theta = 3$). The ring states 
$n_{a1}(2) = 1$, $n_{a1}(3) = 1$, and $n_{a1}(5) = 1$, dominating the band D,
also 
contribute, via the additional excitation $n_\theta = 3$ to the band F. 
In fact, most ring 
excitations listed in Fig.\ \ref{F02res} are found in the higher lying 
absorption bands with $n_\theta > 1$ and $n_R > 0$. The energy stored in
the disappearing modes during photoexcitation is expected to get released into
rotations and translation in the course of dissociation. 

\subsection{15D absorption spectrum: Coordinates $R,\theta,\phi,Q_{a_1}(1,2,5), Q_{a_2}(1,2,3), Q_{b_1}(1,2,3), Q_{b_2}(1,3,5)$}
\label{res_abs_3}

The third calculation discussed in this paper
includes two coupled electronic states, $\widetilde{X}$ and 
$1^1\!A_2(\pi\sigma^*)$. Three modes of each symmetry are
dynamically  active, in 
particular all symmetry-allowed coupling modes $Q_{a_2}(1,2,3)$ 
of the ring are included. 
All other modes are set to their equilibrium values for pyrrolyl.  
The excitation $1^1\!A_2 \leftarrow \widetilde{X}$
is mediated by three TDM components $\mu_x$, $\mu_y$ and 
$\mu_z$. The initial states $\Phi_x(0)$, $\Phi_y(0)$, and $\Phi_z(0)$ 
(cf. Table \ref{T06}) belong
to the irreps $b_2$, $b_1$, and $a_2$, respectively.  The states  
$\Phi_x(0)$ and $\Phi_y(0)$ are linear combinations of 
excitations of the disappearing modes $(\theta,\phi)$ and the one-quantum
excitations of the ring. The state $\Phi_z(0)$ includes only 
the fundamental excitations
of the ring modes $Q_{a_2}$. 

\begin{figure}[h!]
\caption{\scriptsize (a) Quantum mechanical 
absorption spectra for the coupled $1^1\!A_2/\widetilde{X}$ states 
in the 15D calculation. 
The components $\sigma_x$ (red),  $\sigma_y$ (green), $\sigma_z$ (blue), 
and the total
absorption spectrum $\sigma_{\rm tot}$ 
(black) are shown. 
(b) The same spectra evaluated using the convolution approximation.
(c) The \lq spectra' $\bar{\sigma}_Q$ (sticks) and $\bar{\sigma}_R$
(gray shade) used in the convolution.
(d) The spectrogram $S_y(E,\tau)$ as defined in Eq.\ (\ref{vibro}). 
(e) The  
autocorrelation functions $|S_x|$ (red), $|S_y|$ (green) 
and $|S_z|$ (blue) versus time.
(f) The populations in the FC region $P^{\rm FC}_x(t)$ (red), 
$P^{\rm FC}_y(t)$ (green) and $P^{\rm FC}_z(t)$ as functions of time.  
In (c) and (d), the results for 
the $y$-polarized excitation induced by the TDM $\mu_y(\RR)$ are shown. 
The vibrational bands marked with letters A--F in (a) and (b) are assigned
in terms of the quantum numbers $n_\Gamma(i)$ belonging to the irrep $\Gamma$. 
The states mostly contributing to each $\sigma_y$ peak are: 
(A) $n_{a_1}(i) = 0$, $n_{b_1}(2)=1$, $n_{b_1}(3) = 1$, 
and the fundamental excitation of the out-of-plane H bending;
(B) the states of the band A with $n_{a1} = 1$; 
(C, D) 
the states of the band A with $n_{a_1}(2) = 1$, with $n_{a_1}(5) = 1$, 
with $n_{a_1}(2) = 2$, as well as with the combination states
$(n_{a_1}(2) = 1, n_{b_1}(2) = 1)$  and $(n_{a_1}(2) = 1, n_{b_1}(3) = 1)$;  
(E) the states in A with $(n_{a_1}(2) = 1, n_{a_1}(5) = 1)$; 
(F) the states in A with $n_{a_1}(2) = 3$, and 
with $(n_{a_1}(2) = 2, n_{a_1}(5) = 1)$.
The listed ring excitations are accompanied by the 
\lq obligatory' excitations of the disappearing modes.  
}
\includegraphics[scale=0.25]{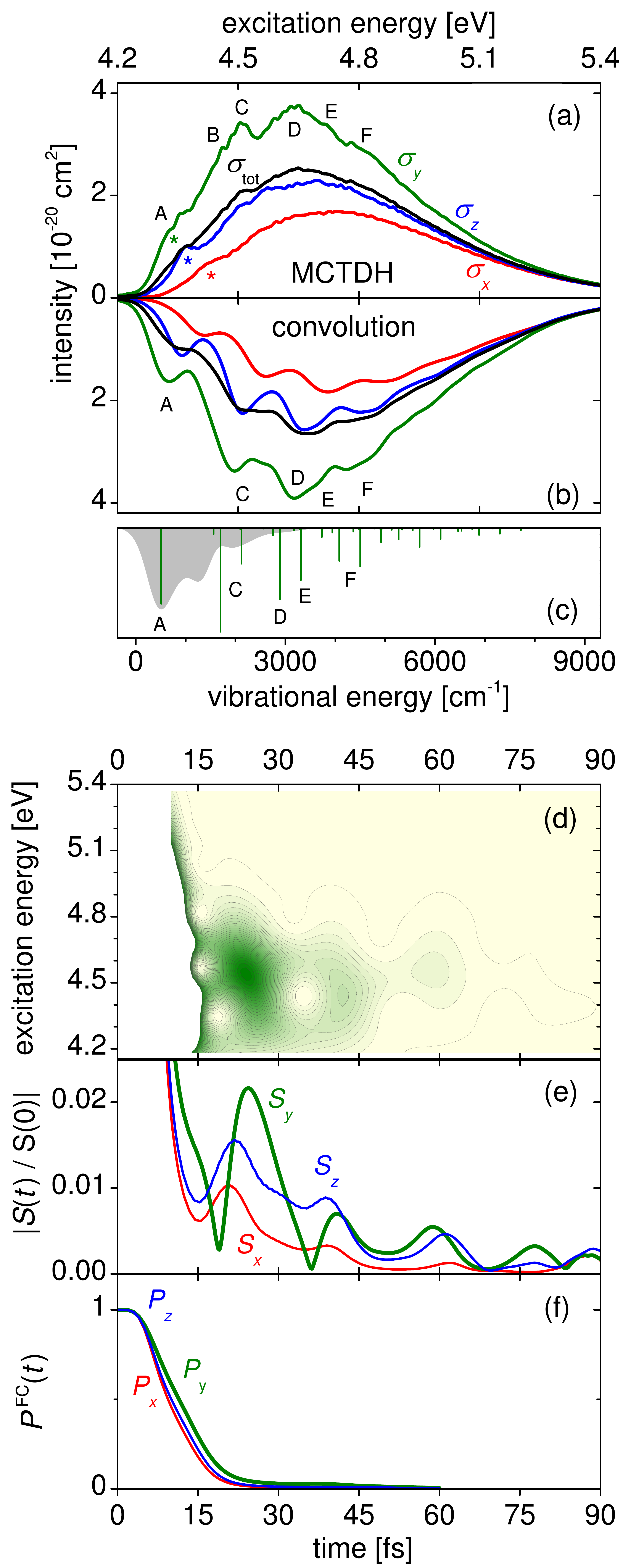}
\label{F04res}
\end{figure}

The spectra for the individual polarizations and the total spectrum 
$\sigma = \left( \sigma_x + \sigma_y + \sigma_z \right) / 3$ are shown in 
Fig. \ref{F04res}(a). As in all preceding calculations, 
the absorption $\sigma_y$ is the strongest (the maximum intensity is 
$3.7 \cdot 10^{-20}\,{\rm cm}^2$). It is about 2 times more intense than 
the weakest spectrum $\sigma_x$ (the maximum is 
$1.7 \cdot 10^{-20}\,{\rm cm}^2$). The contribution $\sigma_z$, absent in the
6D and 11D calculations, is intermediate between the two, with the 
peak intensity of $2.3 \cdot 10^{-20}\,{\rm cm}^2$. The total absorption 
reaches the maximum 
at 4.70\,eV, and the FWHM of the absorption envelope is 0.61\,eV. 
These main absorption parameters are in good agreement with the 11D
calculation. In particular, the true minimum of $V^X$ as well as the 
principal geometrical changes between $V^X$ and $V^{A_2}$ --- which are 
controlled by the totally symmetric modes --- are accurately described 
by the three included modes $Q_{a_1}(1,2,5)$. Further, 
all non-totally symmetric modes with large Herzberg-Teller coefficients in   
the TDM are also included, so that the
maximum total intensity of $2.5 \cdot 10^{-20}\,{\rm cm}^2$
is expected to be accurate. The 15D calculation is our
most reliable estimate of the absorption spectrum of the 
state $1^1\!A_2 (\pi\sigma^*)$. It illuminates one
of the intrinsic features of the $\pi\sigma^*$ photochemistry of model
biochromophores: The total spectra of the $\pi\sigma^*$ states result from the
contributions of several --- in this case three --- absorptions due to 
different spatial components of the TDM vector. 

The broad absorption background is structured by several broad
diffuse bands marked with letters A---F in Fig.\ \ref{F04res}(a). They are 
most conspicuous in the spectrum $\sigma_y$, and even there they are 
less pronounced than the vibrational bands in the 11D
calculation. The marked bands are 
assigned below using the convolution method. 

There is also a second
group of absorption lines in all spectra in Fig.\ \ref{F04res}(a). They
are narrow, densely spaced, and seen as \lq ripples' on the spectral
profiles. These bands have vibronic origin: They are 
found only for the coupled states $A_2/\widetilde{X}$ 
whereas the spectrum of the 
isolated state $1^1\!A_2(\pi\sigma^*)$ 
is smooth. The narrow \lq ripples' in 
Fig.\ \ref{F04res}(a) are Fano resonances which are due to the interference
of two diabatic dissociation pathways, the direct dissociation pathway
in the $A_2$ state and the indirect one involving vibronic 
transitions to the state $\widetilde{X}$ and back to $A_2$ at the CI
$\widetilde{X}/^1\!A_2$.
In fact, the \lq ripples' are associated with the bound vibrational states 
of the state $\widetilde{X}$ which gain intensity via the CI. 
This interference, which has a strong 
impact on the $\pi\sigma^*$ photochemistry
of pyrrole and affects the spectra and the photofragment distributions, 
has been  analyzed in Ref. \onlinecite{GP17A}.

The Fano effect is especially pronounced 
in the dynamics of the second excited state 
$1^1\!B_1(\pi\sigma^*)$. For the state $1^1\!A_2(\pi\sigma^*)$, discussed
in this paper, the coupling at the CI is weak (the population transfer between
$A_2$ and $\widetilde{X}$ is less than 10\%) and the diffuse bands
A---F are not affected by the state crossing and can be 
analyzed in terms of the state $1^1\!A_2(\pi\sigma^*)$ 
alone. The lowest vibrational states excited in the spectra 
$\sigma_x$, $\sigma_y$, and $\sigma_z$ and marked with asterisks in 
Fig.\ \ref{F04res}(a) are associated with one quantum excitations in the 
vibrations promoting the $1^1\!A_2 \leftarrow \widetilde{X}$ transition. 
The largest Herzberg-Teller coefficients are carried by the mode 
$\widetilde{Q}_{b_2}(3)$ (for $\sigma_x$; $\omega = 1321\,{\rm cm}^{-1}$), 
the out-of-plane bending $\theta$ 
(for $\sigma_y$; $\widetilde{\omega} = 616\,{\rm cm}^{-1}$), and the mode 
$\widetilde{Q}_{a_2}(3)$ 
(for $\sigma_z$; $\widetilde{\omega} = 969\,{\rm cm}^{-1}$). The energies
of the absorption origins are arranged according to the indicated frequencies
of the isolated $A_2$ state, i.e.
$E_y < E_z < E_x$. Moreover, the absorption maxima in the three spectra
follow the same order: The larger the
 frequency of the mode promoting the transition, 
the higher the energy of the absorption maximum. 
In the following, we concentrate on the diffuse vibrational bands in the 
spectrum $\sigma_y$ which are strong and which also have clear counterparts
in the total absorption spectrum. However, even this relatively simple 
spectrum is composite and consists of two contributions, 
the major one stemming from the
excitation of the disappearing modes via the TDM $\mu_y(\RR)$ 
and the weak excitation of  the ring via $\mu_y(\QQ)$. 

Photodissociation in the 15D calculation is a fast process. 
In particular, the autocorrelation functions $S_\epsilon(t)$, shown in Fig. 
\ref{F04res}(e), decay over two orders of
magnitude within $\sim 10$\,fs.
Much of this decay is due to the direct
dissociation. This is confirmed with the time dependent survival probabilities
$P_\epsilon^\mathrm{FC}(t)$ in panel (f). 
For the most intense $y$-polarized excitation, about 95\% of all molecules
dissociate directly with the time constant $T_1 = 11.7$\,fs which is 
almost identical with the 11D calculation. The remaining 5\% dissociate
indirectly with the exponential lifetime of $T_2 = 36.0$\,fs, which is 
somewhat longer than in the 11D case. 
These time constants can be cross validated
using the resonance lifetimes: The resonance states corresponding to the
diffuse bands A and F have longer 
lifetimes of 24\,fs and 30\,fs, respectively, while the intense
resonance states found under the bands C and D are short lived, with 
$\tau_{\rm res}(l)< 10$\,fs (see Table \ref{T00}). 
The theoretical time scales are slightly better aligned with 
the experimental values than in the low dimensional calculations. 
In particular, 
the correlations between $T_1$ and $\tau_d=12$\,fs\cite{WNSSAWS15}
 and between $T_2$ and 
$\tau_d=46$\,fs\cite{RWYCYUS13} 
become more apparent. Note that the 
resonance lifetimes, as well as
the measured lifetimes collected in   Table \ref{T00}, 
are non monotonic functions of the photon energy. 
Such fluctuations are often encountered and 
well understood in small polyatomic molecules.\cite{GSCHH03} The lifetimes  
fluctuate because eigenstates excited along different vibrational modes 
experience different effective coupling to the continuum. The observation 
of the lifetime fluctuations in pyrrole can therefore be a 
signature of the state- or even mode
specific photodissociation in this biochromophore.

The isotope effect in pyrrole-$d_1$ is evaluated using the populations
$P_y^\mathrm{FC}(t)$. The resonance states are not used because it becomes
nearly impossible to reliably match the resonance energies in two molecules.
The isotope effect in the 15D case follows the trend of the 
lower 
dimensional calculations. The direct dissociation time in pyrrole-$d_1$ is 
$T_1 = 13.0$\,fs giving $KIE \approx 1.1$. The indirect time is 55.4\,fs
so that in this case $KIE \approx 1.54$. As expected, the isotope effect 
diminishes with increasing dimensionality, and the tunneling contribution is
not very significant.  

Recurrences in the autocorrelation functions in Fig.\ \ref{F04res}(e) 
help to analyze the 
vibrational structure of the spectrum. Three short time recurrences are 
seen in $S_y(t)$ followed by a series of less regular
weak peaks stretching over several hundred femtoseconds
[not shown in panel (e)]. 
The strong short time recurrences are responsible for the 
main absorption peaks A---F while the weak ones are associated with the
long lived Fano resonances\cite{GP17A} 
which are beyond the scope of the present discussion. The relation between
the recurrence times and the absorption bands can be conveniently
visualized using the so-called spectrogram
\begin{equation}
\label{vibro}
S_y(E,\tau) = \int S_y(t) h(t-\tau) e^{iEt}\,dt \, ,
\end{equation}
which is a moving window Fourier transform of the autocorrelation function.
Here $h(t) = \exp(-t^2/2t_0^2)$ is the Gaussian window function.
In the spectrogram $S_y(E,\tau)$, shown in Fig.\ \ref{F04res}(d), 
the time and the frequency domains are represented in the same plot at
the expense of the resolution which is smeared along both 
axes:\cite{JK89,HHG95,VVS96} The time resolution $t_0$ in the plot is 5\,fs 
and corresponds to the energy resolution of 
$\hbar/t_0 = 1100$\,cm$^{-1}$. The spectrogram is dominated by 
three maxima. The largest maximum is found at 
$\tau = 24$\,fs and $E_{\rm ph} = 4.50$\,eV. The energy is close to 
that of the band C which can thus be associated with  
excitations of frequency $\omega \approx 2\pi/\tau \sim 1350$\,cm$^{-1}$. 
Several vibrational modes have fundamentals or overtones close to this
frequency, e.g. one quantum of the totally 
symmetric mode $\widetilde{Q}_{a_1}(2)$ or the out-of-plane bending mode. 
A two quantum excitation of $\widetilde{Q}_{b_1}(2)$ could be another 
possibility, but this non-totally symmetric mode has a zero equilibrium shift 
$\delta \widetilde{Q}$ and a tiny frequency change between  
the ground and the excited electronic states; its excitation is therefore
unlikely. The largest 
maximum in the spectrogram extends over an appreciable energy 
range. With growing energy the recurrence time becomes shorter so that
around 4.62\,eV (band D) $\tau \sim 21$\,fs correlates well with
the frequency of
$\widetilde{Q}_{a_1}(5)$. The other two maxima in the spectrogram are more 
localized in energy. One, seen at  
$\tau = 42$\,fs and $E_{\rm ph} = 4.42$\,eV, lies close to 
the band B, and the associated frequency is 
$\sim 800$\,cm$^{-1}$. One quantum of $\widetilde{Q}_{a_1}(1)$ 
or a two quantum excitation of the
non-totally symmetric mode $\widetilde{Q}_{b_1}(1)$
can account for this frequency. 
The other maximum is found at 
$\tau = 60$\,fs and $E_{\rm ph} = 4.58$\,eV, in the vicinity of the 
broad band D. The corresponding frequency of merely 560\,cm$^{-1}$ is too
small to be related to any pure excitation. 
More likely is that this low frequency is associated with a quantum beat 
between two surviving resonance states. A state with one quantum along
$\widetilde{Q}_{a_1}(5)$ and a state with two quanta along 
$\widetilde{Q}_{b_1}(1)$ is an example of such a pair. 

 The quantum mechanical 
spectra calculated using MCTDH are compared with the 
convolution approximation in Fig.\ 
\ref{F04res}(b). The convolution is applied to the single state $A_2$, and 
the spectrum constructed using the  convolution factors $\bar{\sigma}_R$ and 
$\bar{\sigma}_Q$ describes the diffuse vibrational bands
but not the Fano resonances. Because the
TDMs $\mu_x$ and $\mu_y$ are sums of two terms depending on the  
coordinates of $\RR$- and $\QQ$-space, two convolution terms are needed to 
approximate the spectra $\sigma_x$ and $\sigma_y$, as shown in 
Appendix \ref{appa}. The \lq spectrum' $\bar{\sigma}_R$, exemplified
in \ref{F04res}(c) for the initial state $\Phi_y(0)$, consists of the main
peak and a series of satellites arising as in the 11D case from the 
additional excitations
of the disappearing H atom bending ($n_\theta > 0$) and NH stretch 
($n_R > 0$). 
The widths of $\bar{\sigma}_R$ and $\bar{\sigma}_Q$ are 
$\sim 0.15\,{\rm eV}$ and $\sim 0.4\,{\rm eV}$, respectively: As in the 11D 
case, the total width of 0.65\,eV is mainly due to the progression in the 
ring modes of $a_1$ symmetry, 
whereas the dissociation induces a smaller broadening.

The convolution approximation correctly captures the shapes of the 
quantum mechanical absorption spectra, especially the low resolution 
envelopes. The intensities of the diffuse bands are clearly overestimated,
especially for the spectra $\sigma_x$ and $\sigma_z$. 
As in 11D, the lack of anharmonic coupling leads to
more narrow and better resolved bands. 
 On the other hand, the approximation correctly
reproduces the differences in the spectral resolution between 
the polarizations,
and returns $\sigma_y$
as the most structured and $\sigma_x$ as the least structured spectrum. 

The bands in the convolution and in the MCTDH spectra can be matched and  
related  to the vibrational states in the local $A_2$ minimum. 
The assignment of this high
dimensional spectrum is by no means straightforward, and we concentrate
on the spectrum $\sigma_y$ again. The convolution reproduces all
major bands, only the weak band B appears to be missing. 
This band is probably associated with the excitation of the 
$Q_{a_1}(1)$ mode which is highly correlated with the Jacobi coordinate $R$
and therefore is not well described by the convolution approach.
The vibrational states of the pyrrolyl ring contributing to the bands A---F
and and accompanied by the obligatory excitations in the $\RR$-space are
summarized in 
the caption to Fig.\ \ref{F04res}. 

The band origin A includes the vibrational states with one quantum of 
excitation in the out-of-plane bending and the ring modes of $b_1$ 
symmetry. The vibrational states contributing to the bands ${\rm B}-{\rm F}$ 
belong to several progressions involving excitations out of the
band origin.
Most of the vibrational activity registered in the convolution
approximation is due to the ring modes 
$Q_{a_1}(2)$ and $Q_{a_1}(5)$. We find up to three quanta of excitation in 
$Q_{a_1}(2)$ and up to two quanta in  $Q_{a_1}(5)$.
This is in agreement with the above analysis of 
the spectrogram, and with the assignment of 
the corresponding bands in the convolution factor $\bar{\sigma}_Q$ in panel
(c). The strong bands C and D mainly involve 
the states with one and two quanta of the totally 
symmetric modes $Q_{a_1}(2)$ and $Q_{a_1}(5)$, 
plus a contribution of the out-of-plane bending overtone. 
The weaker neighboring band E is also due to  excitations in 
$Q_{a_1}(2)$ complemented with $Q_{a_1}(5)$. 
Higher excitations in these modes contribute to
the band F. 

As in the 11D spectrum, the ring states are transposed to the higher
energy absorption bands via the additional excitations of the disappearing
modes. An example is the state $n_{a_1} = 1$ providing the main assignment of
the band B. With an additional bending excitation, this state contributes to 
the band C. This general property of the absorption bands of pyrrole acquires 
significance in the analysis of the wavelength resolved  
distributions of the final states of the pyrrolyl radical: 
Absorption bands based on the same ring states and 
differing only in the extent of excitation of the disappearing modes tend to
produce repetitive patterns in the photofragment distributions discussed in
paper II and observed in experiment.

\section{Conclusions}
\label{sum}
This paper analyzes the photodissociation mechanisms of pyrrole
promoted into the lowest excited $1^1\!A_2(\pi \sigma^*)$ 
state. The focus is on 
the absorption spectrum and the dissociation lifetimes. 
The summary our main findings is as follows: 

\begin{enumerate}
\item New 24 dimensional diabatic 
potential energy matrices of the ground electronic 
state $\widetilde{X}$ and the two lowest $\pi \sigma^*$ states are calculated
using CASPT2 method and a local diabatization at conical
intersections. Additionally, the coordinate dependent 
transition dipole moments necessary for the adequate description of the
excitation process into dark electronic states are provided. 
The full dimensional Hamiltonian for the H-atom abstraction
from the NH moiety attached to the pyrrolyl ring is constructed and used in the
quantum mechanical MCTDH calculations of the photodissociation.  

\item The ab initio calculations reproduce the 
known benchmarks for the excitation of the $\pi \sigma^*$ states
in pyrrole with reasonable accuracy. The quantum mechanical calculations
using 15 degrees of freedom predict the absorption of the state 
$1^1\!A_2(\pi \sigma^*)$ to reach the maximum of 
$\sim 3\cdot 10^{-20}$\,cm$^{2}$ near the photon energy of 4.70\,eV. 

\item The 
absorption spectra, including contributions due to 
different TDM components, are mildly structured, and the diffuse bands 
can be assigned to vibrational excitations belonging mainly to 
the pyrrolyl ring combined with excitations of the disappearing modes. 
The prominent diffuse 
structures are due to the resonance states supported
by the shallow local minimum of the state $1^1\!A_2(\pi \sigma^*)$. 

\item The 
calculated dissociation lifetimes, which range from $\sim 12$\,fs for the 
(major) direct decay component to $\sim 36$\,fs for the (minor) indirect 
component, are in good agreement with the experimental observations. 
Resonance calculations explain the experimentally 
observed non monotonic dependence of the lifetime on the excitation energy
as a signature of fluctuations due to 
the state specific photodissociation in pyrrole. 
Resonance states with lifetimes of over 100\,fs are also found, but their
intensity is very low in the calculations.  

\item The 
kinetic isotope effect, evaluated for the pyrrole-$d_1$, is small, 
$KIE \le 2.0$, and the tunneling contribution to the indirect
dissociation is minor. This is because of the low barrier along the minimum
energy path in the state $1^1\!A_2(\pi \sigma^*)$. 

\item For the calculations of the two coupled electronic states $\widetilde{X}$
and $1^1\!A_2(\pi \sigma^*)$, numerous Fano resonances are found even
for 15 dynamically active degrees of freedom illustrating the robustness
of  the Fano interference in pyrrole.\cite{GP17A}

\item A computationally efficient approximation method to calculate 
the absorption spectra is introduced which represents the total absorption
as a convolution of the structureless spectrum due to direct dissociation and
the Franck-Condon stick spectrum of the pyrrolyl ring modes. The convolution
approximation extends the familiar
Franck-Condon approach to the case of 
dissociating molecules.  
The approximation works best for the contributions of
the non-totally symmetric ring modes which have zero displacements between the
ground and the excited electronic state. The largest deviations are found
for the progressions in the totally symmetric ring modes (irrep $a_1$), 
but even in this case the main features of the quantum mechanical
spectra of pyrrole are faithfully reproduced. The convolution approximation
allows one to generate absorption spectra of dissociating molecules with
a minimum ab initio input, and is also an efficient tool to automatically 
assign the diffuse vibrational bands. 

\end{enumerate}

\appendix

\section{Quasi-diabatization based on the adiabatic Hessian matrices}
\label{appb}

In this Appendix we describe the procedure to obtain the diabatic 
states $\widetilde{X}$, $A_2$ and $B_1$ in the $\QQ$-space. 
We specialize the discussion to the definition of the diabatic coupling for 
the  $X/A_2$ conical intersection. 
A similar procedure is applied to the  $X/B_1$ intersection. 

In the $\QQ$-space, the  $X/A_2$ diabatic coupling is linear in the three 
normal modes of $a_2$ symmetry [cf. Eq. (\ref{V_coup})]:
\begin{equation}
W_Q^{X A_2}(\QQ|R) = \sum_{i = 1}^3 \lambda_i^{X A_2}(R) Q_{a_2}(i) \ ,
\end{equation}
where the $R$-dependent coupling strength is given by the Eq. (\ref{loclam})
\begin{equation}
\lambda_i^{X A_2}(R) = \lambda_{{\rm CI},i}^{X A_2} \exp\left(- \left| \frac{R - R_{\rm CI}^{XA_2}}{\Delta} \right|^n \right) \ .
\end{equation}
The diabatization involves the evaluation of five parameters ($\Delta$, $n$ and three pre-factors $\lambda_{{\rm CI},i}^{X A_2}$) and the calculation of the $R$-dependent diabatic Hessian matrices $\left\{ \gamma^\alpha_{ij}(R) \right\}$ in terms of the ab initio adiabatic Hessians $\left\{ \widetilde{\gamma}^\alpha_{ij}(R) \right\}$ ($\alpha = X, A_2$):
\begin{subequations}
\begin{equation}
 \gamma^X_{ij}(R) =  \widetilde{\gamma}^X_{ij}(R) + 2 \frac{\lambda_i^{X A_2}(R) \lambda_j^{X A_2}(R)}{U_\mathrm{Relax}^X(R) - U_\mathrm{Relax}^{A_2}(R)} 
  \label{AdDiaHess_1}
\end{equation}
\begin{equation}
 \gamma^{A_2}_{ij}(R)  =  \widetilde{\gamma}^{A_2}_{ij}(R) - 2 \frac{\lambda_i^{X A_2}(R) \lambda_j^{X A_2}(R)}{U_\mathrm{Relax}^X(R) - U_\mathrm{Relax}^{A_2}(R)}  \ .  
 \label{AdDiaHess_2}
\end{equation} 
 \label{AdDiaHess}
\end{subequations}
In the main text, $\gamma^{\alpha}_{ij}(R)$ are denoted 
$\left.\gamma^{\alpha}_{\Gamma,ij}(R)\right|_{\rm dia}$, and 
$\widetilde{\gamma}^{\alpha}_{ij}(R)$ are denoted 
$\left.\gamma^{\alpha}_{\Gamma,ij}(R)\right|_{\rm adia}$. 
In Eq. (\ref{AdDiaHess}), 
the one-dimensional profiles $U_\mathrm{Relax}^X(R)$ and $U_\mathrm{Relax}^{A_2}(R)$ are evaluated on the MEP of the $1^1\!A_2$ state. 
At the geometries of the MEP, 
the adiabatic and diabatic energies coincide (up to reordering)
whereas the adiabatic and diabatic Hessians with respect to the $a_2$ modes are different. The $R$-dependent Hessians $\left\{ \widetilde{\gamma}^X_{ij}(R) \right\}$ and $\left\{ \widetilde{\gamma}^{A_2}_{ij}(R) \right\}$ 
for the adiabatic states $\widetilde{X}$ and $^1\!A_2$, 
calculated at the geometries of the MEP, diverge in the 
proximity of the $X/A_2$ intersection, located at $R_{\rm CI}^{XA_2} = 5.57\,a_0$. The diabatic Hessians $\left\{ {\gamma}^X_{ij}(R) \right\}$ and $\left\{ {\gamma}^{A_2}_{ij}(R) \right\}$ are constructed by requiring that they are
finite and smooth as functions of $R$. 
This implies that the singularity of the adiabatic Hessians near the intersection is compensated by the divergent term $\sim 2\lambda_i \lambda_j / \left(U^X - U^{A_2} \right)$ of Eq. (\ref{AdDiaHess}).

\begin{figure}[h!]
\caption{The quantities $- \Delta \widetilde{\gamma}_{ii} \Delta U / 4$ of Eq. (\ref{DgDV}) (for $i=j$) calculated for the $a_2$ modes along the MEP on the $1^1\!A_2$ state. The plots of panels (a), (b) and (c) are obtained by using the diagonal adiabatic Hessian matrix elements $\widetilde{\gamma}^\alpha_{ii}$ for the modes $Q_{a_2}(1)$, $Q_{a_2}(2)$ and $Q_{a_2}(3)$, respectively. The red cross indicates the position of the conical intersection, where $- \Delta \widetilde{\gamma}_{ii} \Delta U / 4 = \left( \lambda_{{\rm CI},i}^{X A_2} \right)^2$. The values of $ \lambda_{{\rm CI},i}^{X A_2}$ are reported in red.}
\label{FigApp1}
\bigskip
\includegraphics[scale=0.33]{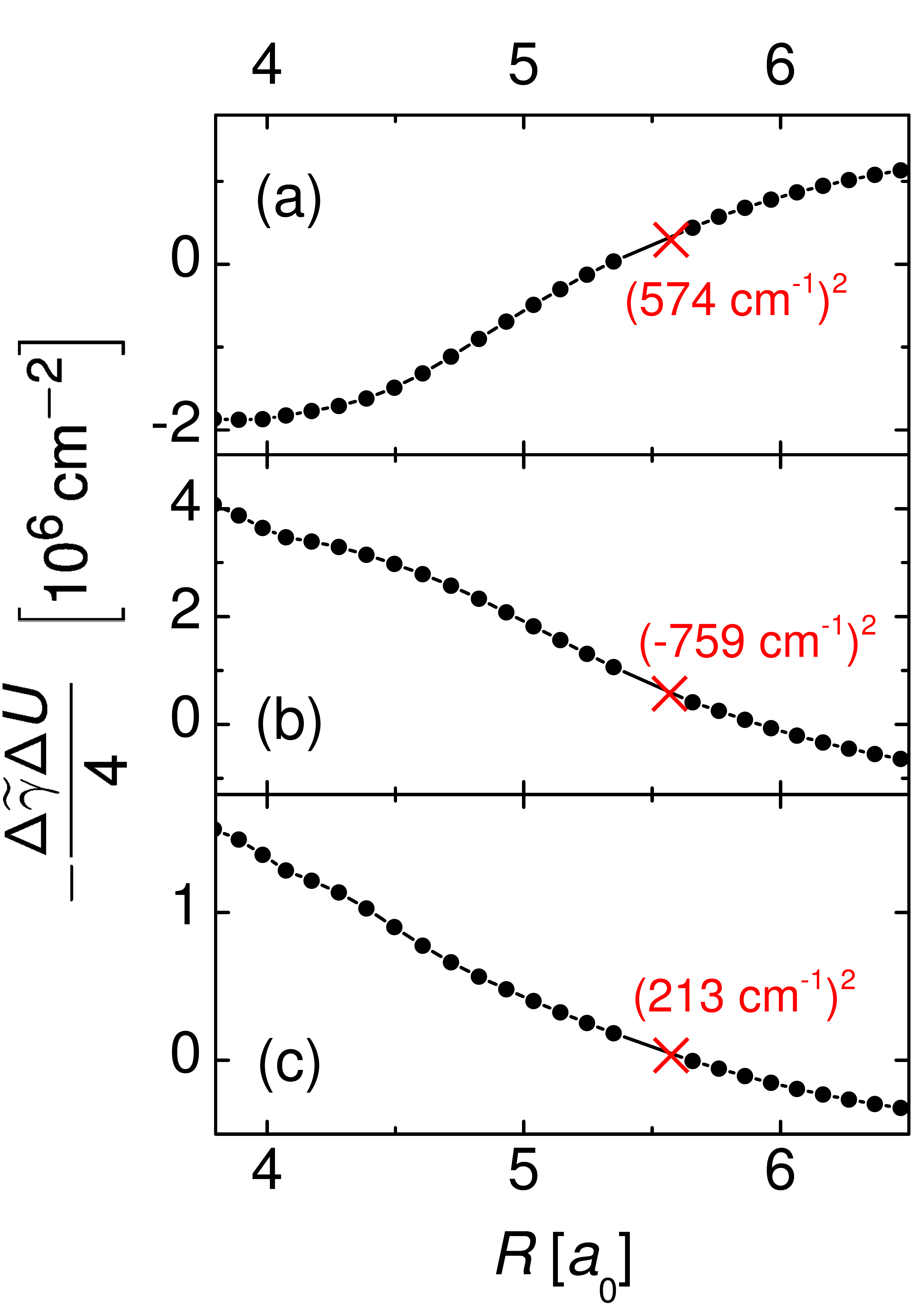}
\end{figure}

In order to find the pre-factors $\lambda_i^{X A_2}(R)$, we combine Eq. (\ref{AdDiaHess_1}) and (\ref{AdDiaHess_2}):
\begin{eqnarray}
-\frac{1}{4}\Delta \widetilde{\gamma}_{ij}(R) \Delta U(R)  & = & - \frac{1}{4} \left( \widetilde{\gamma}^X_{ij}(R) - \widetilde{\gamma}^{A_2}_{ij}(R) \right) \left(U^X_\mathrm{Relax}(R) - U^{A_2}_\mathrm{Relax}(R) \right) \nonumber \\ 
 & = & \lambda^{X A_2}_i(R) \lambda^{X A_2}_j(R) -  \frac{1}{4} \left( {\gamma}^X_{ij}(R) - {\gamma}^{A_2}_{ij}(R) \right) \left(U^X_\mathrm{Relax}(R) - U^{A_2}_\mathrm{Relax}(R) \right) \ . \nonumber \\
 \label{DgDV}
\end{eqnarray}
Since the diabatic Hessians are non-divergent, the parameters $\lambda_{{\rm CI},i}^{X A_2}$ are obtained by the limit
\begin{equation}
- \lim_{R \rightarrow R_{\rm CI}^{X A_2}} \frac{\Delta \widetilde{\gamma}_{ij}(R) \Delta U(R)}{4} = \lambda^{X A_2}_{{\rm CI},i} \lambda^{X A_2}_{{\rm CI},j} \ ,
\end{equation}
which is evaluated by interpolating the quantities $- \Delta \widetilde{\gamma}_{ij}(R) \Delta U(R) / 4$ calculated at the grid points of the MEP. 
Figure \ref{FigApp1} shows the diagonal terms  
$- \Delta \widetilde{\gamma}_{ii}(R) \Delta U(R) / 4$ as functions of $R$. 
Their values at the intersection are marked with red crosses and correspond to the squares of the coupling strengths at the intersection. The relative signs of the coupling strengths $\lambda_{{\rm CI},i}^{X A_2}$ are found by checking the signs of the off-diagonal terms $- \Delta \widetilde{\gamma}_{ij}(R) \Delta U(R) / 4$ at $R = R_{\rm CI}^{XA_2}$.

\begin{figure}[h!]
\caption{Adiabatic (orange dots) and diabatic (blue dots) Hessian matrix elements for the $1^1\!A_2$ electronic state with respect to the $a_2$ modes, calculated at the geometries of the MEP on the  $A_2$ surface. The continuous blue line depicts the Hessian functions used in the wave packet calculations. }
\label{FigApp2}
\bigskip
\includegraphics[scale=0.55]{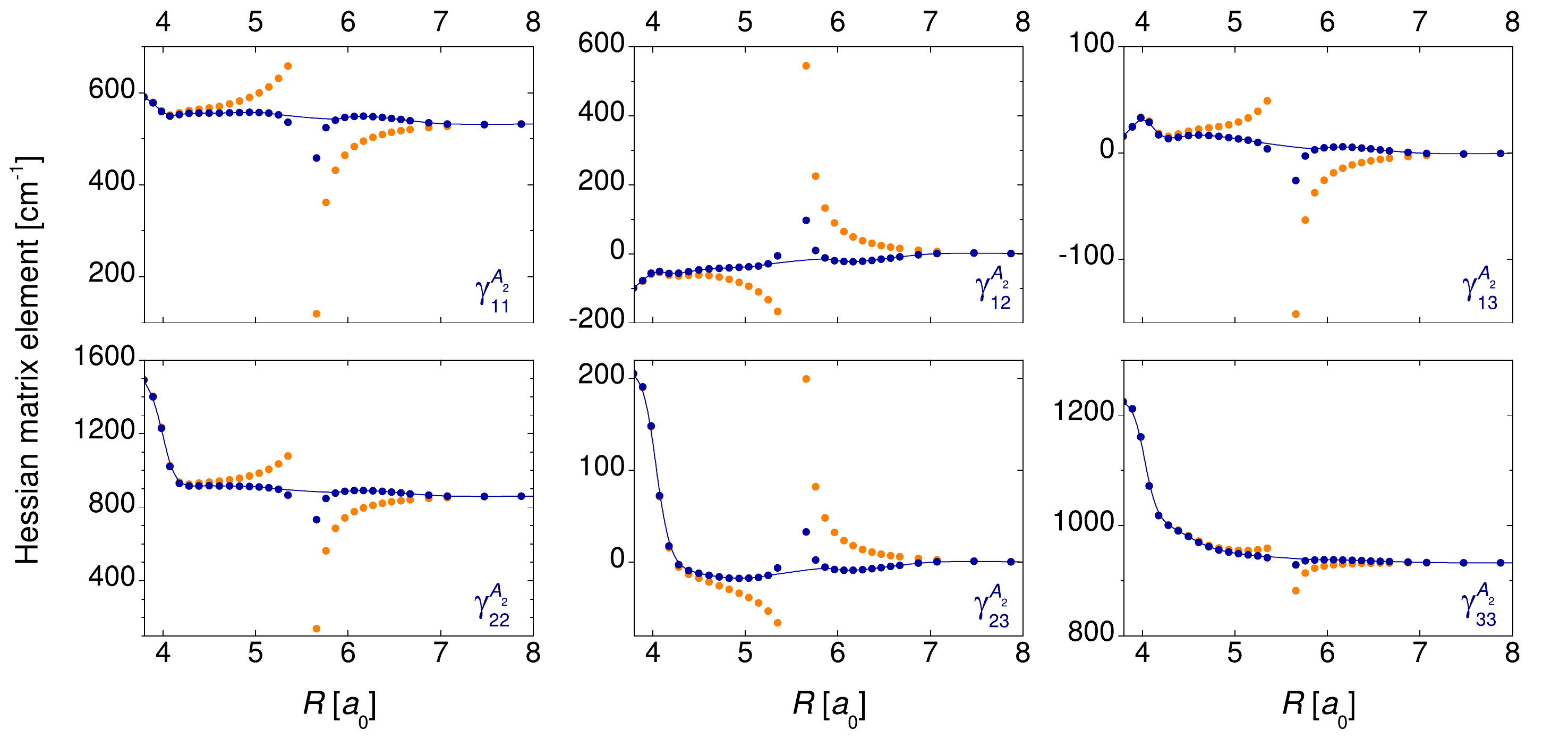}
\end{figure}

Once the pre-factors $\lambda_{{\rm CI},i}^{X A_2}$ are calculated, 
the parameters $\Delta$ and $n$ are tuned \lq by eye', in order to obtain a 
smooth $R$-dependence of the diabatic Hessians, which are calculated using 
Eq. (\ref{AdDiaHess}). Figure \ref{FigApp2} compares the adiabatic (orange dots) and the diabatic (blue dots) Hessian matrix elements for the 
$1^1\!A_2$ state along $R$. The diabatization removes the singularity of the
 ab initio adiabatic Hessian to a good extent. The resulting diabatic Hessian 
matrix elements vary smoothly with respect to $R$, except for a $0.5\,a_0$-wide strip around the intersection. The diabatic Hessians $\left\{ \gamma^X_{ij} \right\}$ and $\left\{ \gamma^{A_2}_{ij} \right\}$ are evaluated for a set of geometries of the MEP and interpolated for use in quantum mechanical calculations. Three points closest to the intersection 
are not included in the interpolation.

\section{Convolution approximation for a general form of the Herzberg-Teller TDM}
\label{appa}
In this Appendix, the convolution approximation described in Section \ref{FCconv} is developed for a more general TDM function,
\begin{equation}
\mu(\RR,\QQ) = \sum_i \mu_{R,i}(\RR) \mu_{Q,i}(\QQ) \ ,
\label{App_TDM}
\end{equation}
which includes Eqs. (\ref{mu_x_A2})-(\ref{mu_y_A2}) as special cases.

The initial state of the parent molecules is taken as in Eq. (\ref{InitState}), $\Psi_0(\RR,\QQ) \approx \Psi_R(\RR) \Psi_Q(\QQ)$, and the photoexcited state $\Phi(\RR,\QQ)$ is obtained by applying the TDM function of Eq. (\ref{App_TDM}),
\begin{eqnarray}
\Phi(\RR,\QQ) & = & \mu(\RR,\QQ) \Psi_0(\RR,\QQ) \nonumber \\
  & = & \sum_i \left(\mu_{R,i}(\RR) \Psi_R(\RR) \right) \left( \mu_{Q,i}(\QQ) \Psi_Q(\QQ) \right) \nonumber \\
  & = & \sum_i F_{R,i}(\RR) f_{Q,i}(\QQ)  \ .
  \label{App_InitState}
\end{eqnarray}
Equation (\ref{App_InitState}) is a
generalization of the simple initial state of 
Eq. (\ref{InitState3}).

Using the separability approximation of Eq. (\ref{Ham1}) for the 
molecular Hamiltonian, an expression akin to Eq. (\ref{auto1}) 
is obtained for the autocorrelation function:
\begin{eqnarray}
S(t) & \approx & \sum_{ij} \left\bra F_{R,i} \left| \exp\left(-i \hat{H}_R t \right) \right| F_{R,j} \right\ket_R \left\bra f_{Q,i} \left| \exp\left(-i \hat{H}_Q t \right) \right| f_{Q,i} \right\ket_Q \nonumber \\
 & = & \sum_{ij} s_R^{ij}(t) s_Q^{ij}(t) \ .
 \label{App_acf}
\end{eqnarray}
where the cross correlation functions $s_R^{ij}(t)$ and $s_Q^{ij}(t)$
are built using the expansion terms $i$ and $j$ in Eq.\ (\ref{App_InitState}).
They are used to define the spectral functions
\begin{eqnarray}
\bar{\sigma}_R^{ij}(E) & = & \int_{-\infty}^\infty s_R^{ij}(t) {\rm e}^{i Et} {\rm d}t \nonumber \\
\bar{\sigma}_Q^{ij}(E) & = & \int_{-\infty}^\infty s_Q^{ij}(t) {\rm e}^{i Et} {\rm d}t \ .
\end{eqnarray}
The absorption spectrum, given by the Fourier transform of the autocorrelation function of Eq. (\ref{App_acf}), is finally given as
\begin{equation}
\sigma(\Eph) = \frac{\Eph}{2 \epsilon_0 c} \sum_{ij} \int_{-\infty}^\infty \bar{\sigma}_R^{ij}(\Eph - \omega) \bar{\sigma}_Q^{ij}(\omega) {\rm d}\omega \ .
\label{App_spectrum}
\end{equation}
Eq. (\ref{App_spectrum}) is the extension of the convolution approximation
to the TDM in the general form of Eq. (\ref{App_TDM}).

The calculations of the contributions stemming from the $\RR$- and
$\QQ-$spaces are performed separately.  
The initial functions $F_{R,i}(\RR)$ are propagated in the $\RR$-space 
under the action of the Hamiltonian $\hat{H}_R$ 
and used to evaluate the cross-correlation functions $s_R^{ij}(t)$ 
whose Fourier transform gives the spectral functions $\bar{\sigma}_R^{ij}(E)$.
The spectral functions $\bar{\sigma}_Q^{ij}(E)$ for the harmonic Hamiltonian 
$\hat{H}_Q$ are given by the expression akin to Eq.\,(\ref{spectrum4}):
\begin{equation}
\bar{\sigma}_Q^{ij}(E) = \sum_{\bm m} \left. \left\bra f_{Q,i}(\QQ) \right| \phi_{\bm m}(\QQ) \right\ket  \left\bra \phi_{\bm m}(\QQ) \left| f_{Q,j}(\QQ) \right\ket \right. \delta(E - E_{\bm m}) \ ,
\end{equation}
and depend on the analytic FC overlap integrals 
$\left\bra f_{Q,i}(\QQ) | \phi_{\bm m}(\QQ) \right\ket$ for the ring modes. 

\vspace{-0.7cm}
\begin{acknowledgments}
S.Yu.G. acknowledges the financial support by the  Deutsche
  Forschungsgemeinschaft. 
\end{acknowledgments}


\begin{thebibliography}{10}
\bibitem{NOTE-PYR01A-01}
{\em {\rm Note that the labeling of the molecular orbitals refers to the
  $C_{\rm 2v}$ geometries only, and the given assignments are, strictly
  speaking, correct in the FC zone.}}
\end{thebibliography}

\clearpage
\newpage
\begin{table}
\caption{
Properties of the CASPT2 potential energy surfaces 
 of the three lowest electronic states of pyrrole compared with the available 
experimental data: Band origins
$T_0$ which include zero-point energies  of the ground
and the excited electronic states; the 
quantum mechanical thresholds $D_0$ for the diabatic
electronic channels; the potential barrier heights for the MEP, 
$E^\ddagger_{\rm MEP}$, and for the ring modes fixed to pyrrole
equilibrium, $E^\ddagger_{\rm pyr}$.  All energies are in eV.} 
\label{T01}
\begin{ruledtabular}
\begin{tabular}{cccccccc}
  &  &  &  &  \vspace{-0.5cm} \\
Diabatic state & $T_0$  & $T_0$ (exp) & $E^\ddagger_{\rm MEP}$& 
$E^\ddagger_{\rm pyr}$& Dissociation channel & $D_0$ & $D_0$ (exp.)  \\
\hline
$\widetilde{X}^1\!A_1(\pi\pi)$ & 0.0 & 0.0 & $-$  & $-$ &  
   ${\rm H}(^1S)/{\rm pyrrolyl}(1 ^2\!A_1)$          & 5.09 & $-$ \\
  &  &  &  &   \\
$1 ^1\!A_2(\pi\sigma^*)$   & 4.32 & $\le 4.88$\tablenotemark[1]& 0.09& 0.22 & 
            ${\rm H}(^1S)/{\rm pyrrolyl}(1 ^2\!A_2)$     & 3.40 & 4.07\tablenotemark[1] \\
  &  &  &  &   \\
$1 ^1\!B_1(\pi\sigma^*)$    & 5.30 & 5.86\tablenotemark[2] &  0.08 & 0.05 & 
           ${\rm H}(^1S)/{\rm pyrrolyl}(1 ^2\!B_1)$ 
& 3.96 & $4.62-4.67$\tablenotemark[3] \\

\tablenotetext[1]{Ref. \onlinecite{CNQA04}.}
\tablenotetext[2]{Ref. \onlinecite{PWG98}.}
\tablenotetext[3]{DFT\cite{GIHKBL04} and MRCI\cite{MLWD06} methods estimate the difference in the threshold energies $D_0(1 ^2\!B_1) - D_0(1 ^2\!A_2)$ to be in the range of 0.55\,eV---0.60\,eV.}
\end{tabular}
\end{ruledtabular}
\end{table}


\begin{sidewaystable}
\caption{
Computed vertical excitation energies $T_v$ 
and classical diabatic dissociation thresholds $D_e$
of the states $1 ^1\!A_2(\pi\sigma^*)$ 
and $1 ^1\!B_1(\pi\sigma^*)$ compared with the 
previously published theoretical data (references are given in square
brackets). Energies are in eV. 
}
\label{T01A}
\begin{ruledtabular}
\begin{tabular}{cccccccccccc}
  &       &      &      &     &      &      &     &      &       &     &     \\
State & Property & 
\scriptsize \shortstack{CASSCF \\ ($8_e,7_o$) \\ AVDZ+\,[\onlinecite{VLMSD05}]} &
\scriptsize \shortstack{CASPT2 \\ ($8_e,7_o$) \\ AVTZ+} &
\scriptsize \shortstack{CASPT2 \\ ($8_e,9_o$) \\ AVTZ+} &
\scriptsize \shortstack{CASPT2 \\ ($8_e,8_o$) \\ AVDZ+\,[\onlinecite{NW14}]} &
\scriptsize \shortstack{MRCI \\ ($8_e,9_o$) \\ AVTZ+} &
\scriptsize \shortstack{MRCI \\ ($6_e,6_o$)  \\ DAVDZ\,[\onlinecite{BVAEML06}]} &
\scriptsize \shortstack{MRCI \\ ($6_e,5_o$) \\ AVDZ\,[\onlinecite{FVSELK11}]} &
\scriptsize \shortstack{EOM-CCSD \\ \  \\ AVTZ+} &
\scriptsize \shortstack{EOM-CCSD \\ \  \\ \,[\onlinecite{LP10}]} &
\scriptsize \shortstack{EOM-CCSD \\ \  \\ \,[\onlinecite{TS97}]} \\
  &       &      &      &     &      &      &     &      &       &     &     \\
\hline

  & $T_v$ &  4.45 &4.80 & 4.87 & 5.06& 4.92 & 5.09& 5.33 &  5.21 & 5.18&4.99 \\
                   
$1^1\!A_2(\pi\sigma^*)$   
  &       &       &     &      &     &      &     &      &       &     &     \\
  & $D_e$ &  3.96 &3.82 & 3.90 & 3.17 & 4.00 &  -  &   -  &   -   &  -  &  -  \\
                   
  &       &       &     &      &     &      &     &      &       &     &     \\
  & $T_v$ &  5.03 &5.45 & 5.67 & 5.86& 5.53 & 5.86& 6.12 &  5.96 & 5.84&5.99 \\
                   
$1 ^1\!B_1(\pi\sigma^*)$  
  &       &       &     &      &     &      &     &      &       &     &     \\
  & $D_e$ &  4.09 &4.31 & 5.11 & - & 4.70 &-  &   -  &   -   &  -  &  -  \\
  &       &       &     &      &     &      &     &      &       &     &     \\
\end{tabular}
\end{ruledtabular}
\end{sidewaystable}

\begin{sidewaystable}
\caption{
Harmonic CASPT2 frequencies ($\mathrm{cm^{-1}}$) of the 
vibrational modes $Q_{\Gamma}(i)$ in the pyrrolyl minimum in increasing order
for four irreps of the $C_{2v}$ group. 
Also shown are the harmonic frequencies of pyrrole in the FC zone including 
the local minimum of
the state $1^1\!A_2(\pi\sigma^*)$ and the global minimum of the 
 state $\widetilde{X}$. Entries denoted $R$, $\theta,\phi = 0^\circ$,  
and $\theta,\phi = 90^\circ$ refer to the disappearing modes of pyrrole 
[NH stretching vibration, out-of-plane bending, and in-plane bending, 
respectively]. Available experimental values, taken from 
Ref.\ \onlinecite{MLH01}, are shown in parenthesis.
Herzberg-Teller coefficients of the TDMs with $\widetilde{X}$
in the FC zone, $\mu_{x,y,z}^{A_2}$ and $\mu_{x,y,z}^{B_1}$, are given 
(in atomic units) in the last three columns for each symmetry block. 
}
\label{T02}
\begin{ruledtabular}
\begin{tabular}{ccccccc|ccccccc}
\hline
 \multicolumn{7}{c}{$a_1$ modes} & \multicolumn{7}{c}{$b_2$ modes} \\
 \hline
Mode & 
\makecell{ pyrrolyl\\minimum}  & 
\makecell{pyrrole local\\ $^1\!A_2$  minimum} & 
 \makecell{pyrrole\\minimum} & 
$\mu_{x,i}^{B_1}$ & $\mu_{y,i}$ & $\mu_{z,i}$  & 
Mode &
\makecell{ pyrrolyl\\minimum}  & 
\makecell{pyrrole local\\ $^1\!A_2$ minimum} & 
 \makecell{pyrrole\\minimum} & 
$\mu_{x,i}^{A_2}$ & $\mu_{y,i}$ & $\mu_{z,i}$  
\\
\hline
$Q_{a_1}(1)$ &   932  &  908  &939 (882)   & 0.0074      &-&-& $Q_{b_2}(1)$ &   710  &   783 &    930 (866)  & -0.0223     &-&-\\ 
$Q_{a_1}(2)$ &   1106 &  1124 &1230 (1148) & 0.0042      &-&-& $Q_{b_2}(2)$ &   997  &  1045 &  1143 (1049)  & -0.0041     &-&-\\ 
$Q_{a_1}(3)$ &   1163 &  1155 &1089 (1017) & 0.0066      &-&-& 
$\substack{\theta \\ \phi=90^\circ}$&    -   &  1029 &  1200 (1134)  & 0.3200 &-&-\\ 
$Q_{a_1}(4)$ &   1268 &  1255 &1164 (1075) & 0.0004      &-&-& $Q_{b_2}(3)$ &  1156  &  1321 &  1585 (1424)  &  0.0311     &-&-\\ 
$Q_{a_1}(5)$ &   1558 &  1584 &1509 (1401) & -0.0127     &-&-& $Q_{b_2}(4)$ &  1408  &  1421 &  1420 (1288)  &  0.0015     &-&-\\ 
$Q_{a_1}(6)$ &   1655 &  1659 &1615 (1472) & -0.0192     &-&-& $Q_{b_2}(5)$ &  1474  &  1545 &  1670 (1519)  &  0.0166     &-&-\\ 
$Q_{a_1}(7)$ &   3369 &  3385 &3383 (3128) & -0.0002     &-&-& $Q_{b_2}(6)$ &  3362  &  3382 &  3383 (3119)  & -0.0111     &-&-\\ 
$Q_{a_1}(8)$ &   3396 &  3409 &3417 (3149) & 0.0045      &-&-& $Q_{b_2}(7)$ &  3376  &  3395 &  3404 (3143)  & 0.0102      &-&-\\ 
$R$         &    -   &  2608 &3915 (3531) & -0.1202     &-&-&             &        &       &               &             & & \\
\hline
 \multicolumn{7}{c}{$b_1$ modes} &  \multicolumn{7}{c}{$a_2$ modes} \\
 \hline
Mode & 
\makecell{ pyrrolyl\\minimum}  & 
\makecell{pyrrole local\\ $^1\!A_2$ minimum} & 
 \makecell{pyrrole\\minimum} & & $\mu_{y,i}^{A_2}$ & $\mu_{z,i}^{B_1}$ & 
Mode &
\makecell{ pyrrolyl\\minimum}  & 
\makecell{pyrrole local\\ $^1\!A_2$  minimum} & 
 \makecell{pyrrole\\minimum} & $\mu_{x,i}$ & $\mu_{y,i}^{B_1}$ & $\mu_{z,i}^{A_2}$ 
\\  
\hline
$\substack{\theta \\ \phi=0^\circ}$ 
&   - & 616 &  362 (475)  &-&-0.33 & -0.35 & $Q_{a_2}(1)$ & 533 &542 & 636 (614)  &-& 0.0046 &  -0.0231 \\  
 $Q_{b_1}(1)$ & 586 & 533 &  634 (620)  &-& 0.0016   &  -0.0360  & $Q_{a_2}(2)$ & 861 &895 & 715 (692)  &-& 0.0033 &  0.0258  \\  
 $Q_{b_1}(2)$ & 757 & 785 &  763 (722)  &-&-0.0263   & -0.0364   & $Q_{a_2}(3)$ & 932 &969 & 909 (864)  &-& 0.0095 &  0.0340  \\  
 $Q_{b_1}(3)$ & 867 & 941 &  878 (827)  &-&-0.0223   &  0.0186   &             &     &    &            & &        &          \\  

\end{tabular}
\end{ruledtabular}
\end{sidewaystable}

\begin{sidewaystable}
\caption{
Summary of the quantum mechanical calculations discussed in this work. 
Different calculations are 
distinguished by the total number of included coordinates. Three Jacobi
coordinates $(R,\theta,\phi)$ are always included and not listed.
For each calculation, the following parameters are specified: The included
electronic states; the included
normal modes of pyrrolyl; the TDMs $\mu_\epsilon$ used in the construction of
the initial state $\Phi_\epsilon(0) 
= \mu_\epsilon \Psi_0$ ($\epsilon = x, y, z$;
$\Psi_0$ 
is the ground vibrational state in $\widetilde{X}$ calculated separately for the
specified set of coordinates); the vertical excitation energy $T_v$ for the
specified set of coordinates; 
the maximum intensity of the calculated spectrum.}
\label{T06}
\begin{ruledtabular}
\begin{tabular}{rccccccc}
\hline
  & \shortstack{\ \\Included\\states} & 
  \shortstack{\ \\Included\\normal modes} & 
\shortstack{$\mu_x$ \\ \ }& 
\shortstack{$\mu_y$ \\ \ }& 
\shortstack{$\mu_z$\\ \ }& 
\shortstack{$T_v$ [eV] \\ \ } & 
\shortstack{\ \\Max. intensity\\ \ [$\mathrm{10^{-20} \ cm^2}$]} \\
 \hline
  &   &    &  &  &  & & \\
6D & $1^1\!A_2(\pi\sigma^*)$ & $Q_{b_1}(1,2,3)$ & 
$\mu_{x,\theta}^{A_2} \sin \theta \sin \phi$ & 
$\begin{array}{r}
\mu_{y,\theta}^{A_2} \sin \theta \cos \phi \\ + 
\displaystyle{\sum_{i=1,2,3}} \mu^{A_2}_{y,i} Q_{b_1}(i)
\end{array}$ 
& -& 4.19 & 4.0 \\
  &   &    &  &  &  & & \\
 \hline
  &   &    &  &  &  & & \\
11D & $1^1\!A_2(\pi\sigma^*)$ &  $Q_{a_1}(1,...,8)$ & 
$\mu_{x,\theta}^{A_2} \sin \theta \sin \phi$  & 
$\mu_{y,\theta}^{A_2} \sin \theta \cos \phi $ 
& -& 4.80 & 1.0  \\
  &   &    &  &  &  & & \\
 \hline
  &   &    &  &  &  & & \\
15D & $\begin{array}{c}\widetilde{X}^1\!A_1(\pi\pi)\\
1^1\!A_2(\pi\sigma^*) 
\end{array}$ & 
$\begin{array}{c}
Q_{a_1}(1,2,5) \\ Q_{a_2}(1,2,3) \\ Q_{b_1}(1,2,3) \\ Q_{b_2}(1,3,5)
\end{array}$ & 
$\begin{array}{r}
\mu_{x,\theta}^{A_2} \sin \theta \sin \phi \\ + 
\displaystyle{\sum_{i=1,3,5}} \mu_{x,i}^{A_2} Q_{b_2}(i)
\end{array}$ &
$\begin{array}{r}
\mu_{y,\theta}^{A_2} \sin \theta \cos \phi \\+ 
\displaystyle{\sum_{i=1,2,3}} \mu_{y,i}^{A_2} Q_{b_1}(i)
\end{array}$ & 
$\displaystyle{\sum_{i=1,2,3}} \mu_{z,i}^{A_2} Q_{a_2}(i)$ 
& 4.72  & 2.5 \\
  &   &    &  &  &  & & \\
\end{tabular}
\end{ruledtabular}
\end{sidewaystable}


\begin{table}[t!]
\caption{Computational details of the MCTDH calculations. 
$N_i,N_j,N_k$ are the number of primitive DVR functions used for each 
particle. $n_X$ and $n_{A_2}$ are the number of single-particle functions 
used for the $\widetilde{X}$ and $^1\!A_2$ states. 
The 6D and 11D calculations include only the $^1\!A_2$ state.
The DVR type HO stands for the harmonic oscillator DVR. 
}
\label{T07}
\begin{ruledtabular}
\begin{tabular}{lcccc}
\hspace{1cm} & \ \ \ \ \ \ Particle \ \ \ \ \ \  &  DVR type & \ \ \ \ \ $N_i,N_j,N_k$ \ \ \ \ \  & \ \ \ $n_X, n_{A_2}$ \ \ \ \\
 \hline
6D & \\
 & $R$ & sine & 98 & 5 \\
 & $(\theta,\phi)$ & 2D Legendre & 71, 21 & 5 \\
 & $Q_{b_1}(1,2,3)$ & HO, HO, HO & 17  & 4 \\
 & & & & \\
11D & \\
 & $R$ & sine & 98 & 9 \\
 & $(\theta,\phi)$ & 2D Legendre & 71, 21 & 7 \\
 & $Q_{a_1}(1,2)$ & HO, HO & 37, 29 & 7 \\
 & $Q_{a_1}(3,4)$ & HO, HO & 21, 21 & 5 \\
 & $Q_{a_1}(5,6)$ & HO, HO & 25, 21 & 4 \\
 & $Q_{a_1}(7,8)$ & HO, HO & 21, 21 & 2 \\  
 & & & & \\
 15D & \\
 & $R,Q_{a_1}(1)$ & sine, HO& 65, 37 & 23, 9 \\
 & $(\theta,\phi)$ & 2D Legendre & 61, 19 & 19, 6 \\
 & $Q_{a_1}(2,5)$ & HO, HO & 29, 25 & 16, 5 \\
 & $Q_{a_2}(1,2,3)$ & HO, HO, HO & 17, 17, 17 & 7, 4 \\
 & $Q_{b_1}(1,2,3)$ & HO, HO, HO & 17, 17, 17 & 4, 3 \\
 & $Q_{b_2}(1,3,5)$ & HO, HO, HO & 17, 17, 17 & 5, 3 \\

\end{tabular}
\end{ruledtabular}
\end{table}

\begin{table}
\caption{Experimental and calculated photodissociation lifetimes of pyrrole. 
For the experimental lifetimes $\tau_{\rm exp}$, 
the excitation wavelength $\lambda$ and the corresponding 
energy $\Delta E_{\rm exp} = E_{\rm ph} - T_0$ 
above the band origin $T_0 = 4.88$\,eV  (254\,nm) 
are given. References to experimental work are given in square brackets. 
The lifetimes $\tau_{\rm res}$, drawn from the resonance
calculations [Eq.\ (\ref{life1})], are shown for the energies
$\Delta E_{\rm calc} = E_{\rm res} - T_0$ above the band origin $T_0$ in the
calculation. In the case of the 
lifetimes $T_1$ and $T_2$ estimated via the
survival probability of Eq.\ (\ref{fit1}) 
for the direct and indirect dissociation, the resonance energy 
$E_{\rm res}$ is replaced with the energy 
of the absorption band maximum.
} 
\label{T00}
\begin{ruledtabular}
\begin{tabular}{ccc|cccc}
\multicolumn{3}{c|}{Experiment} &  
\multicolumn{4}{c}{Calculation} \\
 \hline
$\lambda$ [nm] & $\Delta E_{\rm exp}$ [eV] & $\tau_{\rm exp}$ [fs]& 
$\Delta E_{\rm calc}$ [eV ]& $T_1$ [fs]& $T_2$ [fs]& $\tau_{\rm res}$ [fs]\\
 \hline
250 & 0.08 &\makecell{110 [\onlinecite{LRHR04}]\\126 [\onlinecite{RWYCYUS13}]} 
& $0.0-0.02$ & 6.5\tablenotemark[1] & 6.5\tablenotemark[1] & 
\makecell{8\tablenotemark[1]\\24\tablenotemark[3]}   \\
249.5 & 0.09 & 39 [\onlinecite{KPNWF17}] & 
$0.10$ &  &   & 
8\tablenotemark[2]   \\
245 & 0.18 & 54 [\onlinecite{KPNWF17}] & 
$0.20$ &  &  &  
$< 10 $\tablenotemark[3]   \\
242   & 0.24 & 12  [\onlinecite{WNSSAWS15}] &
$0.27$ &  &  & 
$30 $\tablenotemark[2]   \\
240 & 0.29 & 22 [\onlinecite{KPNWF17}] & 
$0.29$ & 12.8\tablenotemark[2] & 23.8\tablenotemark[2] & 
 \\
238  &  0.33 &  46 [\onlinecite{RWYCYUS13}] &  
$0.33$ & 11.7\tablenotemark[3] & 36.0\tablenotemark[3] & 
$< 10 $\tablenotemark[3]   \\
236   & 0.37  &  19  [\onlinecite{WNSSAWS15}] & 
$0.52$ &  &  & 
$30$\tablenotemark[3] 

\tablenotetext[1]{6D calculation. The band origin is $T_0 = 4.07$\,eV.
The band maximum is 4.1\,eV.}
\tablenotetext[2]{11D calculation. The band origin is $T_0 = 4.39$\,eV.
The band maximum is 4.71\,eV.}
\tablenotetext[3]{15D calculation. The band origin is $T_0 = 4.31$\,eV.
The band maximum is 4.65\,eV.}
\end{tabular}
\end{ruledtabular}
\end{table}


\clearpage
\newpage



\end{document}